\documentclass[preprint,12pt]{elsarticle}

\usepackage{amssymb}
\usepackage{mathtools}
\usepackage{lineno}

\usepackage[margin=2cm]{geometry}
\usepackage{setspace}
\usepackage{hyperref}
\hypersetup{
    colorlinks,
    linkcolor={red!50!black},
    citecolor={blue!50!black},
    urlcolor={blue!80!black}
}
\usepackage[font=footnotesize,labelfont=bf,subrefformat=parens]{subcaption}
\usepackage{enumerate}
\usepackage{siunitx}
\usepackage{gensymb}

\allowdisplaybreaks
\journal{Mathematics and Computers in Simulation}

\begin{document}

\newcommand{\vf}[1]{\underline{#1}}
\newcommand{\tf}[1]{\underline{\underline{#1}}}
\newcommand{\pd}[2]{\frac{\partial #1}{\partial #2}}
\newcommand{\ucd}[1]{\overset{\scriptscriptstyle \triangledown}{\tf{#1}}}
\newcommand{\smashedunderbrace}[2]{\smash{\underbrace{#1}_{#2}}\vphantom{#1}}
\newcommand{\Tr}[1]{#1^{\mathrm{T}}}
\newcommand{\mf}[1]{\mathbf{#1}}

% Dimensionless numbers
\newcommand\Rey{\operatorname{\mathit{Re}}}
\newcommand\Str{\operatorname{\mathit{Sr}}}
\newcommand\Wei{\operatorname{\mathit{Wi}}}
\newcommand\Deb{\operatorname{\mathit{De}}}
\newcommand\Bin{\operatorname{\mathit{Bn}}}

\begin{frontmatter}



% \title{Further investigation of Green-Gauss and Least-Squares gradient reconstruction schemes}
% \title{An overview of gradient reconstruction schemes for Finite Volume methods}
%\title{A large family of gradient reconstruction schemes based on projections}
\title{A unification of least-squares and Green-Gauss gradients under a common projection-based 
gradient reconstruction framework}

\author[1]{Alexandros Syrakos\corref{cor1}}
\ead{syrakos.alexandros@ucy.ac.cy}
\cortext[cor1]{Corresponding author}

\author[2]{Oliver Oxtoby}

\author[2]{Eugene de Villiers}

%\author[3]{Ganesh Natarajan}

\author[3]{Stylianos Varchanis}

\author[3]{Yannis Dimakopoulos}

\author[3]{John Tsamopoulos}

\affiliation[1]{organization={Department of Mechanical and Manufacturing Engineering, University of Cyprus},
            addressline={P.O. Box 20537},
            city={Nicosia},
            postcode={1678},
            state={},
            country={Cyprus}}

\affiliation[2]{organization={Engys},
            addressline={Studio 20, Royal Victoria Patriotic Building, John Archer Way},
            city={London},
            postcode={SW18 3SX},
            state={},
            country={U.K.}}

% \affiliation[3]{organization={Department of Mechanical Engineering, Indian Institute of Technology Palakkad},
%             country={India}}

\affiliation[3]{organization={Laboratory of Fluid Mechanics and Rheology, Dept. of Chemical 
Engineering, University of Patras},
            addressline={},
            city={Patras},
            postcode={26500},
            state={},
            country={Greece}}
            
% \author{}
% 
% \affiliation{organization={},%Department and Organization

\begin{abstract}
We propose a family of gradient reconstruction schemes based on the solution of over-determined 
systems by orthogonal or oblique projections. In the case of orthogonal projections, we retrieve 
familiar weighted least-squares gradients, but we also propose new direction-weighted variants. On 
the other hand, using oblique projections that employ cell face normal vectors we derive variations 
of consistent Green-Gauss gradients, which we call Taylor-Gauss gradients. The gradients are tested 
and compared on a variety of grids such as structured, locally refined, randomly perturbed, 
unstructured, and with high aspect ratio. The tests include quadrilateral and triangular grids, and 
employ both compact and extended stencils, and observations are made about the best choice of 
gradient and weighting scheme for each case. On high aspect ratio grids, it is found that most 
gradients can exhibit a kind of numerical instability that may be so severe as to make the gradient 
unusable. A theoretical analysis of the instability reveals that it is triggered by roundoff 
errors in the calculation of the cell centroids, but ultimately is due to truncation errors of the 
gradient reconstruction scheme, rather than roundoff errors. Based on this analysis, we provide 
guidelines on the range of weights that can be used safely with least squares methods to avoid this 
instability.
\end{abstract}


% \begin{highlights}
% \item Least-squares and Green-Gauss gradients derived from within the same theoretical framework.
% \item The proposed theoretical framework can be used to derive new gradient schemes.
% \item A new weighting scheme, based on direction, is proposed for least-squares gradients.
% \item Several gradient schemes are compared in a variety of test cases.
% \item Occassional numerical instability is observed, and the issue is addressed.
% \end{highlights}

\begin{keyword}



\end{keyword}

\end{frontmatter}

% \linenumbers

\section{Introduction}
\label{sec: intro}

Gradient reconstruction schemes are among the fundamental ingredients of Finite Volume Methods 
(FVMs). They are used in the discretisation of diffusion \cite{Traore_2009, Demirdzic_2015} and 
convection \cite{Jalali_2016} terms, of terms of turbulence closure equations \cite{Ferziger_2002}, 
of terms of non-Newtonian constitutive equations \cite{Afonso_2012, Jalali_2016, Pimenta_2017, 
Syrakos_2019} etc. Although lately a significant amount of research has gone into the development 
of high-order FVMs, the second-order accurate FVMs continue to be the most popular choice, due to 
ease of implementation, low memory requirements, faster execution (for the same number of degrees of 
freedom), better understanding and familiarity with the discretisation and stabilisation schemes 
etc. For such reasons, low-order methods will likely remain a popular choice in the foreseeable 
future (and this holds true even for finite element methods \cite{Varchanis_2019, Chai_2021}, for 
which achievement of high-order accuracy is relatively more straightforward than for FVMs). For 
second-order accurate FVMs, the main families of gradient schemes in use are the Green-Gauss (GG) 
gradients \cite{Barth_1989, Jasak_1996, Lilek_1997, Ferziger_2002, Wu_2014, Moukalled_2016}, which 
are derived from the divergence (Gauss) theorem, and the least-squares (LS) gradients 
\cite{Barth_1991, Muzaferija_1997, Ollivier_2002, Bramkamp_2004, Wu_2014, Moukalled_2016}, which are 
derived from least-squares error minimisation (the latter are also easily applicable to high-order 
FVMs). The performance of each gradient scheme depends on the geometrical characteristics of the 
grid and on the function being differentiated.

In a previous publication \cite{Syrakos_2017} we analysed the properties of both of these gradient 
families, and showed in theory and verified in practice that the GG gradients require special grid 
smoothness in order to be consistent, whereas LS gradients are always at least first-order 
accurate, which is a prerequisite for a second-order accurate FVM on grids of general geometry. 
Despite their shortcomings, GG gradients remain popular and a number of recent publications have 
addressed the problem of their inconsistency \cite{Sozer_2014, Deka_2018, Nishikawa_2018, Wang_2019, 
Athkuri_2020}. One way to remove the inconsistency is through an iterative procedure that couples 
the gradients at neighbouring grid cells \cite{Karimian_2006, Wu_2014, Syrakos_2017, Deka_2018, 
Nishikawa_2018} (equivalently, a large linear system that includes the gradients at all grid cells 
can be solved \cite{Betchen_2010}). This procedure is expensive, but if the gradient iterations are 
interwoven with the solution iterations of an implicit FVM then the gradient computation cost 
becomes almost as small as for the uniterated GG gradient \cite{Syrakos_2017, Deka_2018}. Another 
option for obtaining consistent GG gradient schemes, without the need for iteration, is to apply 
the Gauss theorem to suitably selected auxiliary cells rather than to the actual grid cells 
\cite{Syrakos_2017, Athkuri_2020}.

The aforementioned GG gradient variants achieve consistency at a price, which is either the need 
for iterations or large linear system solving, or the additional complexity of constructing the 
auxiliary cells, which can be significant for three-dimensional grids of general topology. However, 
there does exist another GG gradient variant, named ``quasi-Green'' (QG) gradient by its inventors, 
which is consistent but avoids these additional costs \cite{Brenner_1996, Pont_2017}; it achieves 
this by a form of self-correction. Although not new, the QG gradient seems to have gone mostly 
unnoticed. The QG gradient is of particular interest in the present study because it turns out to 
be identical with a particular member of a new family of gradient schemes whose derivation is 
related to that of the least-squares gradients.

The other major family of gradient schemes is that based on least-squares minimisation. The 
least-squares gradients have no inconsistency, being at least first-order accurate always
\cite{Syrakos_2017}. In the past they acquired a reputation of producing large errors on curved 
geometries gridded with cells of very high aspect ratio \cite{Mavriplis_2003}, but proper 
inverse-distance weighting overcomes the problem in many cases \cite{Mavriplis_2003, Diskin_2008}. 
Nevertheless, in some cases the only way to improve the accuracy is to extend the least-squares 
computational stencil beyond the immediate neighbours of the cell, notably on grids of triangles (or 
tetrahedra) where some spatial directions are under-represented by the nearest-neighbour stencils. 
On such grids, augmented-stencil LS gradients were found in \cite{Diskin_2008} to outperform GG 
gradients, the latter being made consistent by application of corrections calculated with the aid of 
LS gradients. Stencil augmentation is not straightforward when it comes to GG gradients and has not 
been applied, to the best of our knowledge.

In a previous (not peer-reviewed) work \cite{Oxtoby_2019} we proposed a new family of consistent 
gradient methods, with similarities to both the GG and LS gradients, which we called 
``Taylor-Gauss'' (TG) gradients. The starting point of the derivation of the TG gradients is the 
same as for the LS gradients, namely the Taylor series expansion of the values of the function, 
whose gradient is sought, at neighbour cell centres. This results in an overdetermined system for 
the gradient, which in the LS case is solved via an orthogonal projection. In the TG case, it is 
solved by an oblique projection, whose direction is determined by the vectors normal to the cell 
faces -- hence the resemblance to the GG gradient. Therefore, the LS and TG gradients are members of 
a larger family of gradient schemes which express the differentiated variable at neighbour cell 
centres in terms of Taylor series and solve the resulting overdetermined system via some sort of 
projection. Furthermore, after the publication of \cite{Oxtoby_2019}, it has been realised that one 
of the basic members of the TG family can be recast as the QG gradient mentioned above\footnote{This 
was brought to our attention by Ganesh Natarajan of the Indian Institute of Technology Palakkad, one 
of the authors of \cite{Deka_2018}, to whom we are particularly thankful.}. This provides an 
interesting link between least-squares and Green-Gauss gradients.

The present paper aims to present a unified discussion of, and comparison between, the TG, LS, 
and QG gradients. This encompasses the larger part of available schemes for FVMs, as the majority 
of FVM gradient schemes currently in use belong either to the LS family or to the GG family (the 
latter represented here by the QG gradient). The TG gradients constitute the link between the LS 
and GG gradients, as they can be viewed either as a projection method, like the LS gradients, or as 
a family of QG gradients, as will be shown. This theoretical unification of the various gradients 
provides deeper insight that can help to advance the field.

The paper begins in Sec.\ \ref{sec: general framework} with a presentation of the general framework 
for formulating gradient schemes whereby the differentiated variable is expanded in Taylor series 
in the neighbourhood of the cell in question and the resulting overdetermined system is solved via 
some projection. Particular projections that lead to the least-squares and Taylor-Gauss gradients 
are described in Sections \ref{sec: LS} and \ref{sec: TG}, respectively. Several variants of these 
methods are described, that correspond to different choices of weighting scheme. Subsequently, in 
Sec.\ \ref{sec: SCGG}, the Taylor-Gauss gradients are re-derived as self-corrected Green-Gauss 
gradients. The theory is followed by numerical tests on grids of various topologies in Sec.\ 
\ref{sec: tests}, which verify the theoretical orders of accuracy of the schemes. In Section 
\ref{sec: tests hard}, the gradients are tested on the aforementioned challenging case that 
combines a curved geometry and a very high aspect ratio grid. This case brings to the surface issues 
of numerical stability, in addition to accuracy. In particular, it is observed that, as the grid is 
refined, eventually errors related to finite precision arithmetic can become important, depending 
on the weighting scheme. The paper concludes in Sec.\ \ref{sec: conclusions}. 

The scope of the present work is limited to the standalone accuracy of the gradient schemes 
themselves, i.e.\ their ability to compute the gradient of a function. Although there are 
applications for which this knowledge suffices, e.g.\ \cite{Correa_2011}, in the majority of 
applications the gradient schemes would be part of an overall composite discretisation scheme used 
to solve a set of differential equations -- a finite volume method. The effect of the gradient 
scheme on the discretisation error of a FVM is complex and not determined solely by the accuracy of 
the gradient scheme per se. A study of the latter, however, which is the topic of the present paper, 
is a prerequisite for obtaining a more complete understanding of the former. We also note that all 
of the presently examined gradient schemes are ``explicit'' in the sense that they compute the 
gradient using only the values of the differentiated variable at nearby cell centres. As mentioned, 
there exist also ``implicit'' gradient schemes where the gradient at a cell depends additionally 
on the gradients at nearby cells; hence these schemes require iteration or solution of domain-wide 
linear systems. Given the much higher computational cost of the implicit gradients, they do not 
appear to be advantageous compared to explicit gradients if all that is required is to compute the 
gradient of a function. However, it is possible that they have advantages when used as a component 
of a FVM, e.g.\ they may offer better stability or iterative convergence. The effect of the choice 
of gradient scheme on the FVM solution deserves a separate study, which we plan for the future, on 
a variety of continuum mechanics problems.

It is also noted that all of the presented gradient schemes are formally of first-order accuracy 
(in practice they are second-order accurate on smooth grids); thus they can serve as components of 
second-order accurate FVMs. Least-squares gradients are straightforward to extend to higher-order 
accuracy (see e.g.\ \cite{Ollivier_2002, Jalali_2013}), but not so for gradients that employ face 
normal vectors (TG and GG gradients). Finally, to keep things simple, the present work is limited 
to two-dimensional geometries.

\section{General framework: Taylor expansion and projection}
\label{sec: general framework}

\begin{figure}[tb]
  \centering
  \includegraphics[scale=0.90]{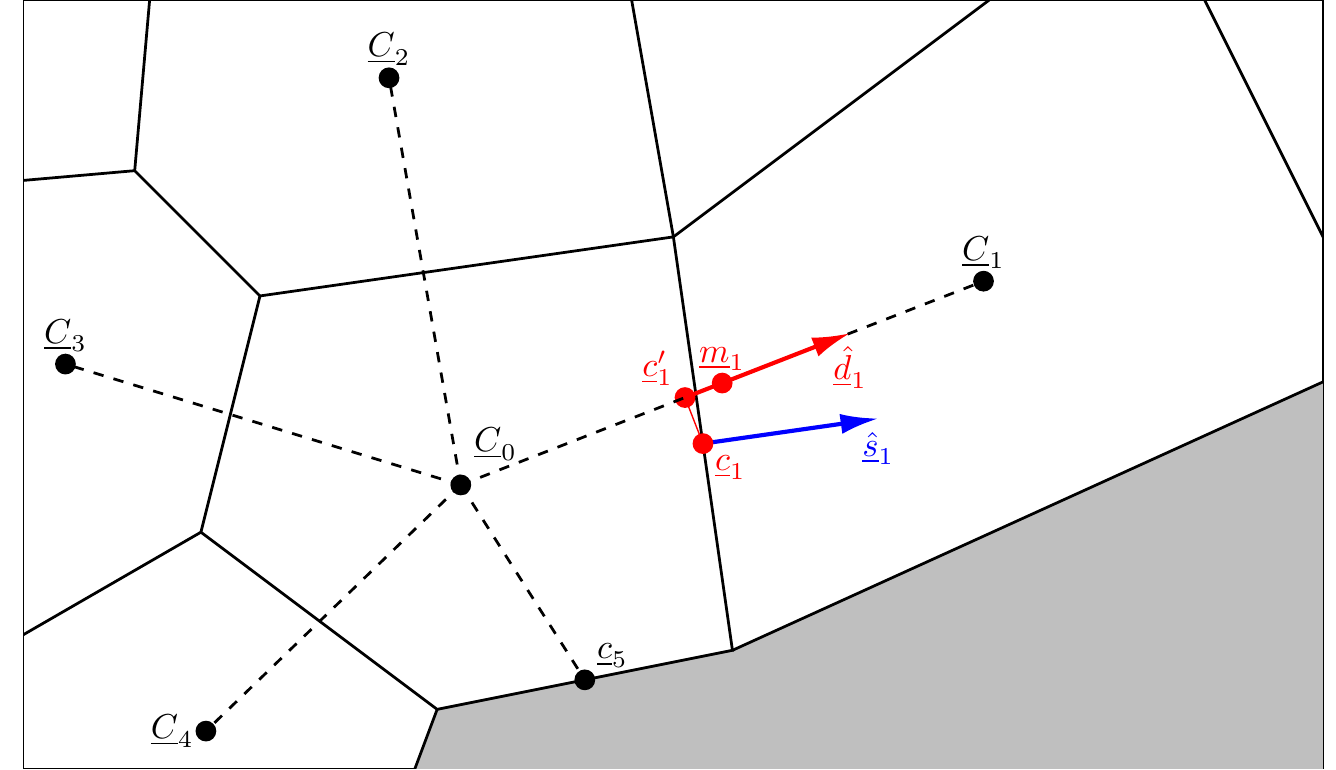}
  \caption{The notation adopted in the present paper (see text for details).}
  \label{fig: notation}
\end{figure}

Suppose we wish to calculate the gradient of a function $\phi$ at the centre $\vf{C}_0$ of a cell 
under consideration (Fig.\ \ref{fig: notation}). Let us first introduce some definitions and 
notation. The cell has $F$ faces, each of which either separates it from a single neighbour cell, 
or is a boundary face. We will use a local indexing system where index 0 is assigned to the cell 
where the gradient is sought, and indices $f > 0$ are assigned to nearby cells involved in the 
calculation. For $f \leq F$, index $f$ is assigned to the neighbour cell that shares face $f$. The 
centroids of these cells are denoted as $\vf{C}_0$ and $\vf{C}_f$, respectively (geometric vectors 
are denoted with an underline). If face $f$ is a boundary face then its centroid $\vf{c}_f$ will be 
used instead, assuming the value of $\phi$ is known there. Gradient schemes will reconstruct the 
gradient at $\vf{C}_0$ from the values of $\phi$ at $\vf{C}_0$ and at $N$ additional points, where 
$N = F$ for nearest-neighbour schemes and $N>F$ for extended-stencil schemes. Using only the 
nearest neighbours allows an implementation that uses simpler data structures; however, sometimes 
these points do not carry sufficient information and additional points are needed. Some of the 
gradient schemes examined here, namely the least-squares schemes, can use any set of neighbour 
points, but others, namely those making use of the cell's faces such as the Green-Gauss gradient, 
are limited to nearest neighbours only, and cannot be extended to additional points in a 
straightforward manner.

Some geometric definitions are as follows. An interior face $f$ has centroid $\vf{c}_f$, whose 
projection on the line joining $\vf{C}_0$ and $\vf{C}_f$ is denoted as $\vf{c}'_f$. The point 
$\vf{m}_f = (\vf{C}_0 + \vf{C}_f)/2$ lies midway between $\vf{C}_0$ and $\vf{C}_f$. The distance 
vector from $\vf{C}_0$ to $\vf{C}_f$ is denoted as $\vf{D}_f = \vf{C}_f - \vf{C}_0$, and the unit 
vector in the same direction as $\hat{\vf{d}} = \vf{D}_f / \|\vf{D}_f\|$. The unit vector normal to 
face $f$ is denoted by $\hat{\vf{s}}_f$, and if we multiply this by the face area $S_f$ we get the 
face vector $\vf{S}_f = S_f \hat{\vf{s}}_f$. We can define the following important grid quality 
metrics \cite{Syrakos_2017}: Skewness is the deviation of the centroid $\vf{c}_f$ from the line 
joining $\vf{C}$ and $\vf{C}_f$, and can be quantified as $\| \vf{c}_f - \vf{c}'_f \| / \| \vf{D}_f 
\|$; non-orthogonality is the angle between $\hat{\vf{d}}_f$ and $\hat{\vf{s}}_f$; and unevenness is 
the asymmetrical distancing of points $\vf{C}$ and $\vf{C}_f$ from face $f$, which can be quantified 
as $\| \vf{c}'_f - \vf{m}_f \| / \| \vf{D}_f \|$.

Eventually, the reconstruction scheme must involve only values of $\phi$ at cell centres, but in a 
first step we can express $\phi$ in terms of Taylor series at a number of points $\vf{P}_1, 
\vf{P}_2, \ldots, \vf{P}_N$ that are not necessarily the cell centroids. If the points $\vf{P}_f$ 
are different from $\vf{C}_f$ then a further step is required where the (approximate) values 
$\phi(\vf{P}_i)$ are interpolated from the values $\phi(\vf{C}_j)$. In any case, the change of 
$\phi$ between $\vf{C}_0$ and the points $\vf{P}_f$, $\Delta \phi_f = \phi(\vf{P}_f) - 
\phi(\vf{C}_0)$, is given by (using Taylor series):
%^c
\begin{equation} \label{eq: Taylor series}
 \Delta \phi_f \;=\; \vf{R}_f \cdot \nabla\phi(\vf{C}_0)
               \;+\; \frac{1}{2} \vf{R}_f \vf{R}_f : \nabla \nabla \phi(\vf{C}_0)
               \;+\; O(h^3)
\end{equation}
%^c
where $\vf{R}_f = \vf{P}_f - \vf{C}_0$, and $h$ is a typical cell dimension. Vectors written next to 
each other, such as $\vf{R}_f \vf{R}_f$ above, denotes the tensor product.

If we drop the second- and higher-order terms of the above equations, then we are left with $N$ 
equations with D=2 or D=3 unknowns (the components of $\nabla \phi(\vf{C}_0)$), in two or three 
dimensional space, respectively. Thus we have an over-determined system with no solution. From it, 
we can obtain a full-rank D$\times$D system with unique solution by the following procedure. We 
first weigh (left-multiply) each equation \eqref{eq: Taylor series} by a \textit{vector} $\vf{V}_f$ 
(the choice of vector will be discussed later), to convert it into a vector equation:
%^c
\begin{equation} \label{eq: Taylor series x Vf}
 \vf{V}_f \Delta \phi_f \;=\; \vf{V}_f \vf{R}_f \cdot \nabla\phi(\vf{C}_0)
               \;+\; \frac{1}{2} \vf{V}_f \vf{R}_f \vf{R}_f : \nabla \nabla \phi(\vf{C}_0)
               \;+\; O(\vf{V}_f) \cdot O(h^3)
\end{equation}
%^c
Then, we sum all of the equations \eqref{eq: Taylor series x Vf}:
%^c
\begin{equation} \label{eq: new gradient with O(h^3)}
 \sum_f \vf{V}_f \Delta \phi_f 
 \;=\;
 \left[ \sum_f \vf{V}_f \vf{R}_f \right] \cdot \nabla \phi(\vf{C}_0)
 \;+\;
 \frac{1}{2} \left[ \sum_f \vf{V}_f \vf{R}_f \vf{R}_f \right] : \nabla \nabla \phi(\vf{C}_0)
 \;+\;
 O(\vf{V}_f) \cdot O(h^3)
\end{equation}
%^c
Because $\vf{R}_f = O(h)$, we have $\vf{V}_f \vf{R}_f \vf{R}_f = O(\vf{V}_f) \cdot O(h^2)$.  
Grouping together all terms of order 2 or higher and solving for $\nabla \phi (\vf{C}_0)$ we obtain
%^c
\begin{equation} \label{eq: new gradient with O(h^2)}
 \nabla \phi (\vf{C}_0)
 \;=\;
 \left[ \sum_f \vf{V}_f \vf{R}_f \right]^{-1}
 \!\cdot
 \left[ \sum_f \vf{V}_f \Delta \phi_f \right]
 \;+\;
 \underbrace{
 \left[ \sum_f \vf{V}_f \vf{R}_f \right]^{-1}
 \!\cdot
 O(\vf{V}_f) \cdot O(h^2)
 }_{
 =\; [O(\vf{V}_f) O(h)]^{-1} \cdot O(\vf{V}_f) O(h^2) \;=\; O(h)
 }
\end{equation}
%^c
or, summarising:
%^c
\begin{equation} \label{eq: generic gradient}
 \nabla \phi (\vf{C}_0)
 \;=\;
 \left[ \sum_f \vf{V}_f \vf{R}_f \right]^{-1}
 \!\cdot
 \left[ \sum_f \vf{V}_f \Delta \phi_f \right]
 \;+\;
 O(h)
\end{equation}
%^c
We have thus arrived at a gradient scheme that is at least first-order accurate (it is also exact 
for linear functions, as can be seen from Eq.\ \eqref{eq: new gradient with O(h^3)}, where in the 
case of a linear function $\nabla \nabla \phi$ and all the higher derivatives are zero), and 
therefore appropriate for second-order accurate FVMs \cite{Syrakos_2017}. It must be noted that 
this analysis assumes that the values $\Delta \phi_f$ are exact, which will be the case if the 
points $\vf{P}_f$ are cell centroids, $\vf{C}_f$. If not, then the interpolation used to obtain 
approximate values $\tilde{\phi}(\vf{P}_f) = \phi(\vf{P}_f) + O(h^p)$, where $p$ is the order of 
interpolation, will introduce additional errors. In this case, the exact differences $\Delta\phi_f$ 
will not be available to us, but rather approximate values $\Delta \tilde{\phi}_f = \Delta \phi_f + 
O(h^p)$. Substituting in \eqref{eq: generic gradient} we obtain:
%^c
\begin{equation}
 \nabla \phi (\vf{C}_0)
 \;=\;
 \left[ \sum_f \vf{V}_f \vf{R}_f \right]^{-1}
 \!\cdot
 \left[ \sum_f \vf{V}_f \Delta \tilde{\phi}_f \right]
 \;+\;
 \underbrace{
 \left[ \sum_f \vf{V}_f \vf{R}_f \right]^{-1}
 \!\cdot
 \left[ \sum_f \vf{V}_f O(h^p) \right]
 }_{
 =\; [O(\vf{V}_f) O(h)]^{-1} \cdot O(\vf{V}_f) O(h^p) \;=\; O(h^{p-1})
 }
 \;+\;
 O(h)
\end{equation}
%^c
Therefore, the order of accuracy of the gradient in this case is $\min\{p-1,\,1\}$. In order to 
have a gradient that is at least first-order accurate, the order of interpolation must be $p \geq 
2$. This makes good candidates for points $\vf{P}_f$ the points that lie on the lines joining 
$\vf{C}_0$ to $\vf{C}_f$, such as $\vf{c}'_f$ and $\vf{m}_f$, as $\phi$ can be obtained there with 
linear interpolation ($p=2$).

It is noteworthy that, to this point, we have particularised neither the weighting vectors 
$\vf{V}_f$ (possible choices include $\vf{R}_f$, $\hat{\vf{d}}_f$, $\vf{D}_f$ $\hat{\vf{s}_f}$, 
$\vf{S}_f$, and many others), nor the points $\vf{P}_f$ (and therefore the vectors $\vf{R}_f$ 
appearing in \eqref{eq: generic gradient}). The scheme \eqref{eq: generic gradient} is, therefore, 
quite general. Despite appearing as something new, it will soon be shown that very familiar 
gradient schemes, such as least-squares gradients and a self-corrected variation of the Green-Gauss 
gradient, can be cast in this form.

The scheme \eqref{eq: generic gradient} is very general, but nevertheless there are some minimal 
restrictions on the vectors $\vf{R}_f$ and $\vf{V}_f$ involved in the calculation. In order for the 
D$\times$D matrix $\sum_f \vf{V}_f \vf{R}_f$ to be invertible, it must be of full rank, i.e.\ it 
must have linearly independent columns (and linearly independent rows). Obviously each 
of the component matrices $\vf{V}_f \vf{R}_f$ has rank 1 and is singular, but their sum may be of 
full rank. The columns of $\sum_f \vf{V}_f \vf{R}_f$ are linear combinations of the vectors 
$\vf{V}_f$, and therefore we need at least D linearly independent vectors $\vf{V}_f$ (in fact we 
can't have more than D in D-dimensional space). Similarly, for $\sum_f \vf{V}_f \vf{R}_f$ to have D 
linearly independent rows we need D linearly independent direction vectors $\vf{R}_f$. These 
requirements are hardly surprising.

\subsection{Some linear algebra}
\label{ssec: linear algebra}

Before discussing particular instances of the general scheme \eqref{eq: generic gradient} it is 
useful to take another look at it from a linear algebra perspective. In matrix form, the 
over-determined system \eqref{eq: Taylor series}, after dropping the second- and higher-order 
terms, is (for simplicity the two-dimensional case is assumed):
\begin{equation} \label{eq: overdetermined system ME}
    \begin{bmatrix}
        R_{1,x} & R_{1,y} \\
        R_{2,x} & R_{2,y} \\
        \vdots  & \vdots  \\
        R_{N,x} & R_{N,y}
    \end{bmatrix}
    \cdot
    \begin{bmatrix}
        \phi_{.x} \\
        \phi_{.y}
    \end{bmatrix}
    \;=\;
    \begin{bmatrix}
        \Delta \phi_1 \\
        \Delta \phi_2 \\
        \vdots        \\
        \Delta \phi_N
    \end{bmatrix}
\end{equation}
where $\vf{R}_f = (R_{f,x}, R_{f,y})$ and $\nabla \phi(\vf{C}_0) = (\phi_{.x}, \phi_{.y})$ in 
Cartesian component form. We can write this in compact matrix notation as
\begin{equation} \label{eq: overdetermined system M}
    \mf{Rx} = \mf{b}
\end{equation}
where $\mathbf{R}$ is $N \times 2$, $\mathbf{x}$ is $2 \times 1$ and contains the sought 
derivatives, and $\mathbf{b}$ is $N \times 1$. To turn this into a solvable $2\times2$ system, we 
left-multiply it by the transpose of a full-rank $N \times 2$ matrix $\mathbf{V}$:
\begin{equation} \label{eq: projected system M}
    \Tr{\mf{V}} \mf{Rx} = \Tr{\mf{V}} \mf{b}
\end{equation}
This is a $2 \times 2$ system which can be solved for $\mathbf{x}$. The solution is:
\begin{equation} \label{eq: projected system solution M}
    \mf{x} \;=\; (\Tr{\mf{V}} \mf{R})^{-1} \Tr{\mf{V}} \mf{b}
\end{equation}
A little contemplation reveals that Eq.\ \eqref{eq: projected system M} is equivalent to \eqref{eq: 
Taylor series x Vf} (if we omit the higher-order terms), and Eq.\ \eqref{eq: projected system 
solution M} is equivalent to Eq.\ \eqref{eq: generic gradient}; the rows of the matrix $\mf{V}$ are 
the vectors $\vf{V}_f$.

\begin{figure}[tb]
    \centering
    \begin{subfigure}[b]{0.48\textwidth}
        \centering
        \includegraphics[scale=0.85]{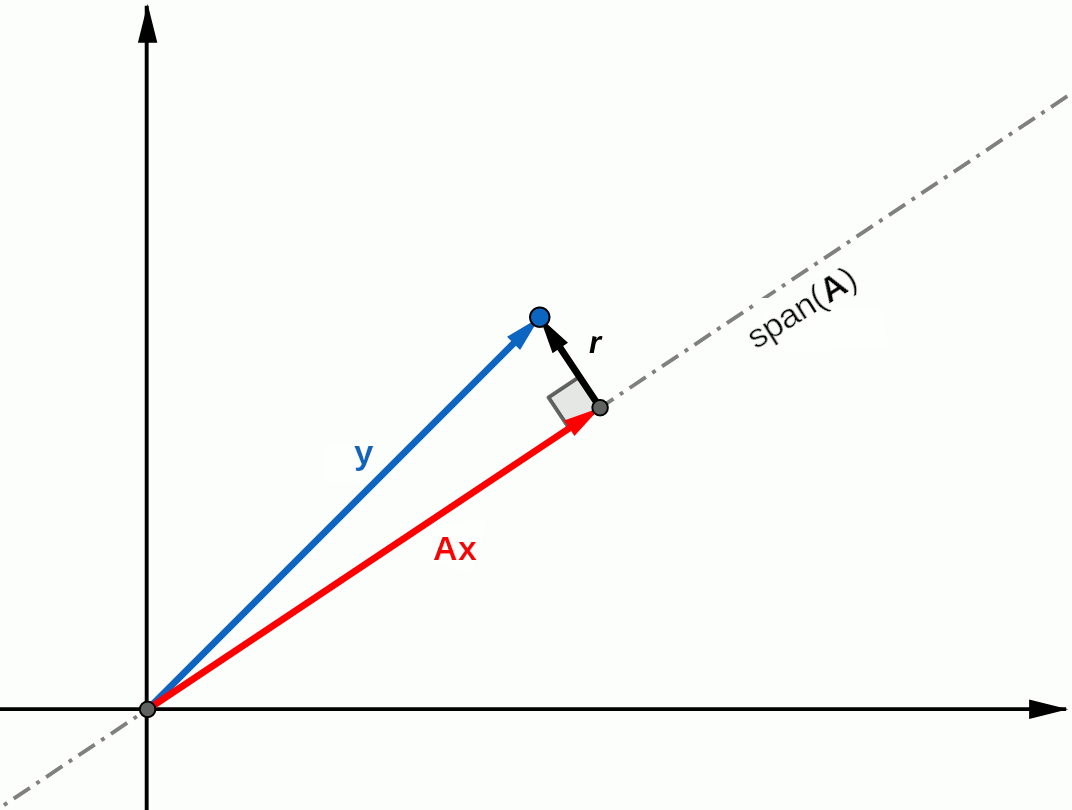}
        \caption{Orthogonal projection}
        \label{sfig: orthogonal projection}
    \end{subfigure}
    \begin{subfigure}[b]{0.48\textwidth}
        \centering
        \includegraphics[scale=0.85]{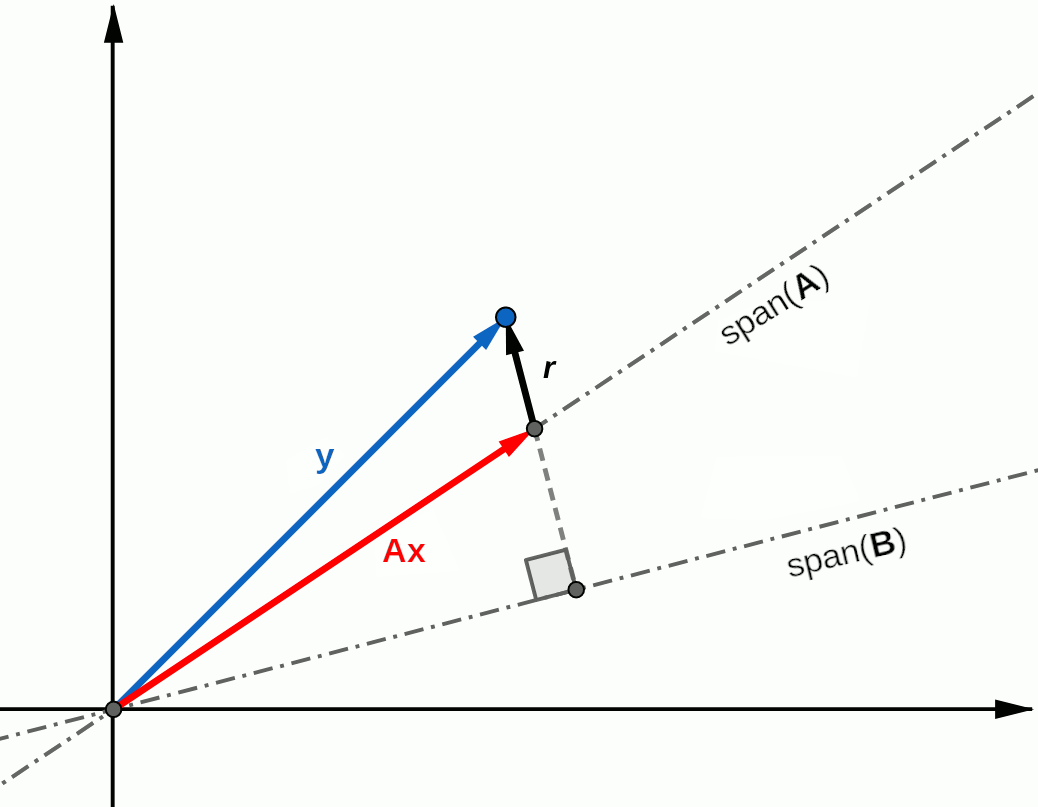}
        \caption{Oblique projection}
        \label{sfig: oblique projection}
    \end{subfigure}
    \caption{Orthogonal \subref{sfig: orthogonal projection} and oblique \subref{sfig: oblique 
projection} projections $\mf{Ax}$ of the vector $\mf{y}$ onto the subspace $\mathrm{span}(\mf{A})$. 
The difference $\mf{r} = \mf{y}-\mf{Ax}$ is orthogonal to either $\mathrm{span}(\mf{A})$ 
\subref{sfig: orthogonal projection} or to another subspace $\mathrm{span}(\mf{B})$ \subref{sfig: 
oblique projection}.}
  \label{fig: projections}
\end{figure}

We know from linear algebra (e.g.\ \cite{Trefethen_1997}) that if $\mf{Ax} = \mf{y}$ is an 
overdetermined system then the solution $\mf{x}$ that minimises the L2 norm of the error $\mf{r} = 
\mf{y} - \mf{Ax}$ satisfies
\begin{equation} \label{eq: projection orthogonal}
 \Tr{\mf{A}} (\mf{y} - \mf{Ax}) \;=\; \mf{0}
\end{equation}
In this case, the error $\mf{r}$ is orthogonal to the column space of $\mf{A}$ and $\mf{Ax}$ is the 
orthogonal projection of $\mf{y}$ onto $\mathrm{span}(\mf{A})$. However, we could instead choose 
$\mf{x}$ such that $\mf{r} = \mf{y} - \mf{Ax}$ is orthogonal to another subspace, spanned by the 
columns of a matrix $\mf{B}$, in which case $\mf{Ax}$ is an oblique projection of $\mf{y}$ onto 
$\mathrm{span}(\mf{A})$:
\begin{equation} \label{eq: projection oblique}
 \Tr{\mf{B}} (\mf{y} - \mf{Ax}) \;=\; \mf{0}
\end{equation}
These cases are illustrated in Fig.\ \ref{fig: projections}. Equation \eqref{eq: projected system 
M} is of the form \eqref{eq: projection oblique}, but sometimes can be cast in the form \eqref{eq: 
projection orthogonal} which is preferable from a theoretical standpoint because it allows us to 
discern a quantity that is minimised by the gradient scheme. Choices of $\mf{V}$ that allow this 
are, for example:
\begin{itemize}
 \item $\mf{V} = \mf{R}$. Equation \eqref{eq: projected system M} can be rearranged in the form 
\eqref{eq: projection orthogonal} as $\Tr{\mf{R}} (\mf{b} - \mf{Rx}) = \mf{0}$, whose solution 
$\mf{x}$ minimises $\|\mf{b}-\mf{Rx}\|_2$. This is the unweighted least squares method.
 \item $\mf{V} = \mf{DR}$, where $\mf{D}$ is a diagonal matrix. Equation \eqref{eq: projected 
system 
M} can be recast in the form \eqref{eq: projection orthogonal} as 
$\Tr{(\mf{D}^{\frac{1}{2}}\mf{R})} (\mf{D}^{\frac{1}{2}}\mf{b} - \mf{D}^{\frac{1}{2}}\mf{Rx}) = 
\mf{0}$. The quantity minimised is $\| \mf{D}^{\frac{1}{2}}\mf{b} - \mf{D}^{\frac{1}{2}}\mf{Rx} 
\|_2 = \| \mf{D}^{\frac{1}{2}}(\mf{b} - \mf{Rx}) \|_2$. So, it is again the magnitude of the 
discrepancy array $\mf{r} = \mf{b} - \mf{Rx}$ that is minimised, but this time not all components 
carry equal weight; the weights are stored in $\mf{D}^{\frac{1}{2}}$. Usually, components that come 
from farther neighbours carry less weight. Weighted least-squares gradients have this form.
 \item $\mf{V} = \Tr{\mf{W}}\mf{WR}$, where $\mf{W}$ is a full matrix. This is the natural 
generalisation of the previous case: Eq.\ \eqref{eq: projected system M} can be cast as 
$\Tr{(\mf{W} 
\mf{R})} (\mf{W} \mf{b} - \mf{W} \mf{Rx}) = \mf{0}$, and the norm of $\mf{W}(\mf{b}-\mf{Rx})$ is 
minimised.
\end{itemize}

\section{Least-squares gradients}
\label{sec: LS}

Least-squares gradients are those that can be cast in the form \eqref{eq: projection orthogonal}. 
Here we will list a few particular schemes resulting from particular choices of the weights (matrix 
$\mf{D}$; full weight matrices $\mf{W}$, although offering more control, will not be examined here) 
-- one of these schemes is new, to the best of our knowledge. Least squares gradients usually 
employ the cell centroids $\vf{C}_f$ as the points $\vf{P}_f$, avoiding any interpolation, which is 
the practice followed here as well.

\subsection{Unweighted least-squares}
\label{ssec: LS0}

The unweighted least-squares gradient is the scheme that comes from setting $\vf{V}_f = \vf{R}_f$ 
in Eq.\ \eqref{eq: generic gradient}, or $\mf{V} = \mf{R}$ in \eqref{eq: projected system solution 
M}. It minimises the sum $\sum_f (\Delta\phi_f - \vf{R}_f \cdot \tilde{\nabla}(\vf{C}_0))^2$ (where 
$\tilde{\nabla}$ denotes the approximate gradient). This sum can also be written as
\begin{equation}
 \sum_f \left[ \|\vf{R}_f\| \left(
 \frac{\Delta\phi_f}{\|\vf{R}_f\|}  \;-\;
 \hat{\vf{d}}_f \cdot \tilde{\nabla}\phi(\vf{C}_0)
 \right) \right]^2
\end{equation}
Thus although ``unweighted'', this scheme preferentially tries to minimise the difference between 
the approximation of the directional derivative, $\hat{\vf{d}}_f \cdot \tilde{\nabla}(\vf{C}_0)$, 
and the finite difference approximation of that derivative, $\Delta\phi_f/\|\vf{R}_f\|$, mostly in 
the directions $f$ of more distant neighbours (larger $\|\vf{R}_f\|$). This is unnatural, as 
normally one would want to assign equal weights to the directional derivatives in all directions, or 
even to assign larger weights to the directions of \textit{closer} neighbours, since the closer we 
are to $\vf{C}_0$ the more linearly $\phi$ varies. Therefore it is not surprising that this scheme 
often performs poorly, and that inverse-distance-weighted least squares schemes are more popular.

\subsection{Inverse-distance-weighted least-squares}
\label{ssec: LSq}

To remedy the aforementioned problem, the weight vectors $\vf{V}_f$ can instead be defined as
\begin{equation} \label{eq: LS}
%  \vf{V}_f \;=\; \|\vf{R}_f\|^{-q} \: \vf{R}_f
 \vf{V}_f \;=\; \frac{1}{\|\vf{R}_f\|^q} \, \vf{R}_f
\end{equation}
where $q$ is an exponent usually in the range $q \in [0,4]$ ($q=0$ corresponds to the unweighted 
least squares gradient of Sec.\ \ref{ssec: LS0}). For brevity, henceforth the scheme with weights 
\eqref{eq: LS} will be denoted as the LS($q$) gradient.

\subsection{Area-weighted least-squares}
\label{ssec: LSA}

Inverse-distance-weighting accounts for the fact that the neighbour points $\vf{P}_f$ may not be 
equidistant from $\vf{C}_0$. Another factor that may cause the performance of the gradient to 
suffer is that the points $\vf{P}_f$ may not be distributed equally in all directions around 
$\vf{C}_0$. In \cite{Syrakos_2017} it was shown that when some points are clustered in a particular 
direction, the scheme \eqref{eq: LS} may not achieve optimal accuracy because this direction is 
over-represented in the calculation.

If only nearest neighbours are included in the calculation, then a possible remedy is to include 
the face areas as factors in the respective weights:
\begin{equation} \label{eq: LSA}
%  \vf{V}_f \;=\; \|\vf{S}_f\| \, \|\vf{R}_f\|^{-q} \: \vf{R}_f
 \vf{V}_f \;=\; \frac{\|\vf{S}_f\|}{\|\vf{R}_f\|^q} \, \vf{R}_f
\end{equation}
This is because usually (especially if the grid is orthogonal, $\hat{\vf{d}}_f = \hat{\vf{s}}_f$) 
face area is roughly inversely proportional to the degree of clustering in that direction: a large 
area means that there are no other neighbours around, while a small area means the opposite. This 
is not completely accurate, but it is easy to implement as the face areas $\|\vf{S}_f\|$ are 
readily available. Henceforth we will refer to this scheme as LSA($q$) gradient.

\subsection{Direction-weighted least-squares}
\label{ssec: LSD}

A more diligent quantification of the degree of clustering in each direction is the following. We 
define
\begin{equation} \label{eq: LSD directional cosine}
  \theta_{ij} \;=\; \theta_{ji} \;=\;
  \begin{dcases}
    \hat{\vf{d}}_i \cdot \hat{\vf{d}}_j  &  
                             \quad \text{if} \;\;\; \hat{\vf{d}}_i \cdot \hat{\vf{d}}_j \geq 0  \\
    0                                    &
                             \quad \text{if} \;\;\; \hat{\vf{d}}_i \cdot \hat{\vf{d}}_j < 0
  \end{dcases}
\end{equation}
$\theta_{ij}$ is a measure of the alignment of directions $\hat{\vf{d}}_i$ and $\hat{\vf{d}}_j$, 
ranging from 0, if $\hat{\vf{d}}_i$ and $\hat{\vf{d}}_j$ are perpendicular or point in opposite 
directions, to 1 if they point in the same direction. Now consider the point $\vf{P}_f$, which lies 
in the direction $\hat{\vf{d}}_f$ with respect to $\vf{C}_0$. If other points, $\vf{P}_j$, are also 
roughly aligned in the same direction, then the contribution of point $\vf{P}_f$ to the computed 
gradient should be reduced accordingly. One possibility for quantifying this is through the 
following factor:
\begin{equation} \label{eq: LSD directional factor}
 \Theta_f \;=\; \frac{1}{\sum_j \theta_{fj}}
\end{equation}
This compares the contribution of point $\vf{P}_f$ to direction $\hat{\vf{d}}_f$ (this contribution 
is $\theta_{ff} = \hat{\vf{d}}_f \cdot \hat{\vf{d}}_f = 1$, the numerator) to the total 
contribution from all points (the sum in the denominator). The smaller $\theta_{ff}$ is compared to 
$\sum_j \theta_{fj}$, the less point $\vf{P}_f$ should contribute to the gradient. 

Therefore, a reasonable choice of weights seems to be:
\begin{equation} \label{eq: LSD}
%  \vf{V}_f \;=\; \Theta_f \, \|\vf{R}_f\|^{-q} \: \vf{R}_f
 \vf{V}_f \;=\; \frac{\Theta_f}{\|\vf{R}_f\|^q} \, \vf{R}_f
\end{equation}
With this choice, the quantity minimised by the least-squares gradient is:
\begin{equation} \label{eq: LSD minimisation goal}
 \sum_f \left[ 
 \left( \frac{\Theta_f}{\|\vf{R}_f\|^q} \right)^{\frac{1}{2}}
 \|\vf{R}_f\|\left(
 \frac{\Delta\phi_f}{\|\vf{R}_f\|}  \;-\;
 \hat{\vf{d}}_f \cdot \tilde{\nabla}\phi(\vf{C}_0)
 \right) \right]^2
\end{equation}
as follows from the fact that using $\mf{V}=\mf{DR}$ minimises $\mf{D}^{\frac{1}{2}}(\mf{b} - 
\mf{Rx})$ (Sec.\ \ref{ssec: linear algebra}).

To ensure that this scheme works reasonably, let us perform a couple of checks. To isolate the 
effect of directionality, let us assume that all neighbour points $\vf{P}_f$ are equidistant from 
$\vf{C}_0$. First, if the points are distributed equally in all directions, then due to symmetry 
the factors $\Theta_f$ will be equal for all points, and all of them will receive equal weight, as 
is desirable. As another check, consider an extreme case where two points $\vf{P}_i$ and $\vf{P}_j$ 
coincide, and none of the other points contribute in that particular direction ($\theta_{ij} = 1$ 
and $\theta_{ik} = 0$ for $k \neq i,j$). The reader can verify that the contributions of these two 
points to the quantity \eqref{eq: LSD minimisation goal} sum up to what would be the contribution 
of a single point at that location, had there been only one point there.

The scheme \eqref{eq: LSD}, henceforth referred to as LSD($q$) gradient, is more expensive than the 
simpler scheme \eqref{eq: LSA}, but has the advantage that the only geometric data it employs are 
the locations of the points $\vf{P}_f$, and thus it does not even require a well-defined grid. This 
is in line with the popular traditional scheme \eqref{eq: LS}. The LSA scheme \eqref{eq: LSA}, on 
the other hand, requires a well-defined grid with cell faces. The use of the cell faces entails that 
only nearest neighbours can be involved in the calculation (as for GG gradients). The LSD gradient 
has no such restriction.

\section{Taylor-Gauss gradients}
\label{sec: TG}

Another option that naturally comes to mind concerning the weight vectors $\vf{V}_f$ is to base 
them on the face area vectors $\vf{S}_f$. 
\begin{equation} \label{eq: TG}
  \vf{V}_f \;=\; \frac{1}{\|\vf{R}_f\|^q} \, \vf{S}_f
\end{equation}
This scheme, called ``Taylor-Gauss'' (TG) gradient in \cite{Oxtoby_2019} is more difficult to 
analyse from a theoretical perspective because it is an oblique projection scheme and does not 
perform a task of minimisation. Nevertheless, it will be shown in the next section that it can be 
interpreted as a consistent form of Green-Gauss gradient. The scheme \eqref{eq: TG} with the choice 
$\vf{P}_f = \vf{C}_f$ will henceforth be referred to as TG($q$) gradient.

Since Green-Gauss gradients typically employ face-centre values of $\phi$, let us briefly examine 
the implications of choosing the points $\vf{P}_f$ to be the points $\vf{c}'_f$ rather than 
$\vf{C}_f$. It turns out that the resulting scheme is equivalent to that of using the cell 
centroids $\vf{C}_f$ but with altered weights. In particular, for $\vf{P}_f = \vf{c}'_f$ we have 
$\vf{R}_f = \vf{c}'_f - \vf{C}_0$ and $\Delta \phi_f = \phi(\vf{c}'_f) - \phi(\vf{C}_0)$ in Eq.\ 
\eqref{eq: generic gradient}, but let us define also $\vf{R}^*_f = \vf{C}_f - \vf{C}_0$ and $\Delta 
\phi^*_f = \phi(\vf{C}_f) - \phi(\vf{C}_0)$ (these are the corresponding values that would be used 
if we had chosen $\vf{P}_f = \vf{C}_f$). The value $\phi(\vf{c}'_f)$ is interpolated linearly. If
\begin{equation} \label{eq: a_f}
 a_f \;=\; \frac{\|\vf{c}'_f-\vf{C}_0\|}{\|\vf{C}_f-\vf{C}_0\|}
\end{equation}
is the linear interpolation factor, then $\vf{R}_f = a_f \vf{R}^*_f$ and $\Delta\phi_f = a_f \Delta 
\phi^*_f$. Equation \eqref{eq: generic gradient}, with weight vectors \eqref{eq: TG}, can then be 
written as (omitting the truncation error):
%^c
\begin{equation} \label{eq: generic gradient with interpolation}
 \nabla \phi (\vf{C}_0)
 \;=\;
 \left[ \sum_f \frac{a^{1-q}_f}{\|\vf{R}^*_f\|^q} \vf{S}_f \vf{R}^*_f \right]^{-1}
 \!\cdot
 \left[ \sum_f \frac{a^{1-q}_f}{\|\vf{R}^*_f\|^q} \vf{S}_f \Delta \phi^*_f \right]
\end{equation}
%^c
Obviously, the weight vectors in \eqref{eq: generic gradient with interpolation} differ from 
\eqref{eq: TG} in that they have an extra factor $a^{1-q}_f$. So, choosing $\vf{P}_f = \vf{c}'_f$ 
is 
equivalent to choosing $\vf{P}_f = \vf{C}_f$ except that the weight vectors \eqref{eq: TG} are 
multiplied by $a^{1-q}_f$. We will refer to the scheme \eqref{eq: TG} with $\vf{P}_f = \vf{c}'_f$ 
as TGI($q$) gradient (TG with interpolation). For $q = 1$, we have $a^{1-q}_f = 1$ and hence 
gradients TG(1) and TGI(1) are identical.

\section{Self-corrected Green-Gauss gradients}
\label{sec: SCGG}

In this section we briefly review the ``quasi-Green'' (QG) gradient of \cite{Brenner_1996, 
Pont_2017} and show that it is identical to a member of the Taylor-Gauss gradient family. 
Furthermore, we will show that all members of the TG family can be viewed as self-corrected GG 
gradients.

The starting point for most Green-Gauss gradients is the following equation, derived from 
application of the Gauss (divergence) theorem to cell 0 ($\Omega_0$ denotes its volume) -- see 
\cite{Syrakos_2017} for details of the derivation:
\begin{equation} \label{eq: GG precursor}
    \nabla \phi(\vf{C}_0) \;=\; \frac{1}{\Omega_0} \sum_f \vf{S}_f \phi(\vf{c}_f) \;+\; O(h)
\end{equation}
To proceed further, one must express the values of $\phi$ at the face centroids, $\phi(\vf{c}_f)$, 
as a function of the values at the cell centroids $\vf{C}_i$, via interpolation. The most popular 
schemes substitute $\phi(\vf{c}_f)$ in \eqref{eq: GG precursor} with the value of $\phi$ at 
$\vf{c}'_f$ or at $\vf{m}_f$ or at some other point on the line joining $\vf{C}_0$ to $\vf{C}_f$ 
where $\phi$ can be linearly interpolated. However, this is only an $O(h)$ approximation to 
$\phi(\vf{c}_f)$ and introduces an $O(1)$ error into Eq.\ \eqref{eq: GG precursor}, making the 
resulting scheme inconsistent -- see \cite{Syrakos_2017} for details. Alternatively, a second-order 
accurate interpolation for $\phi(\vf{c}_f)$ has been employed:
\begin{equation} \label{eq: interpolation usual}
    \phi(\vf{c}_f) \;=\; (1-a_f) \phi_0 \;+\; a_f \phi_f \;+\; 
    (\vf{c}_f - \vf{c}'_f) \cdot \nabla \phi(\vf{c}'_f) \;+\; O(h^2)
\end{equation}
where $\phi_i = \phi(\vf{C}_i)$ and $a_f$ is the interpolation factor \eqref{eq: a_f} (thus the sum 
of the first two terms of the right-hand side of \eqref{eq: interpolation usual} is the linearly 
interpolated value of $\phi$ at point $\vf{c}'_f$). The gradient in the right-hand side is usually 
linearly interpolated from the gradients computed at $\vf{C}_0$ and $\vf{C}_f$. In this case, 
substituting \eqref{eq: interpolation usual} into \eqref{eq: GG precursor} results in a consistent 
GG gradient where the gradient at each cell is coupled to the gradients at all neighbouring cells, 
thus requiring either iterations \cite{Karimian_2006, Traore_2009, Syrakos_2017} or solution of a 
large linear system \cite{Betchen_2010} to obtain the values of the gradient at all grid cells. 
This 
significant computational overhead makes the scheme unattractive.

However, it is not hard to see that linear interpolation for the gradient in \eqref{eq: 
interpolation usual} is superfluous. Linear interpolation is used for 2nd-order accuracy, but the 
gradients at $\vf{C}_0$ and $\vf{C}_f$ are likely only 1st-order accurate anyway. Furthermore, a 
first-order accurate approximation of the gradient at $\vf{c}'_f$ is all that is needed in 
\eqref{eq: interpolation usual} because the $O(h)$ error of the gradient, multiplied by $(\vf{c}_f 
- \vf{c}'_f)$, contributes only a $O(h^2)$ error to \eqref{eq: interpolation usual}. The most 
convenient 1st-order approximation to $\nabla\phi(\vf{c}'_f)$ is just $\nabla \phi (\vf{C}_0)$, 
because this avoids any coupling between the gradients of neighbouring cells\footnote{A  
consequence 
of this choice is that $\phi(\vf{c}_f)$ is calculated differently for the two cells sharing face 
$f$.}. So, replacing $\nabla \phi (\vf{C}_0)$ for $\nabla \phi (\vf{c}'_f)$ in \eqref{eq: 
interpolation usual} and substituting the latter expression in \eqref{eq: GG precursor} gives:
\begin{equation}
    \nabla \phi( \vf{C}_0 ) 
    \;=\;
    \frac{1}{\Omega_0} \sum_f \vf{S}_f \left[ (1-a_f) \phi_0 \;+\; a_f \phi_f 
\right]
    \;+\;
    \frac{1}{\Omega_0} \sum_f \vf{S}_f (\vf{c}_f - \vf{c}'_f) \cdot \nabla \phi(\vf{C}_0) 
    \;+\; O(h)
\end{equation}
which can be solved for the gradient $\nabla\phi(\vf{C}_0)$ to obtain:
\begin{equation} \label{eq: QG}
 \nabla \phi(\vf{C}_0)
 \;=\; 
 \frac{1}{\Omega_0} \left[
   \tf{I} \;-\; \frac{1}{\Omega_0} \sum_f \vf{S}_f (\vf{c}_f - \vf{c}'_f) \right]^{-1}
 \sum_f \vf{S}_f \bar{\phi}(\vf{c}'_f)
 \;+\; O(h)
\end{equation}
where $\bar{\phi}(\vf{c}'_f) = (1-a_f) \phi_0 + a_f \phi_f$ is the linearly interpolated value of 
$\phi$ at $\vf{c}'_f$, and $\tf{I}$ is the identity tensor. It is noted that if $\vf{c}_f - 
\vf{c}'_f = 0$ at all faces then the scheme \eqref{eq: QG} reduces to the common  GG gradient; 
however, in the presence of skewness ($\vf{c}_f - \vf{c}'_f \neq 0$), the gradient \eqref{eq: QG} 
uses its own self (through Eq.\ \eqref{eq: interpolation usual}) to apply a correction that retains 
consistency. So, scheme \eqref{eq: QG} is a self-corrected, consistent, first-order accurate GG 
gradient. It has been used in \cite{Brenner_1996, Pont_2017} where it is termed ``quasi-Green'' (QG) 
gradient. 

Interestingly, the scheme \eqref{eq: QG} can be re-written in a familiar form. Consider first the 
tensor being inverted (in square brackets in \eqref{eq: QG}). By writing $\vf{c}_f - \vf{c}'_f = 
(\vf{c}_f - \vf{C}_0) - (\vf{c}'_f - 
\vf{C}_0)$ it can be seen to equal
\begin{equation} \label{eq: QG manipulation 1}
    \tf{I} \;-\; \frac{1}{\Omega_0} \sum_f \vf{S}_f (\vf{c}_f - \vf{c}'_f)
    \;=\;
    \tf{I} \;-\; \frac{1}{\Omega_0} \sum_f \vf{S}_f (\vf{c}_f - \vf{C}_0)
           \;+\; \frac{1}{\Omega_0} \sum_f \vf{S}_f (\vf{c}'_f - \vf{C}_0)
\end{equation}
The second term on the right-hand side happens to equal $\tf{I}$, as shown in the Appendix, so it
cancels out with the first term. In the remaining term, if we define $\vf{R}_f = \vf{C}_f - 
\vf{C}_0$, we can express $\vf{c}'_f - \vf{C}_0 = a_f \vf{R}_f$. Finally, we can manipulate the 
vector multiplied by this inverse tensor in \eqref{eq: QG} as follows:
\begin{equation} \label{eq: QG manipulation 2}
    \sum_{f=1}^F \vf{S}_f \left[(1-a_f) \phi_0 + a_f \phi_f\right]
    \;=\; 
    \sum_{f=1}^F \vf{S}_f \left[ \phi_0 -a_f \phi_0 + a_f \phi_f \right]
    \;=\;
    \phi_0 \sum_{f=1}^F \vf{S}_f \;+\; \sum_{f=1}^F \vf{S}_f a_f \Delta \phi_f
\end{equation}
where $\Delta \phi_f = \phi_f - \phi_0$. The first term of the right-hand side of \eqref{eq: QG
manipulation 2} is zero because $\sum_f \vf{S}_f = \vf{0}$.

Putting everything together, the QG gradient \eqref{eq: QG} can also be expressed as
\begin{equation} \label{eq: QG as TG}
 \nabla \phi(\vf{C}_0)
 \;=\; 
 \left[ \sum_f a_f \vf{S}_f \vf{R}_f \right]^{-1}
 \sum_f a_f \vf{S}_f \Delta \phi_f
\end{equation}
($1/\Omega_0$ cancels with $(1/\Omega_0)^{-1}$, and the truncation error has been dropped). But 
this is just the TGI(0) gradient (compare with \eqref{eq: generic gradient with interpolation}, for 
$q=0$). 

In fact, all members of the Taylor-Gauss family can be viewed as self-corrected Green-Gauss 
gradients. This is because, although we have assumed that point $\vf{c}'_f$ is the projection of 
$\vf{c}_f$ onto the line through $\vf{C}_0$ and $\vf{C}_f$, there is nothing in the above analysis 
that forces this choice. The derivation of the QG gradient \eqref{eq: QG}, or equivalently 
\eqref{eq: QG as TG}, is equally valid wherever we choose to place $\vf{c}'_f$ on that line; it 
could even be placed beyond point $\vf{C}_f$, in which case $a_f > 1$ and $\phi(\vf{c}'_f)$ is 
linearly extrapolated rather than interpolated. Taylor-Gauss gradients are characterised by the 
fact that the weight vectors $\vf{V}_f$ are parallel to the face vectors $\vf{S}_f$, and what 
distinguishes between different TG variants is the magnitudes of $\vf{V}_f$. The same effect can be 
accomplished from the QG perspective by varying the locations of the $\vf{c}'_f$ points. For 
example, comparing Eq.\ \eqref{eq: TG} with Eq.\ \eqref{eq: QG as TG} we see that the TG($q$) 
gradient is equivalent to a QG gradient with the interpolation points placed so that $a_f = 
1/\| \vf{R}_f \|^q$ (or $a_f = c/\| \vf{R}_f \|^q$, where the value $c$ is the same for all 
faces).

\section{Some comments}
\label{sec: comments}

\subsection{Inverse-distance weighting and face-area weighting}
\label{ssec: distance and area weighting}

Inverse-distance weighting is quite beneficial for least-squares gradients, but somewhat less so 
for Taylor-Gauss gradients, as the numerical results that follow will show. This is because the 
latter's weights include the face areas $\|\vf{S}_f\|$, which can achieve a similar effect as 
inverse-distance weighting. Usually, neighbours $\vf{C}_f$ that lie across large faces are 
relatively closer to $\vf{C}_0$ than neighbours across small faces, because a large face ``pushes'' 
other neighbours to lie at greater distances.

It is noteworthy that, as inspection of the minimisation quantity \eqref{eq: LSD minimisation goal} 
reveals, ``inverse-distance'' weighting is actually in effect only for $q > 2$; for $q < 2$, it is 
in fact the farthest points (larger $\|\vf{R}_f\|$) that are prioritised for the minimisation of 
the finite difference error $\Delta \phi_f / \|\vf{R}_f\| - \hat{\vf{d}}_f \cdot \tilde{\nabla} 
\phi(\vf{C}_0)$. This holds for all least-squares gradients (replace $\Theta_f$ in \eqref{eq: LSD 
minimisation goal} with $1$ or $\|\vf{S}_f\|$ for the LS($q$) and LSA($q$) gradients, respectively).

\subsection{Order of accuracy: special cases}
\label{ssec: accuracy special cases}

In Sec.\ \ref{sec: general framework} it was shown that all the gradients that fall into the 
general framework examined here are at least first-order accurate (Eq.\ \eqref{eq: generic 
gradient}). However, in some special cases the order of accuracy may increase to 2; in particular, 
this happens when the gradients are applied on smooth structured grids, because under these 
circumstances the leading term of the truncation error becomes zero, or converges to zero at a 
second-order rate rather than first-order.

Considering the terms of Eq.\ \eqref{eq: new gradient with O(h^3)} that are truncated in order to 
arrive at the generic gradient \eqref{eq: generic gradient}, we see that for the leading term to be 
zero for all functions $\phi$ the tensor $\sum_f \vf{V}_f \vf{R}_f \vf{R}_f$ must equal zero. The 
$(i,j,k)$ component of this third order tensor is $\sum_f V_{f,i} R_{f,j} R_{f,k}$, where $V_{f,i}$ 
is the $i$-th component of $\vf{V}_f$ etc. For the least-squares gradients, the components of this 
tensor become
\begin{equation} \label{eq: least squares error components}
 \sum_f V_{f,i} R_{f,j} R_{f,k} \;=\;
 \sum_f \frac{\beta_f}{\|\vf{R}_f\|^q} R_{f,i} R_{f,j} R_{f,k}
\end{equation}
where $\beta_f$ is either $1$, for LS($q$), or $\|\vf{S}_f\|$, for LSA($q$), or $\Theta_f$, for 
LSD($q$). Likewise, the components for the Taylor-Gauss gradients are
\begin{equation} \label{eq: Taylor Gauss error components}
  \sum_f V_{f,i} R_{f,j} R_{f,k}
  \;=\;
  \sum_f \frac{\beta_f S_f}{\|\vf{R}_f\|^q} (\hat{\vf{s}_f} \cdot \hat{\vf{e}_i}) R_{f,j} R_{f,k}
\end{equation}
where $\hat{\vf{e}}_i$ is the unit vector in the $i$-th coordinate direction, and $\beta_f$ is 
either $1$, for TG($q$), or $a_f$, for TGI($q$).

One situation where all these tensor components are zero is when the contributions from opposite 
faces or neighbours of a cell cancel out. Suppose that faces $f = 1$ and $f = 2$ are opposite to 
each other and geometrically similar so that $\vf{R}_1 = -\vf{R}_2$, $\beta_1 = \beta_2$, and 
$\vf{S}_1 = -\vf{S}_2$ (if used). Then it is easy to see that the contributions of these two faces 
to the sums \eqref{eq: least squares error components} and \eqref{eq: Taylor Gauss error components} 
cancel out. If all faces of the cell can be paired this way, then the first-order error will vanish 
and the gradient will be second-order accurate. This situation occurs on Cartesian grids, but also 
on smooth structured grids, if they are refined in a way that skewness and unevenness diminish on 
finer grids. In the latter case, the conditions  $\vf{R}_1 = -\vf{R}_2$, $\beta_1 = \beta_2$, and 
$\vf{S}_1 = -\vf{S}_2$ do not hold perfectly, but hold in the limit of infinite grid fineness. This 
topic is discussed in detail in \cite{Syrakos_2017}.

Unfortunately, even if the grid is structured, the conditions for 2nd-order accuracy will, in 
general, not be met at boundary cells, because if face 1 is a boundary face, and the opposite face 2 
is an internal face, then $\vf{R}_1 = \vf{c}_1 - \vf{C}_0 \neq -\vf{R}_2 = -(\vf{C}_2 - \vf{C}_0)$. 
The two vectors, although parallel and facing in opposite directions, have different lengths. 
Nevertheless, there are particular choices of the exponent $q$ that make the lengths of the $\vf{R}$ 
vectors irrelevant. These choices are $q=3$ for least squares gradients\footnote{This special 
choice was noted also in \cite{Syrakos_2017}, but there it was the ``$q=3/2$'' choice, because of a 
different definition of $q$. In particular, as noted in Sec.\ \ref{ssec: linear algebra}, using 
$\mf{V}=\mf{DR}$ minimises the norm of $\mf{D}^{1/2}(\mf{b}-\mf{Rx})$. In the present paper, $q$ 
is the exponent that appears in $\mf{D}$, while in \cite{Syrakos_2017} it is the exponent that 
appears in $\mf{D}^{1/2}$. If $q$ is the present exponent and $q'$ is the corresponding one of 
\cite{Syrakos_2017}, then $q'=q/2$.}, and $q=2$ for Taylor-Gauss gradients. For $q=3$, the component 
\eqref{eq: least squares error components} can be written as $\sum_f \beta_f 
(R_{f,i}/\|\vf{R}_f\|)(R_{f,j}/\|\vf{R}_f\|)(R_{f,k}/\|\vf{R}_f\|)$. Each of the ratios 
$R_{f,i}/\|\vf{R}_f\|$ etc.\ is the cosine of an angle related to the direction of $\vf{R}_f$, and 
is therefore independent of the length of $\vf{R}_f$. So, if $\vf{R}_1$ and $\vf{R}_2$ are parallel 
but point in opposite directions, then these ratios for $\vf{R}_1$ will be equal and opposite to 
those for $\vf{R}_2$, and the two contributions will cancel each other out in the sum \eqref{eq: 
least squares error components}. This situation will occur at boundary cells of structured grids. An 
inspection of component \eqref{eq: Taylor Gauss error components} shows that the same effect occurs 
for Taylor-Gauss gradients if $q=2$, but not for the interpolated version: for the TGI(2) gradient, 
$\beta_1 = a_1 \neq \beta_2 = a_2$ and the contributions of faces 1 and 2 do not cancel each other 
out.

Summarising, on smooth structured grids, all the gradients examined herein that are described by 
the generic formula \eqref{eq: generic gradient} are 2nd-order accurate at the interior cells and 
1st-order accurate on boundary cells, except for LS(3), LSA(3), LSD(3) and TG(2), which are 
2nd-order accurate even there.

\subsection{Subtle difference between the TG and QG gradients}
\label{ssec: TG and QG difference}

Since the Gauss theorem converts a volume integral into a surface integral over the whole surface 
of the cell, Green-Gauss gradients must normally include all faces of the given cell in the 
calculation. However, viewed from the Taylor-Gauss perspective, as a projection scheme, the 
gradient \eqref{eq: QG as TG} has no such restriction. We can use only a subset of the cell's 
faces.  For example, if the gradient is used for extrapolating a variable (e.g.\ pressure or stress) 
to a boundary, then the boundary face itself may be omitted from the gradient calculation.

\section{Numerical tests}
\label{sec: tests}

\subsection{Consistency and order of accuracy}
\label{ssec: results consistency}

We begin testing the gradients on the grids that were employed in \cite{Syrakos_2017}, which 
consist of quadrilateral cells (Fig.\ \ref{fig: grids}). The grid types considered differ with 
respect to the effect that grid refinement has on qualities such as skewness, unevenness and 
non-orthogonality. The design of the experiments is based on the fact that the observed order of 
accuracy may depend on these qualities; in particular, the common Green-Gauss gradient's 
inconsistency is triggered by skewness \cite{Syrakos_2017} (also by unevenness \cite{Sozer_2014} or 
non-orthogonality \cite{Deka_2018} for some variants). The types of grids employed are:

\begin{figure}[tb!]
    \centering
    \begin{subfigure}[b]{0.32\textwidth}
        \centering
        \includegraphics[width=0.98\linewidth]{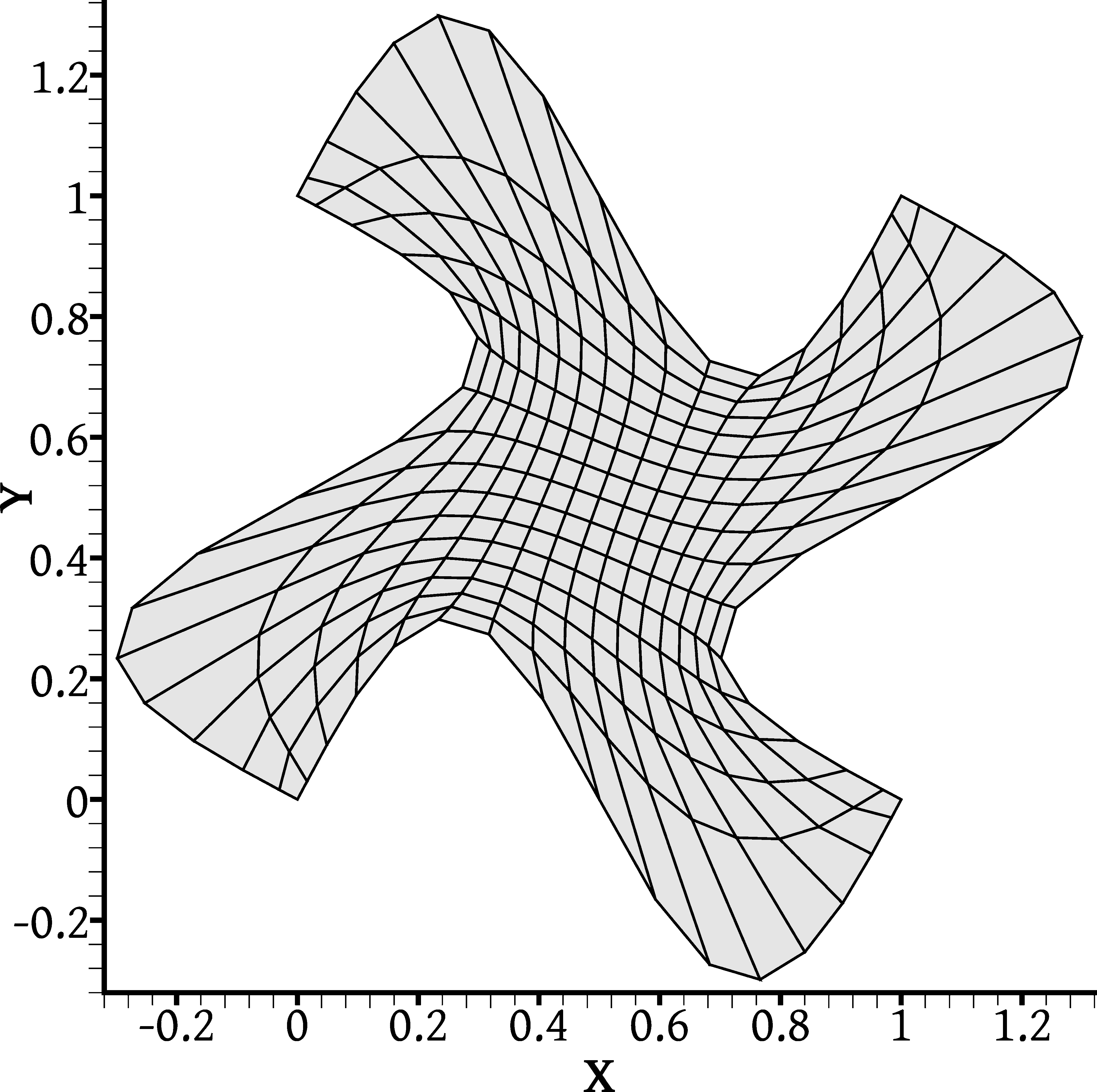}
        \caption{Elliptic grid, $l=3$}
        \label{sfig: grid elliptic}
    \end{subfigure}
    \begin{subfigure}[b]{0.32\textwidth}
        \centering
        \includegraphics[width=0.98\linewidth]{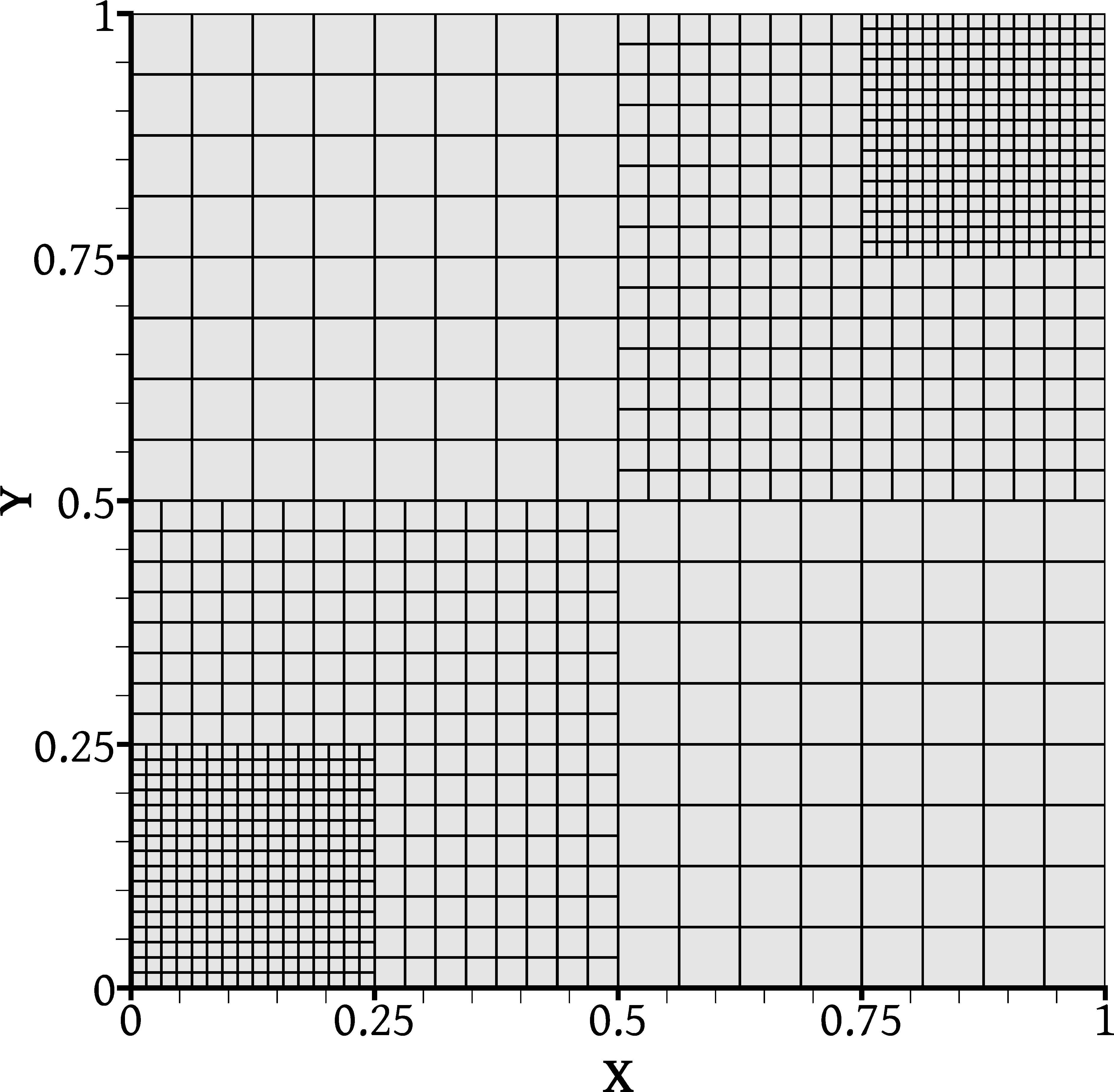}
        \caption{Refined grid, $l=3$}
        \label{sfig: grid refined}
    \end{subfigure}
    \begin{subfigure}[b]{0.32\textwidth}
        \centering
        \includegraphics[width=0.98\linewidth]{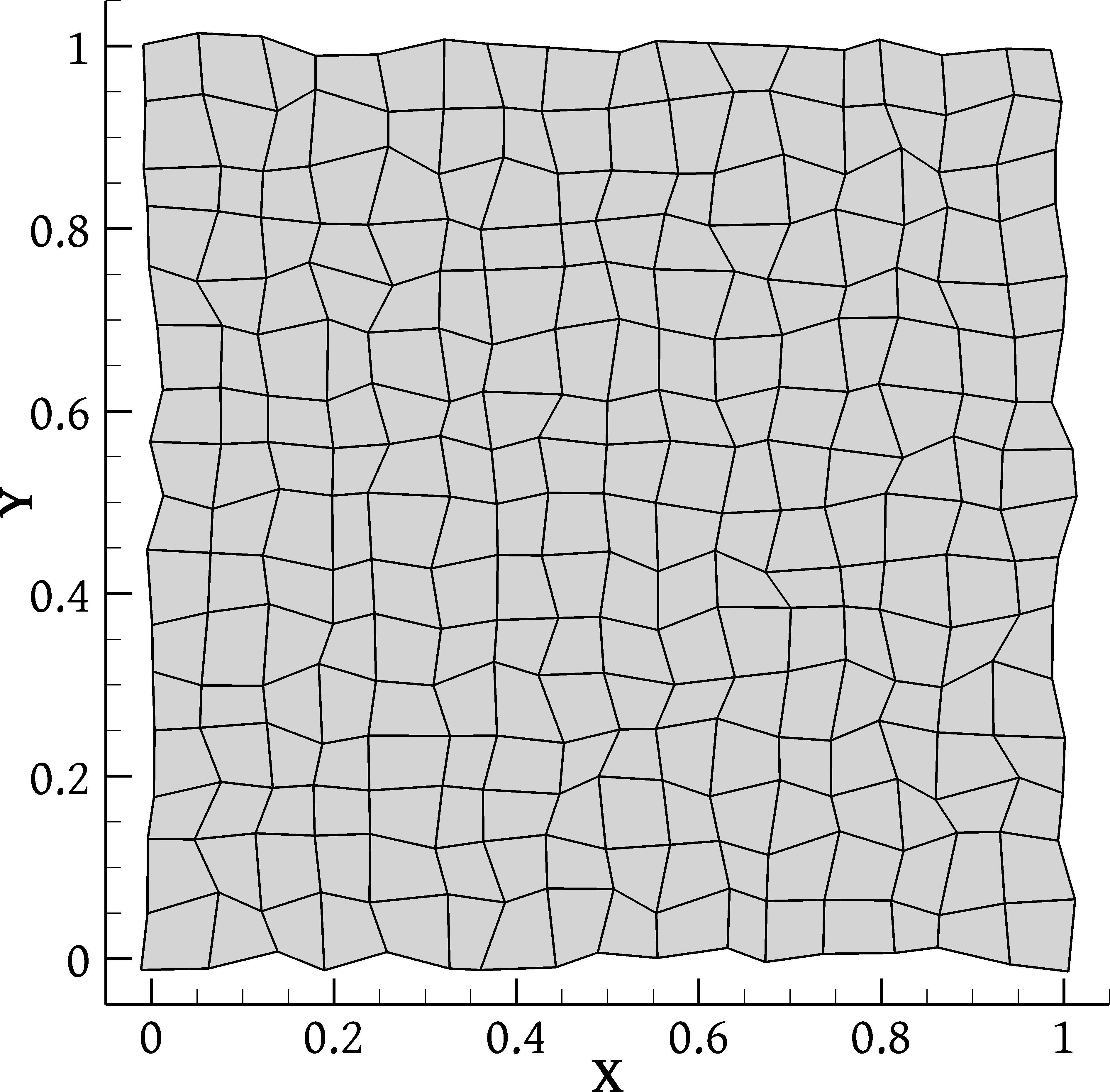}
        \caption{Perturbed grid, $l=3$}
        \label{sfig: grid random}
    \end{subfigure}
    \caption{Grids of different kinds, at the $l = 3$ level of refinement.}
  \label{fig: grids}
\end{figure}

\begin{figure}[tb!]
    \centering
    \begin{subfigure}[b]{0.49\textwidth}
        \centering
        \includegraphics[width=0.99\linewidth]{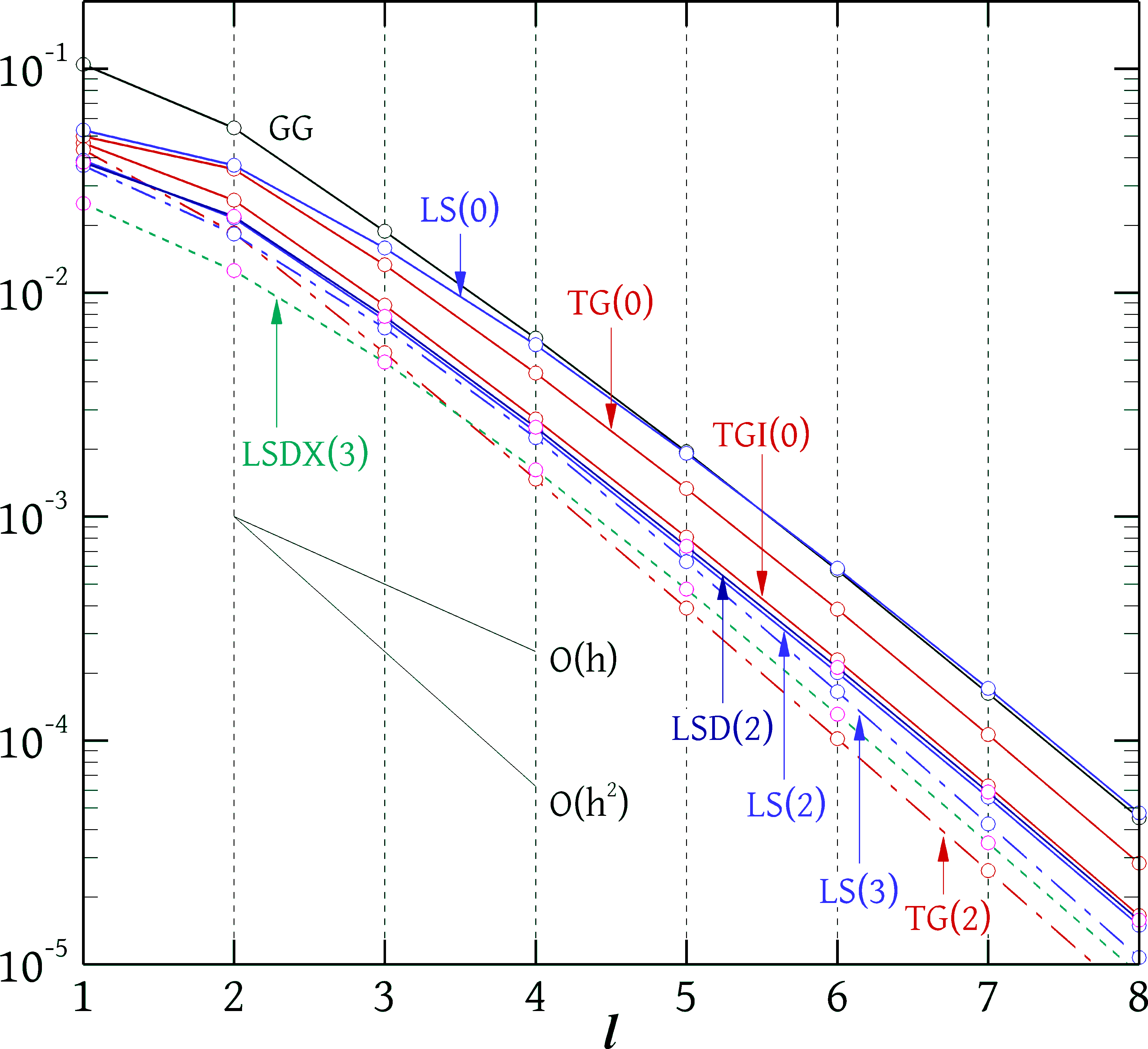}
        \caption{Mean errors}
        \label{sfig: mean elliptic}
    \end{subfigure}
    \begin{subfigure}[b]{0.49\textwidth}
        \centering
        \includegraphics[width=0.99\linewidth]{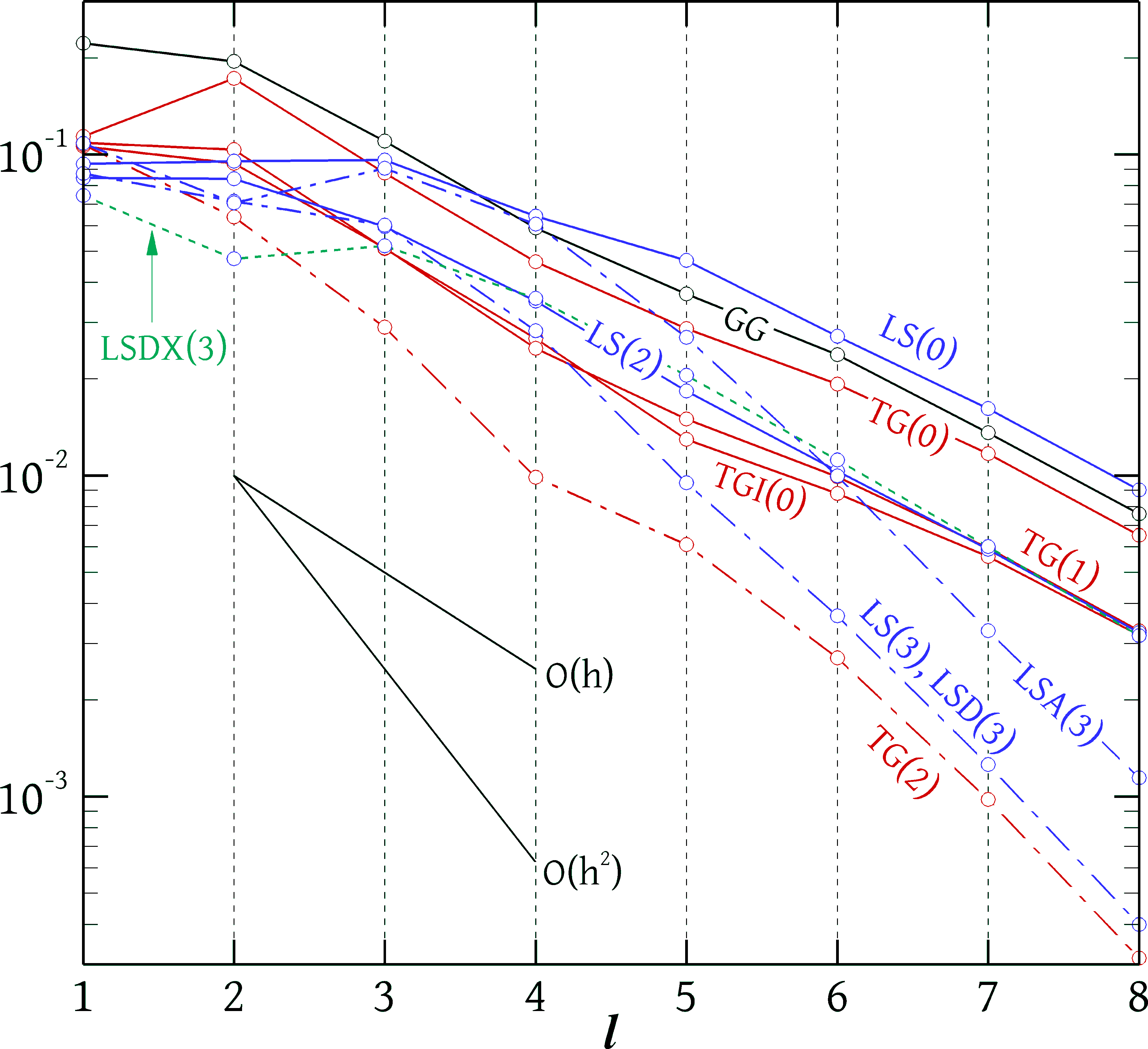}
        \caption{Maximum errors}
        \label{sfig: max elliptic} 
    \end{subfigure}
    \caption{Minimum and maximum errors of gradient schemes versus refinement level $l$, for the 
elliptic grids (Fig.\ \ref{sfig: grid elliptic}).}
  \label{fig: errors elliptic}
\end{figure}

\begin{figure}[tb!]
  \centering
  \includegraphics[width=0.99\textwidth]{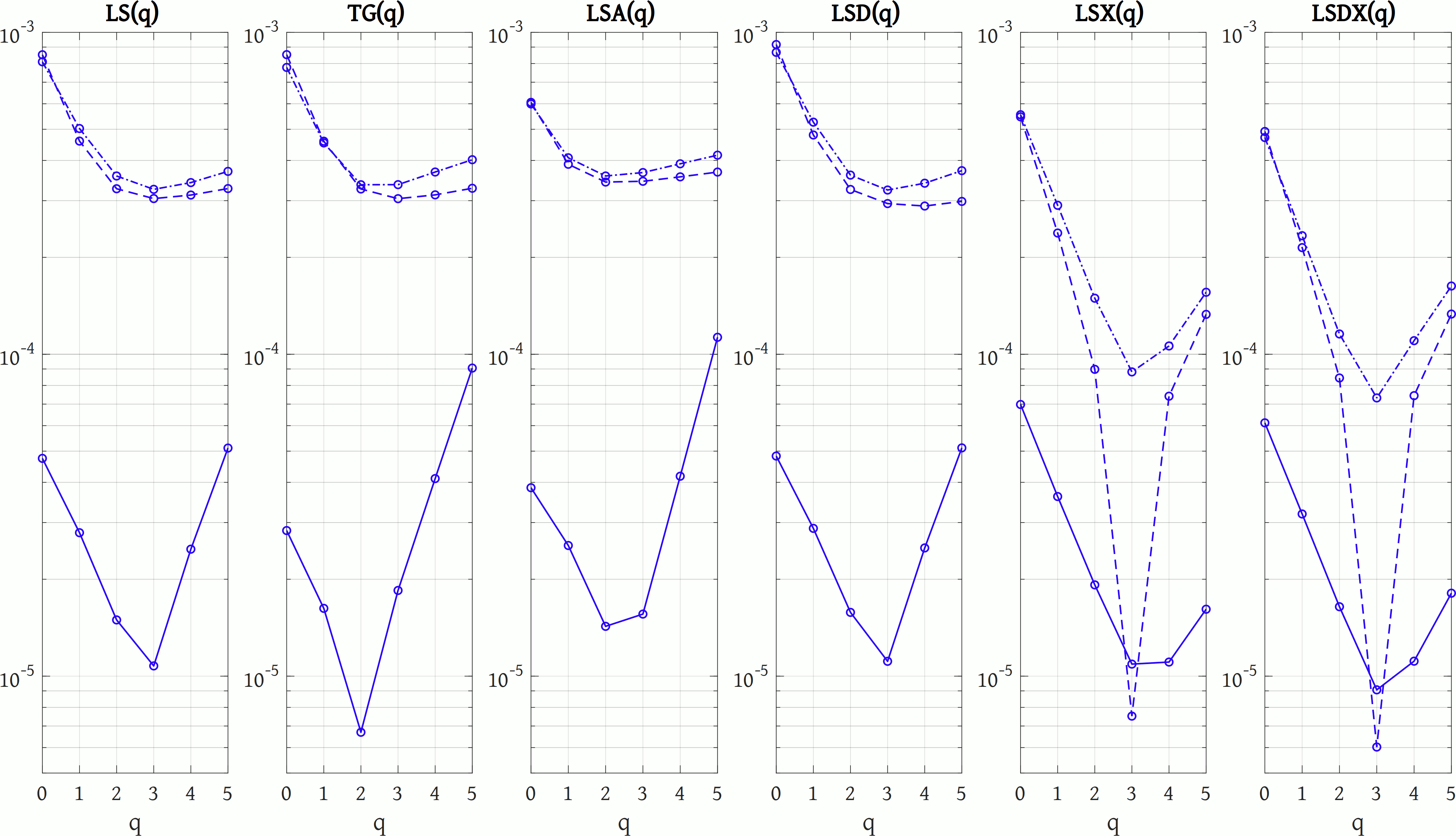}
  \caption{Average errors of various gradient schemes as a function of the exponent $q$, on the 
finest ($l=8$) elliptic grids. Continuous lines: original structured grid (Fig.\ \ref{sfig: grid 
elliptic}). Dashed lines: orderly triangulated grid (Fig.\ \ref{sfig: grid elliptic tri ordered}). 
Dash-dot lines: randomly triangulated grids (Fig.\ \ref{sfig: grid elliptic tri random}).}
  \label{fig: l8 errors elliptic mean}
\end{figure}

\begin{figure}[tb!]
  \centering
  \includegraphics[width=0.99\textwidth]{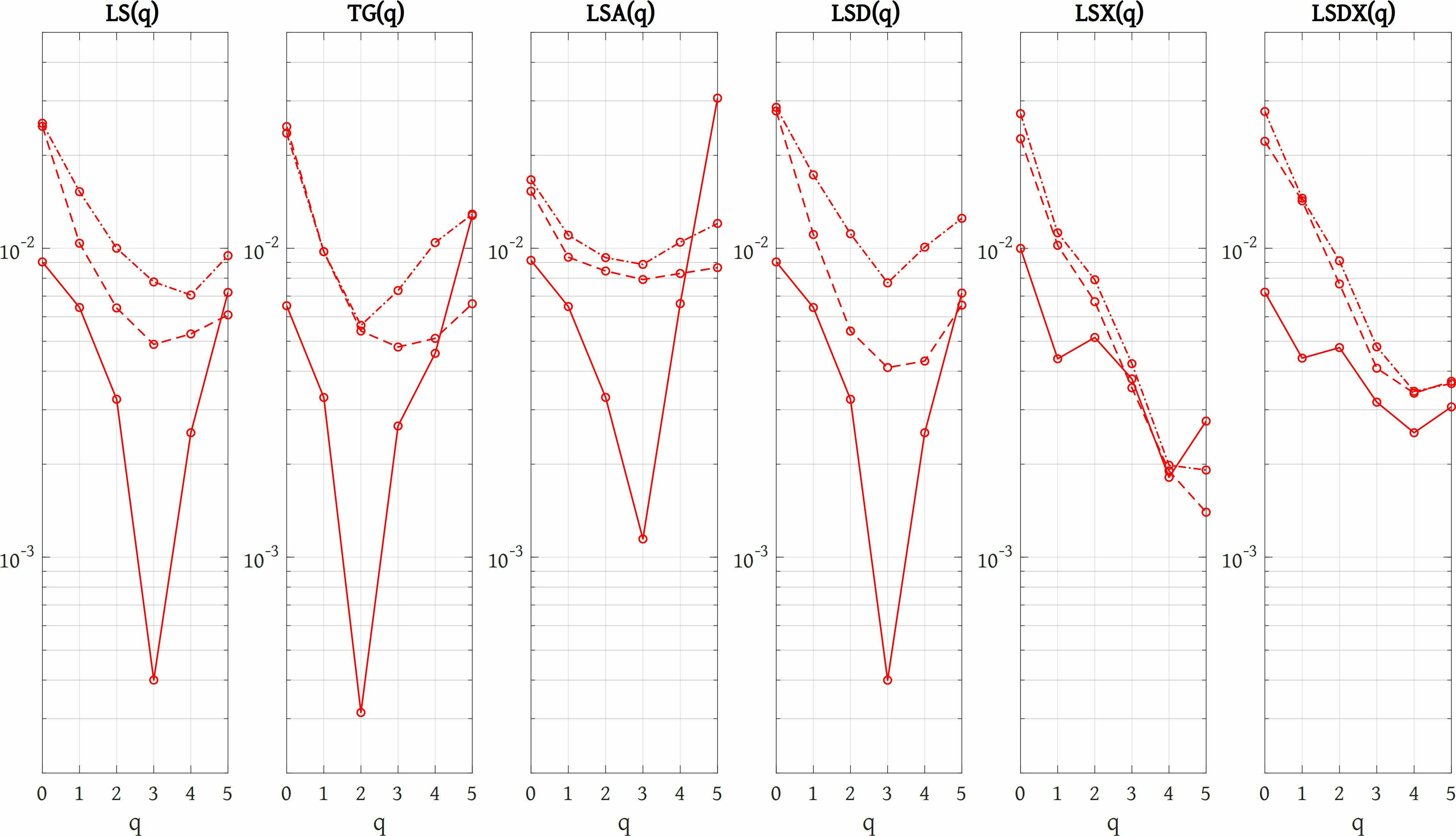}
  \caption{Maximum errors of various gradient schemes as a function of the exponent $q$, on the 
finest ($l=8$) elliptic grids. Continuous lines: original structured grid (Fig.\ \ref{sfig: grid 
elliptic}). Dashed lines: orderly triangulated grid (Fig.\ \ref{sfig: grid elliptic tri ordered}). 
Dash-dot lines: randomly triangulated grids (Fig.\ \ref{sfig: grid elliptic tri random}).}
  \label{fig: l8 errors elliptic max}
\end{figure}

\begin{figure}[tb!]
    \centering
    \begin{subfigure}[b]{0.49\textwidth}
        \centering
        \includegraphics[width=0.99\linewidth]{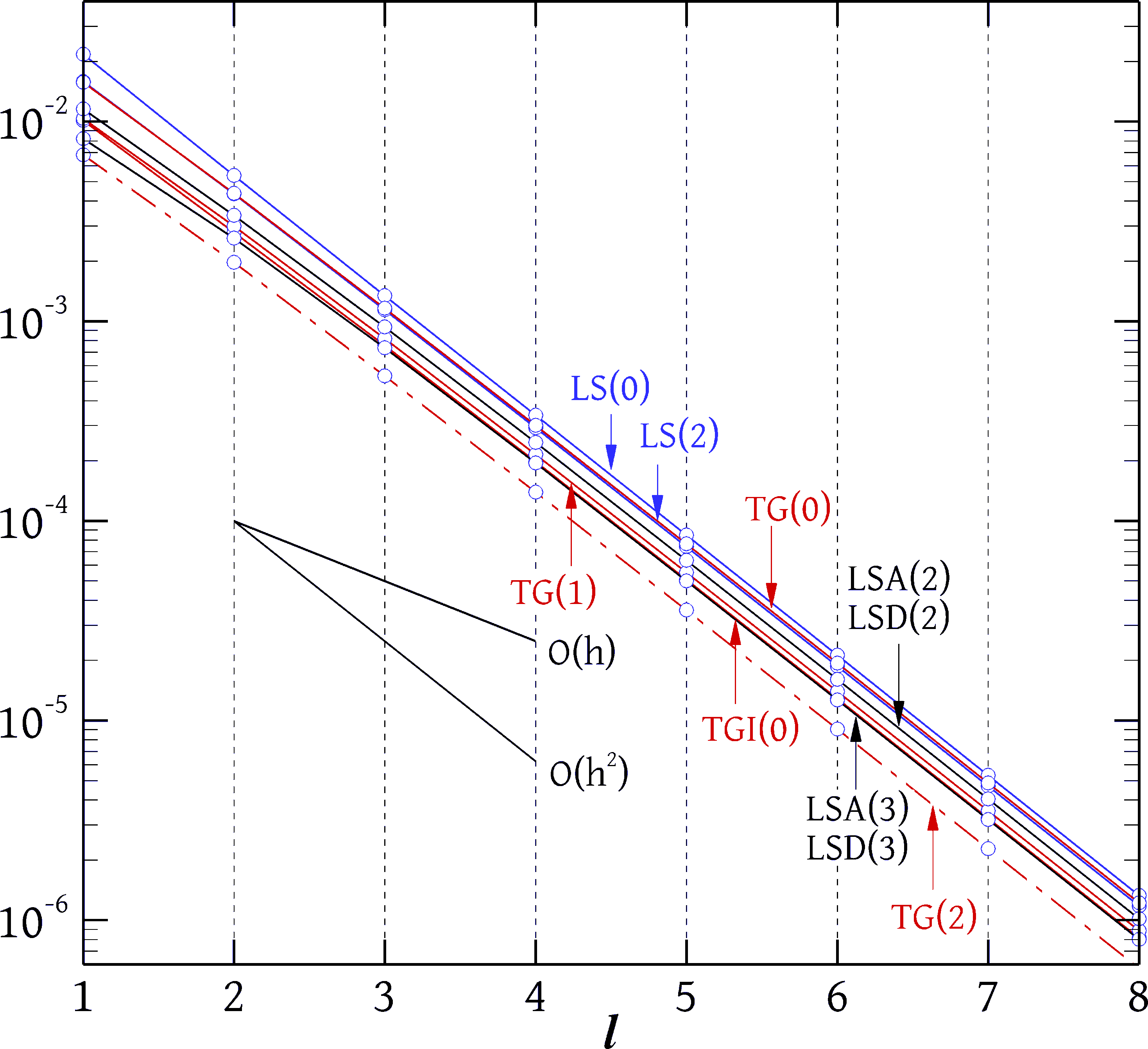}
        \caption{Mean errors}
        \label{sfig: mean refined}
    \end{subfigure}
    \begin{subfigure}[b]{0.49\textwidth}
        \centering
        \includegraphics[width=0.99\linewidth]{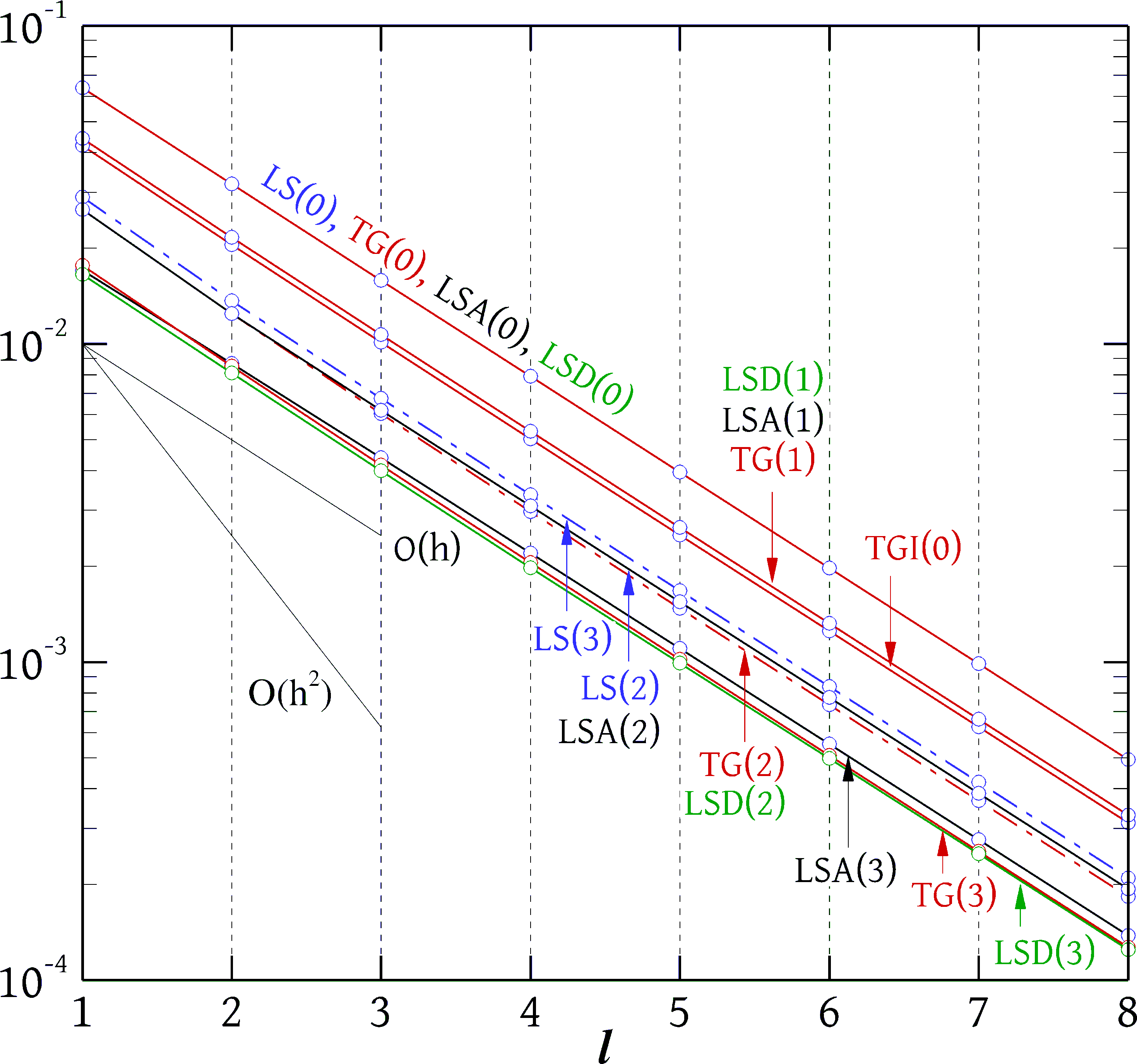}
        \caption{Maximum errors}
        \label{sfig: max refined} 
    \end{subfigure}
    \caption{Minimum and maximum errors of various gradients versus refinement level $l$, for the 
locally refined grids (Fig.\ \ref{sfig: grid refined}).}
  \label{fig: errors refined}
\end{figure}

\begin{figure}[tb!]
  \centering
  \includegraphics[width=0.75\textwidth]{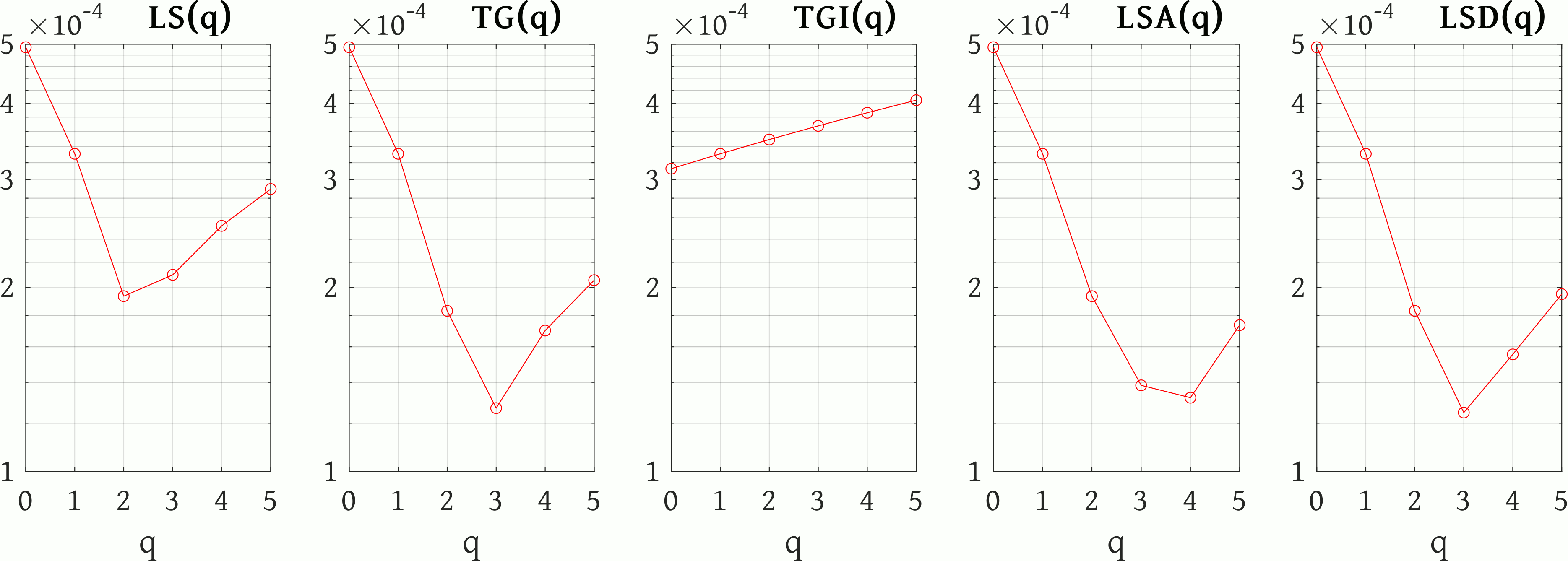}
  \caption{Maximum errors of various gradient schemes as a function of the exponent $q$, on the 
finest ($l=8$) locally refined grid.}
  \label{fig: l8 errors quad refined max}
\end{figure}

\begin{figure}[tb!]
  \centering
  \includegraphics[width=0.99\textwidth]{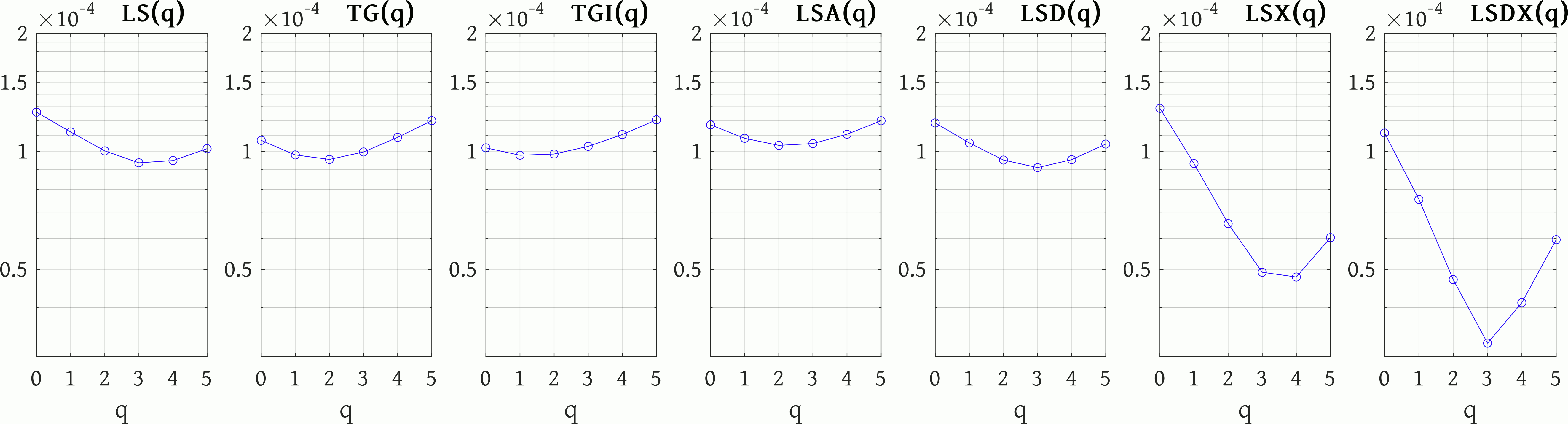}
  \caption{Average errors of various gradient schemes as a function of the exponent $q$, on the 
finest ($l=8$) randomly perturbed grid.}
  \label{fig: l8 errors quad distorted mean}
\end{figure}

\begin{figure}[tb!]
  \centering
  \includegraphics[width=0.99\textwidth]{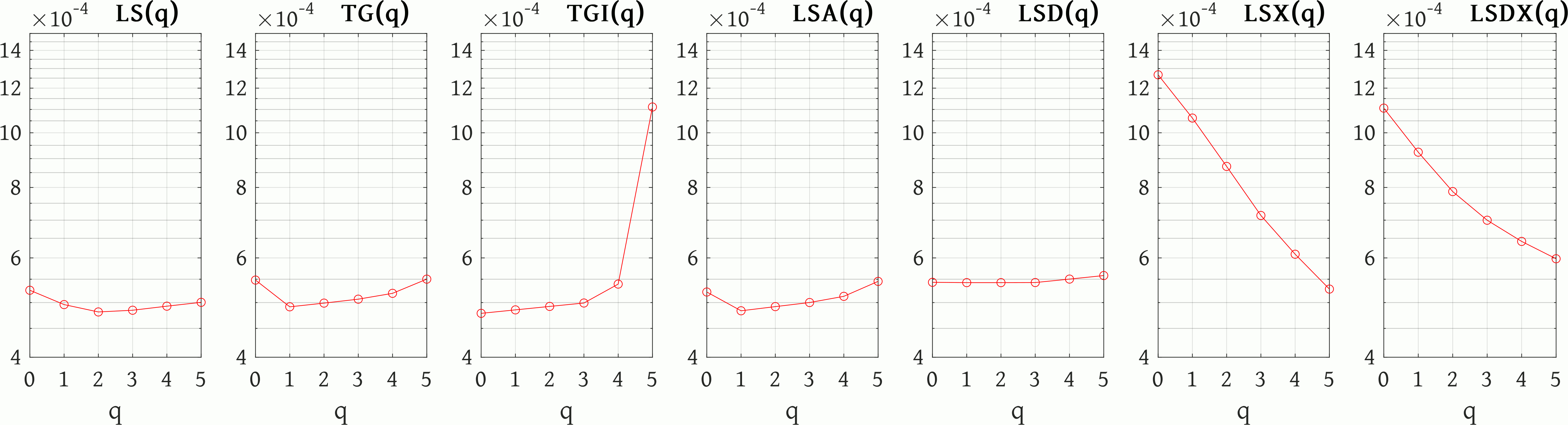}
  \caption{Maximum errors of various gradient schemes as a function of the exponent $q$, on the 
finest ($l=8$) randomly perturbed grid.}
  \label{fig: l8 errors quad distorted max}
\end{figure}

\begin{enumerate}[(a)]
 \item \label{enum: structured grids} Structured grids generated by solving a set of elliptic 
partial differential equations, such as that depicted in Fig.\ \ref{sfig: grid elliptic} (see 
\cite{Syrakos_2017} for construction details). Such grids exhibit skewness, unevenness and 
non-orthogonality, but the first two qualities diminish with refinement.
 \item \label{enum: refined grids} Cartesian grids with local refinement patches, such as that 
depicted in Fig.\ \ref{sfig: grid refined}. Finer grids are obtained by splitting every cell, 
including those of the patches, into four smaller cells. Therefore, all finer grids are similarly 
patched, and grid qualities are not affected by refinement. These grids are characterised by large, 
localised skewness and unevenness (confined near the patch interfaces) that do not diminish with 
refinement.
\item \label{enum: distorted grids} Grids derived from Cartesian ones by random perturbation of 
their nodes, such as that of Fig.\ \ref{sfig: grid random} -- see \cite{Syrakos_2017} for details. 
Skewness, unevenness and non-orthogonality remain large throughout the grid irrespective of 
refinement.
\end{enumerate}
For each kind of grid we employ 8 levels of refinement, $l = 1, 2, \ldots 8$, with each successive 
grid having 4 times as many cells as the previous one; the grids of level $l=3$ are shown in Fig.\ 
\ref{fig: grids}. The tests concern the calculation of the gradient of the function $\phi(x,y) = 
\tanh(x) \cdot \tanh(y)$. The effect of grid refinement on the mean and maximum errors 
$\|\tilde{\nabla} \phi(\vf{P}) - \nabla \phi (\vf{P})\|$ across all grid cells, where 
$\tilde{\nabla}\phi$ is the approximate gradient and $\nabla \phi$ is the exact gradient, are 
plotted in Figs.\ \ref{fig: errors elliptic}, \ref{fig: errors refined} and \ref{fig: errors 
random}. In order to not clutter the diagrams, the errors of only a subset of the schemes are 
plotted. Following \cite{Syrakos_2017}, for grids (\ref{enum: structured grids}) and (\ref{enum: 
distorted grids}) the average error is calculated simply by dividing the total sum by the number of 
grid cells, whereas for grids (\ref{enum: refined grids}) the error of each cell is weighed by the 
cell volume to account for the patches of different fineness. The slopes of the lines in these 
figures reveal the order of accuracy of the gradients. Additionally, the errors on the finest grid 
levels ($l = 8$) as a function of the inverse-distance weighting exponent $q$ are plotted in Figs.\ 
\ref{fig: l8 errors elliptic mean}, \ref{fig: l8 errors elliptic max}, \ref{fig: l8 errors quad 
refined mean}, \ref{fig: l8 errors quad refined max}, \ref{fig: l8 errors quad distorted mean} and 
\ref{fig: l8 errors quad distorted max} (the gradients LSX and LSDX included in these figures will 
be defined and discussed in Sec.\ \ref{ssec: results extended stencils}). At this level of fineness 
all methods have reached their asymptotic rate of convergence.

On the smooth structured grids, refinement causes the mean errors (Fig.\ \ref{sfig: mean elliptic}) 
to decrease at a second-order rate, despite the gradients being nominally only first-order accurate 
(Eq.\ \eqref{eq: generic gradient}). This is due to the property that structured grids generated by 
the solution of differential equations possess, that their skewness and unevenness diminish through 
refinement (see \cite{Syrakos_2017} for an explanation). This causes truncation error component 
cancellations of the sort described in Sec.\ \ref{ssec: accuracy special cases}, increasing the 
order of the leading-order truncation error term. The same happens even with the (nominally 
inconsistent) common Green-Gauss (GG) gradient which is included for comparison purposes in Fig.\ 
\ref{fig: errors elliptic}. However, the performance of the GG gradient is amongst the poorest. The 
unweighted LS gradient also performs poorly, while the gradients' performance improves with increase 
of the weighting exponent $q$ up to $q=2$ for the TG gradients and $q=3$ for the LS gradients 
(Fig.\ \ref{fig: l8 errors elliptic mean}). For larger $q$ the performance deteriorates. The most 
accurate gradient in this case appears to be the TG(2). This gradient, together with LS(3), LSA(3) 
and LSD(3), benefit from the even more special situation described in Sec.\ \ref{ssec: accuracy 
special cases}, namely they retain their second-order accuracy even at  boundary cells, as verified 
by Fig.\ \ref{sfig: max elliptic}; it is at these cells that the maximum errors occur for the other 
gradients, seen to decrease at only a first-order rate in Fig.\ \ref{sfig: max elliptic}. Despite 
this advantageous property, the performance of the LSA gradient is comparatively poor -- see also 
Figs.\ \ref{fig: l8 errors elliptic mean}, \ref{fig: l8 errors elliptic max}. It is worth noting 
that the TGI(0) gradient (the QG gradient) performs quite well, much better than its inconsistent GG 
counterpart, and even significantly better than the un-interpolated TG(0) gradient.


\begin{figure}[tb!]
  \centering
  \includegraphics[width=0.75\textwidth]{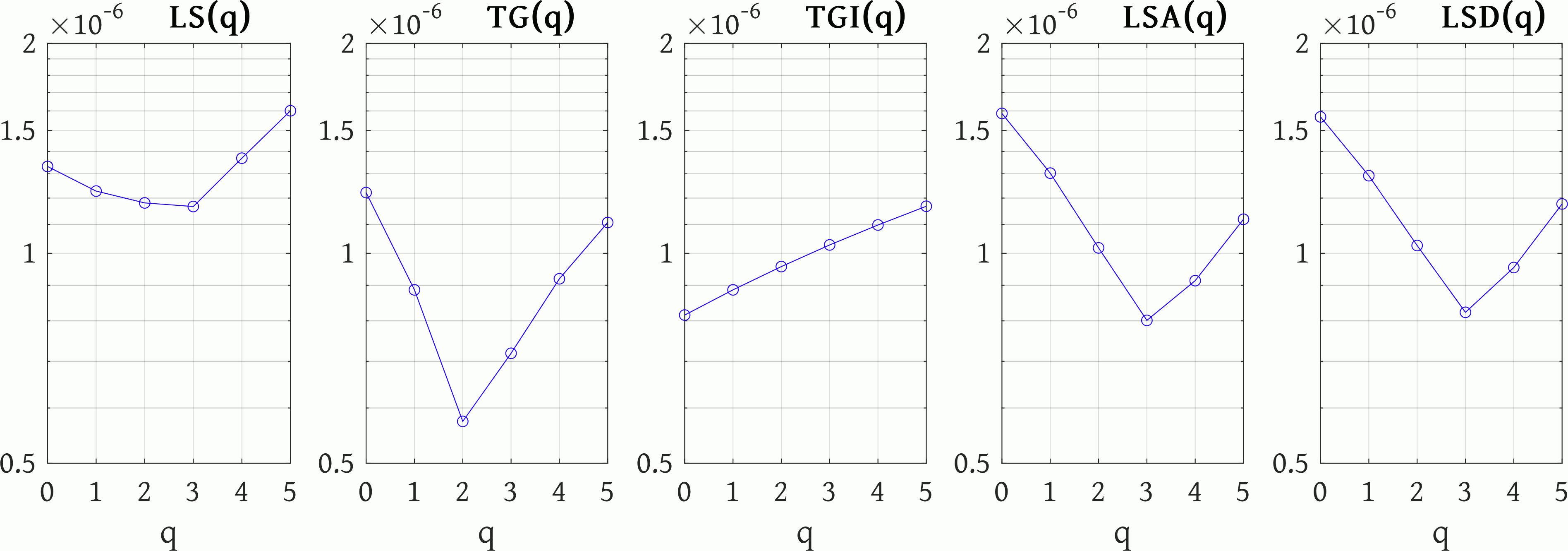}
  \caption{Average errors of various gradient schemes as a function of the exponent $q$, on the 
finest ($l=8$) locally refined grid.}
  \label{fig: l8 errors quad refined mean}
\end{figure}

The locally refined grids (Fig.\ \ref{sfig: grid refined}) are essentially uniform Cartesian over 
most of the domain, except near the patch interfaces. Such grids could, for example, be used in 
flow simulations in combination with immersed boundary methods. On Cartesian grids, all gradient 
schemes reduce to the same (second-order accurate) simple formula. Hence the differences between 
the mean errors plotted in Fig.\ \ref{sfig: mean refined} are due solely to cells at patch 
interfaces (details on the topology of these cells, which induces skewness and unevenness, can be 
found in \cite{Syrakos_2017}) and boundaries. Since these cells are a small minority, the mean 
errors of the various gradient schemes do not differ by much. As can be seen in Fig.\ \ref{sfig: 
mean refined}, the mean error of all gradient schemes that fall into the presently proposed general 
framework decreases at a second-order rate. However, at the cells near the patch interfaces, where 
there are large skewness and unevenness, the gradients are only first-order accurate, yet the number 
of these cells is too small to have a effect on the rate of convergence of the mean error (see 
\cite{Syrakos_2017} for a more quantitative discussion). The mean errors of the LS(3), LSA(3), 
LSD(3) and TG(2) gradients are affected favourably by the second-order accuracy that these schemes 
enjoy at boundary cells, but in this particular experiment we are mostly interested in what happens 
at the patch interfaces. The maximum errors, plotted in Fig.\ \ref{sfig: max refined}, are more 
helpful in this respect. They all decrease at a first-order rate (including for the LS(3), LSA(3), 
LSD(3) and TG(2) gradients, which means that they do not occur at boundary cells but at patch 
interface cells). Additionally, again Figs.\ \ref{fig: l8 errors quad refined mean} and \ref{fig: 
l8 errors quad refined max} show the variation of the mean and maximum errors on the finest grid 
with respect to the inverse-distance weighting exponent $q$. From these results, we can see that 
inverse distance weighting can help substantially, up to $q = 3-4$. Furthermore, the fact that 
coarse cells at patch interfaces have two neighbours on the fine patch side creates a directional 
clustering situation, and the accuracy suffers in the absence of any mitigative measures. LSA, LSD 
and TG gradients do incorporate mitigation strategies (area or directional weighting), unlike LS 
gradients whose accuracy obviously suffers. Interestingly, the TGI(0) (QG) gradient again performs 
decently, better than TG(0) or TG(1). In fact, Figs.\ \ref{fig: l8 errors quad refined mean} and 
\ref{fig: l8 errors quad refined max} show that within the TGI($q$) family, inverse distance 
weighting is not beneficial, and the best performance is obtained for $q=0$.


\begin{figure}[tb!]
    \centering
    \begin{subfigure}[b]{0.49\textwidth}
        \centering
        \includegraphics[width=0.99\linewidth]{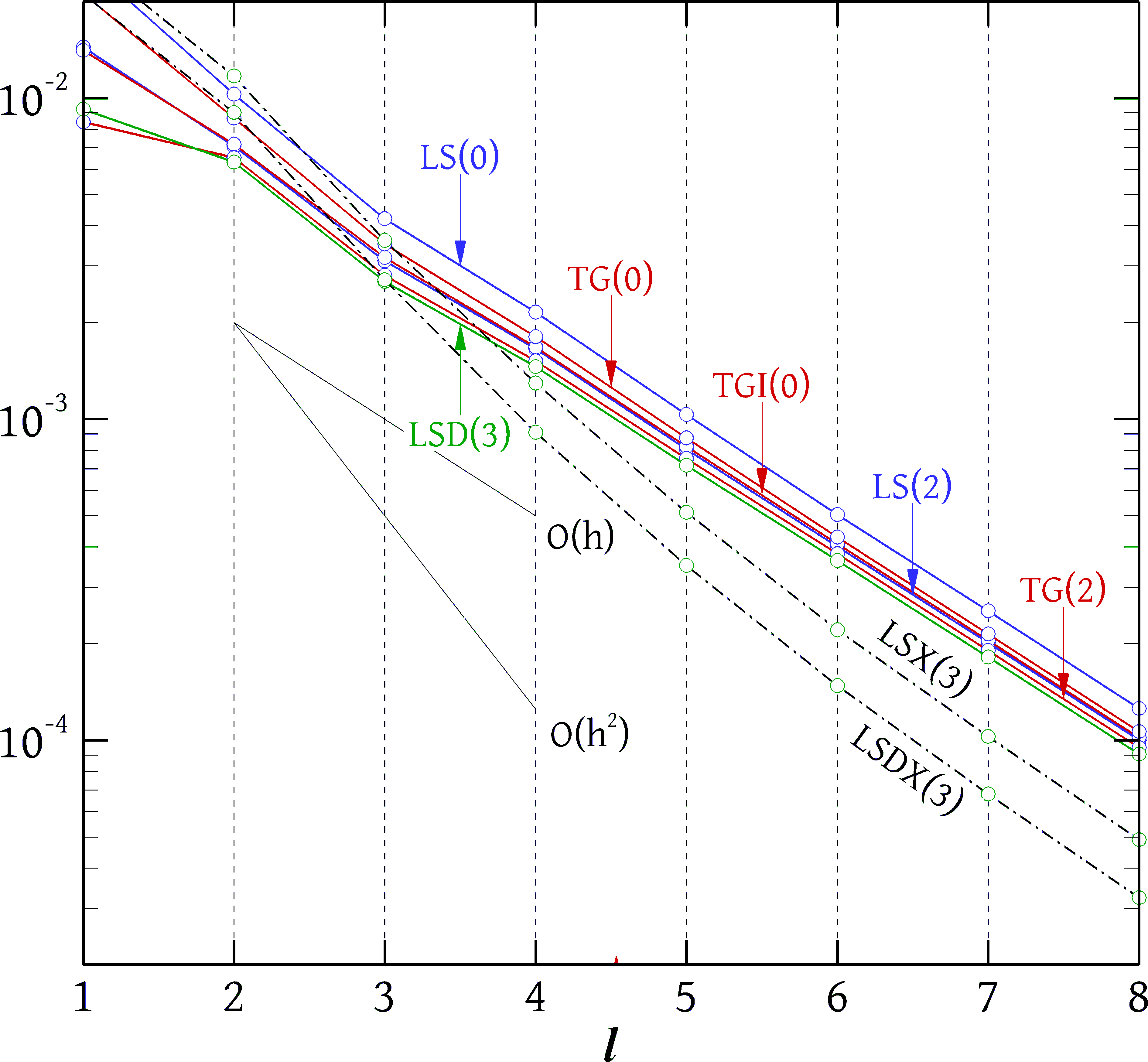}
        \caption{Mean errors}
        \label{sfig: mean random}
    \end{subfigure}
    \begin{subfigure}[b]{0.49\textwidth}
        \centering
        \includegraphics[width=0.99\linewidth]{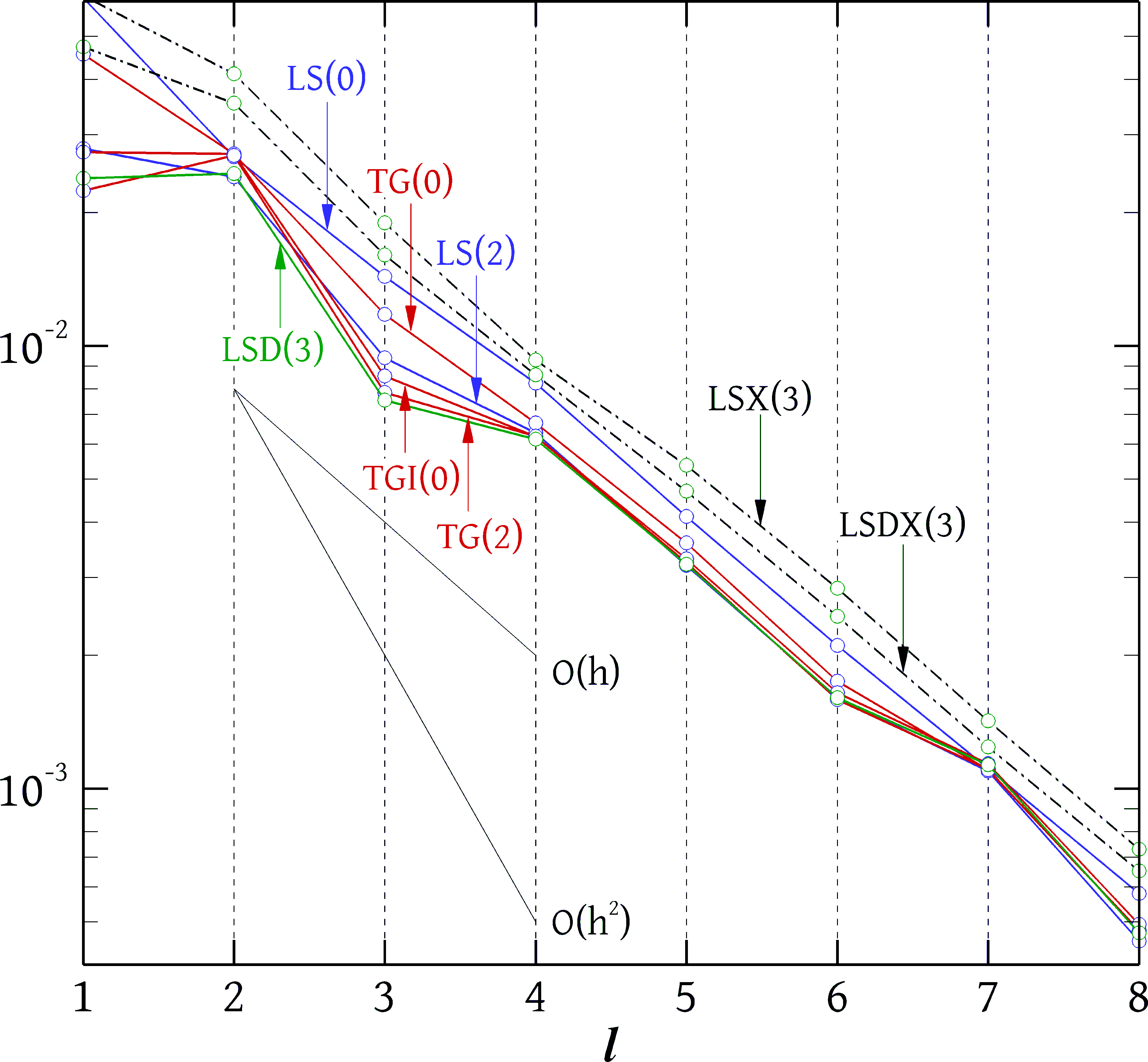}
        \caption{Maximum errors}
        \label{sfig: max random} 
    \end{subfigure}
    \caption{Minimum and maximum errors of various gradients versus refinement level $l$, for the 
randomly perturbed grids (Fig.\ \ref{sfig: grid random}).}
  \label{fig: errors random}
\end{figure}

Moving on to the randomly perturbed grids, Fig.\ \ref{fig: errors random} shows that all methods are 
first-order accurate with respect to the mean error as well as the maximum error. This behaviour is 
expected, as these grids exhibit arbitrary, non-diminishing skewness and unevenness throughout the 
domain and there are no special conditions that could increase the order of accuracy beyond the 
nominal first-order of Eq.\ \eqref{eq: generic gradient} through error cancellations. Figures 
\ref{fig: errors random}, \ref{fig: l8 errors quad distorted mean} and \ref{fig: l8 errors quad 
distorted max} show that the best and worst performers are the LS(0) and LSD(3) gradients, 
respectively, but the differences are quite small and no gradient has a decisive advantage. Again, 
inverse distance weighting helps up to $q = 2 - 3$, but only slightly in the present case.


\subsection{Grids of triangles}
\label{ssec: results triangles}

\begin{figure}[tb]
    \centering
    \begin{subfigure}[b]{0.32\textwidth}
        \centering
        \includegraphics[width=0.98\linewidth]{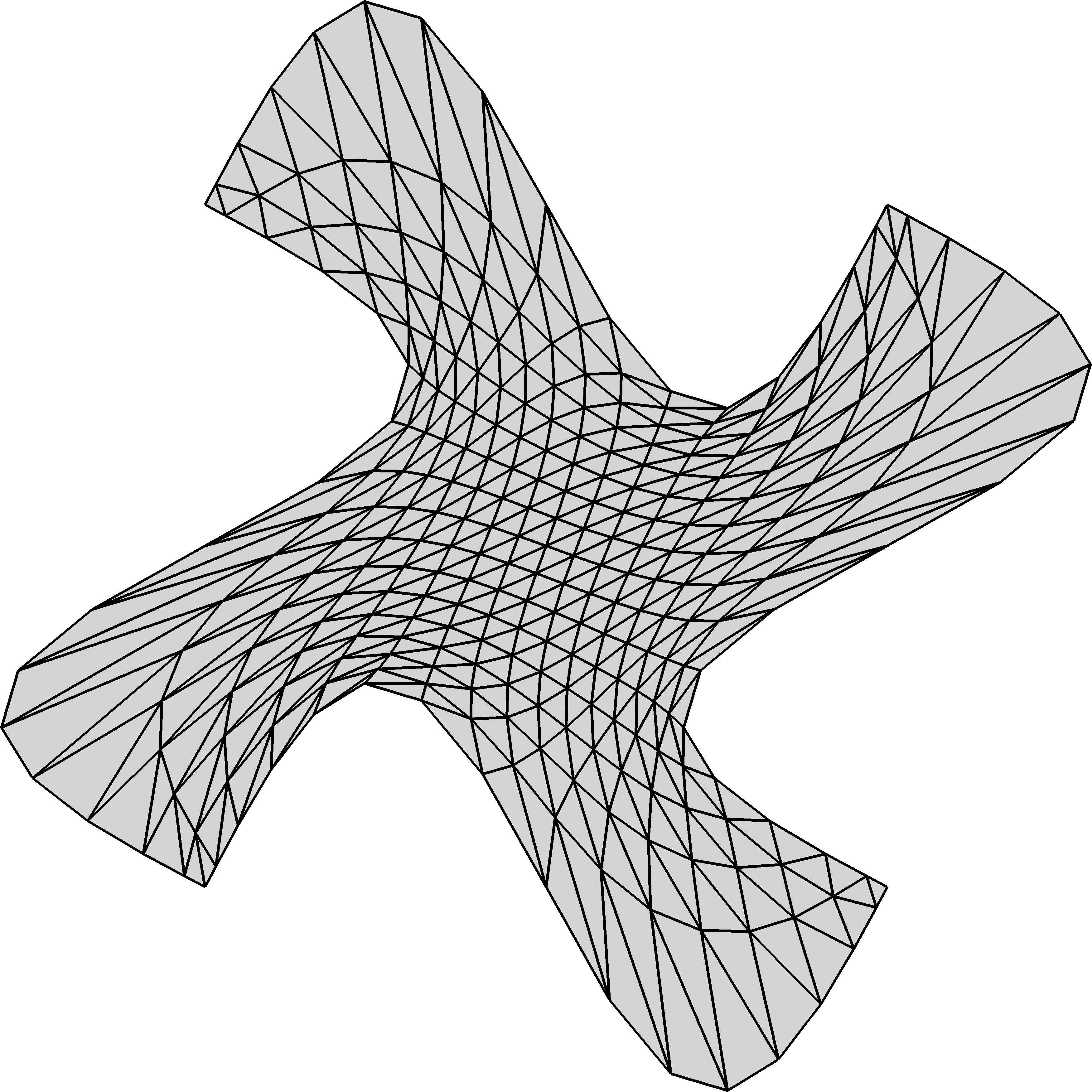}
        \caption{Orderly triangulated elliptic grid}
        \label{sfig: grid elliptic tri ordered}
    \end{subfigure}
    \begin{subfigure}[b]{0.32\textwidth}
        \centering
        \includegraphics[width=0.98\linewidth]{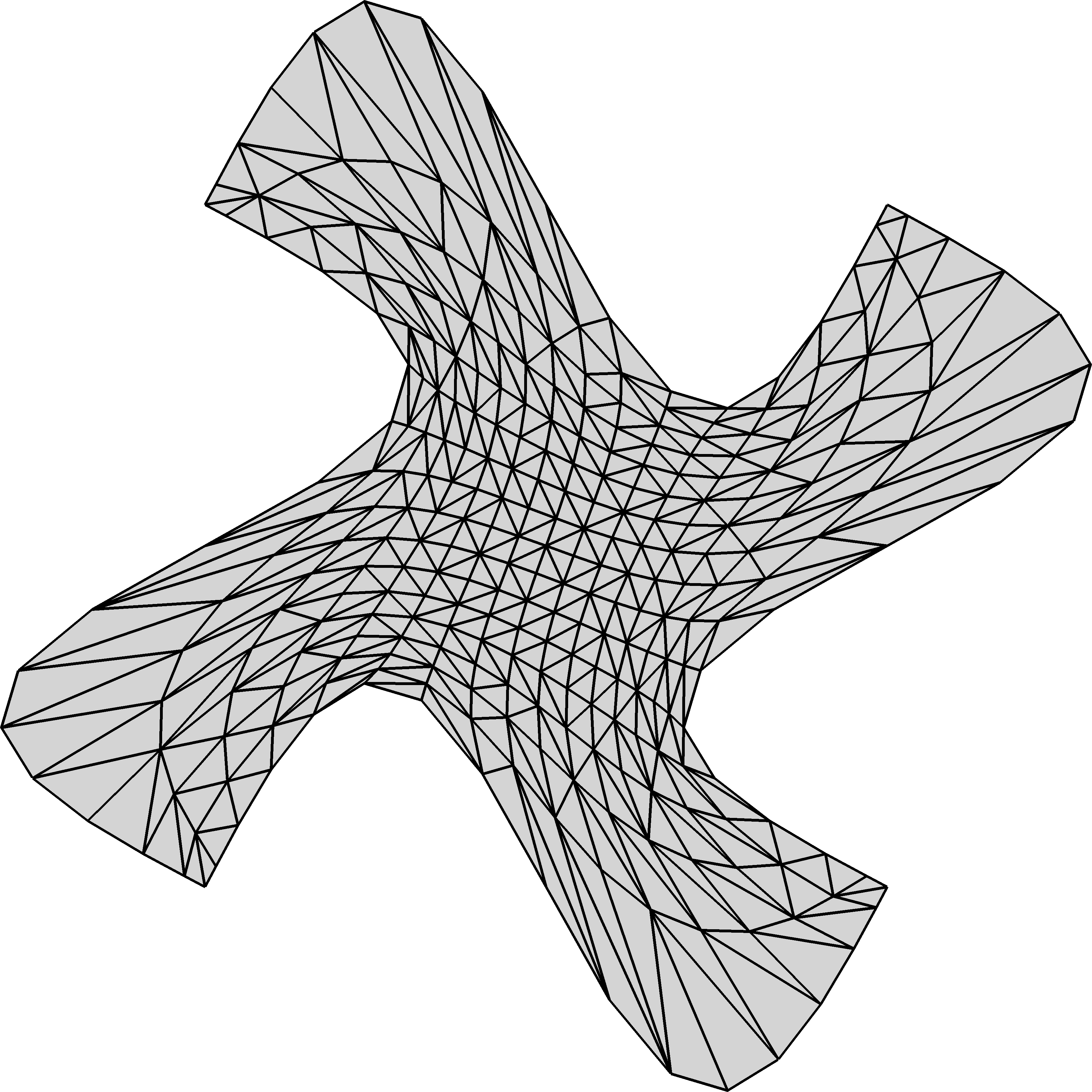}
        \caption{Randomly triangulated elliptic grid}
        \label{sfig: grid elliptic tri random}
    \end{subfigure}
    \begin{subfigure}[b]{0.32\textwidth}
        \centering
        \includegraphics[width=0.98\linewidth]{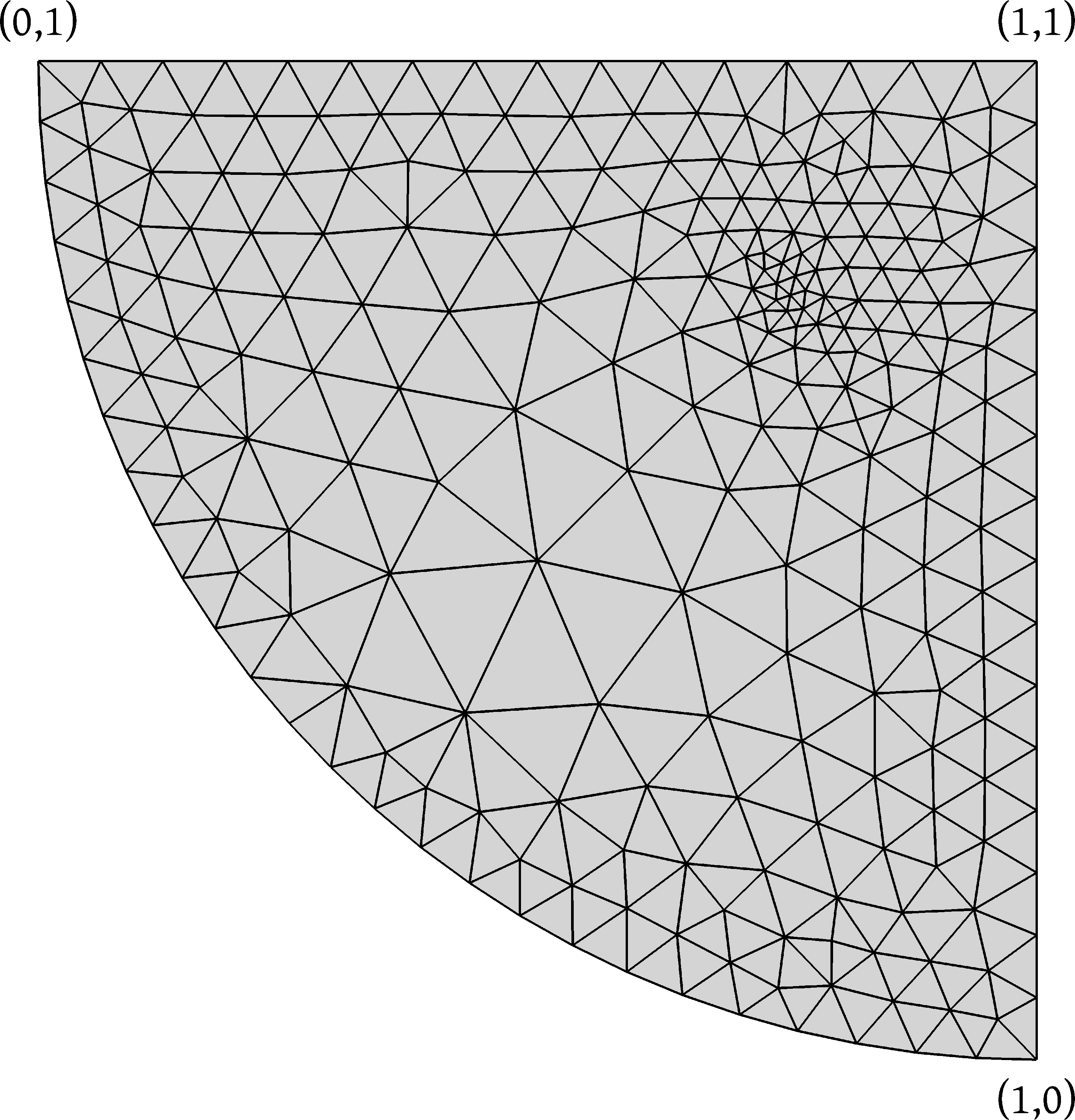}
        \caption{Gmsh grid}
        \label{sfig: grid tri gmsh}
    \end{subfigure}
    \caption{Grids of triangles, at the $l = 3$ level of refinement.}
  \label{fig: grids tri}
\end{figure}

Grids consisting of triangles are quite popular due to the availability of corresponding automated 
meshing algorithms, which greatly reduce the pre-processing time required for solving a problem. 
This advantage usually comes at a price of a loss of accuracy compared to structured grids or grids 
of quadrilaterals in general (similar observations hold also in three-dimensional space). In the 
present Section, we test the performance of the gradient schemes on triangular meshes. Figure 
\ref{fig: grids tri} shows the three kinds of meshes used: the first two are based on the 
elliptically generated structured meshes of Sec.\ \ref{ssec: results consistency} (Fig.\ \ref{sfig: 
grid elliptic}) which have been triangulated by splitting each quadrilateral into two triangles, 
either along the same diagonal, as in Fig.\ \ref{sfig: grid elliptic tri ordered}, or along a random 
diagonal, as in Fig.\ \ref{sfig: grid elliptic tri random}. The third group of meshes were 
constructed using the Gmsh mesh generator \cite{Geuzaine_2009}, v.\ 4.8.4, which was used to 
tessellate a quarter-disc-shaped domain, of unit radius, as in Fig.\ \ref{sfig: grid tri gmsh}.

\begin{figure}[tb!]
    \centering
    \begin{subfigure}[b]{0.49\textwidth}
        \centering
        \includegraphics[width=0.99\linewidth]{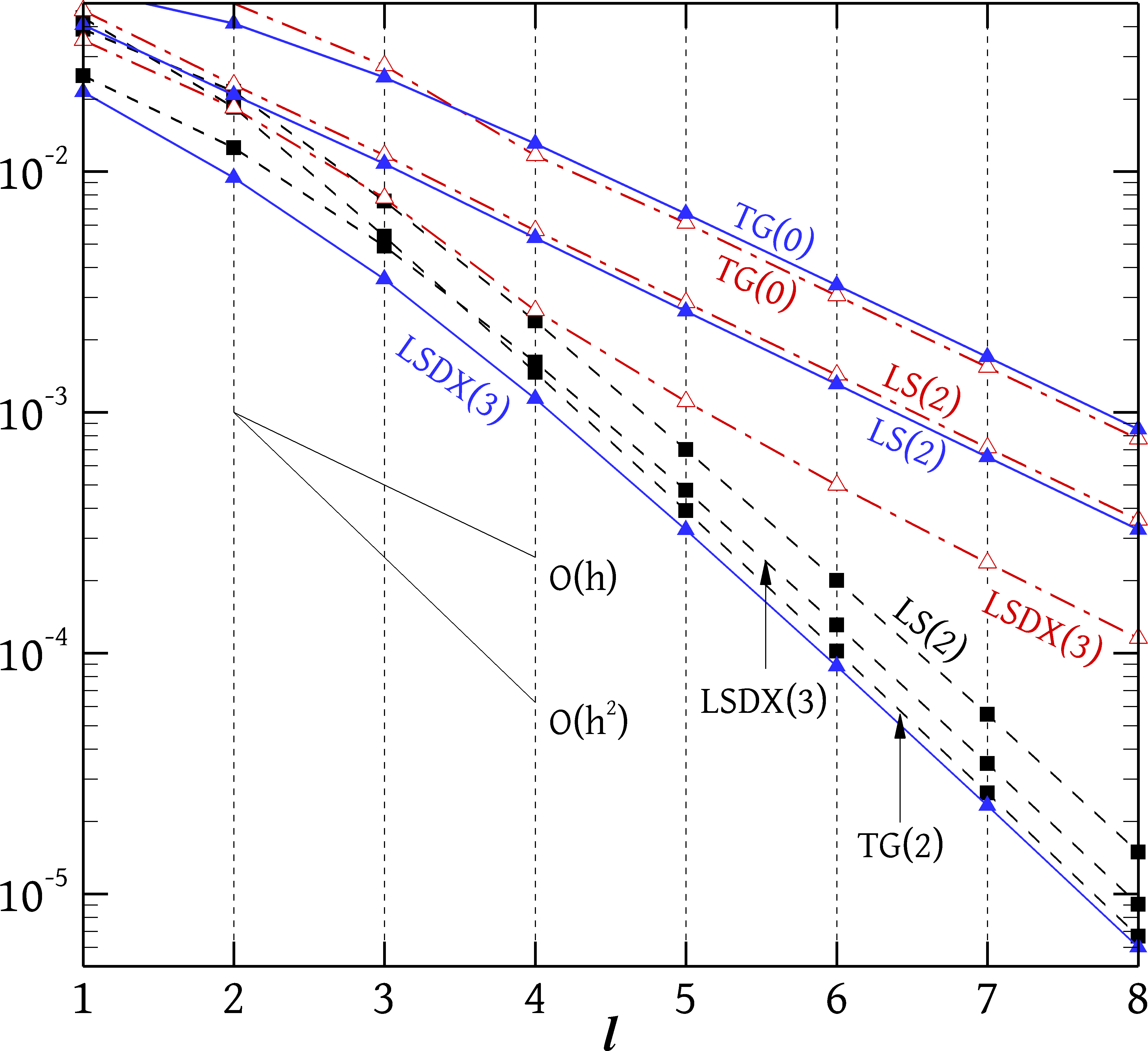}
        \caption{Mean errors}
        \label{sfig: mean elliptic tri}
    \end{subfigure}
    \begin{subfigure}[b]{0.49\textwidth}
        \centering
        \includegraphics[width=0.99\linewidth]{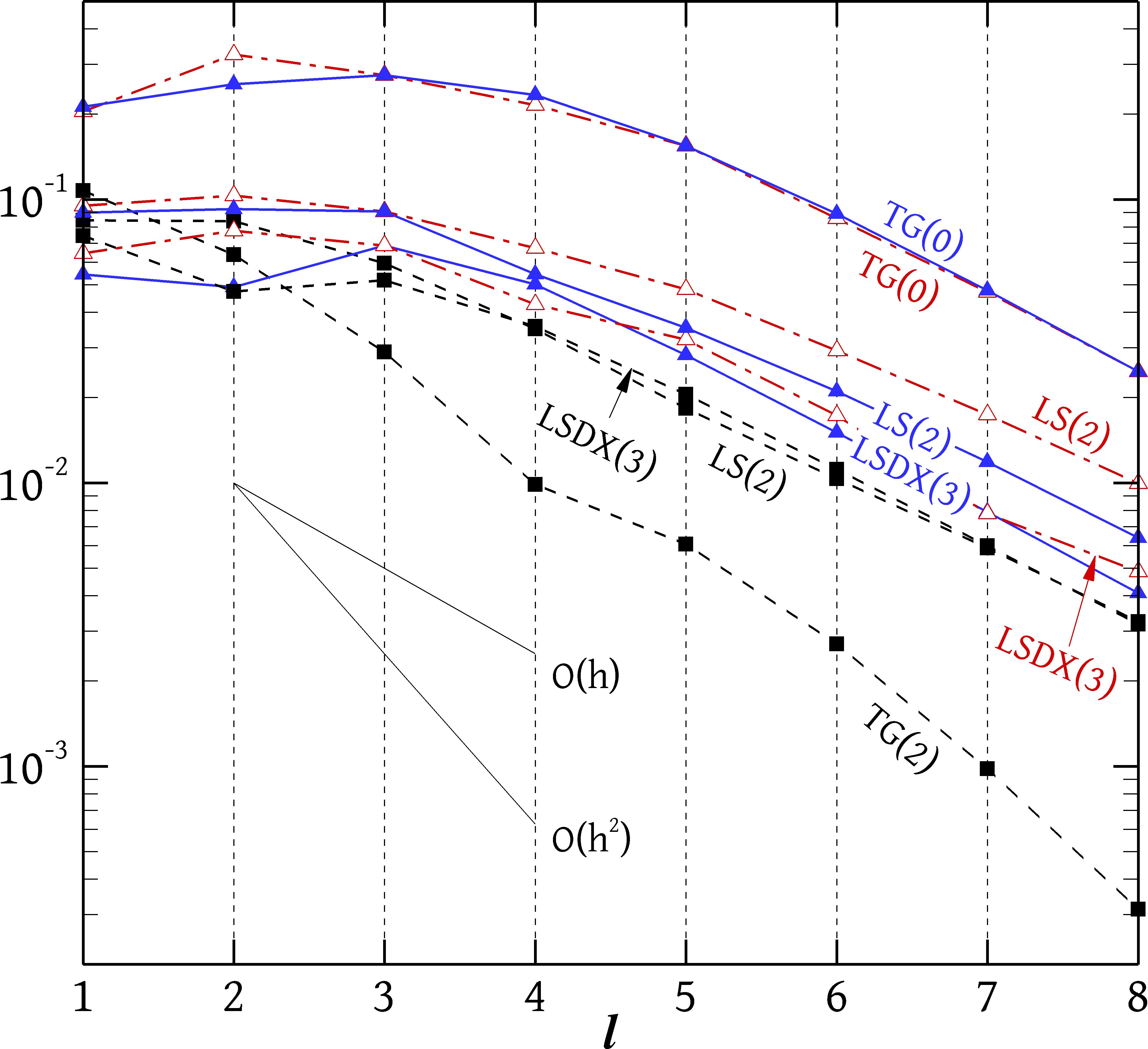}
        \caption{Maximum errors}
        \label{sfig: max elliptic tri} 
    \end{subfigure}
    \caption{Minimum and maximum errors of gradient schemes versus refinement level $l$, for the 
triangulated elliptic grids (Fig.\ \ref{sfig: grid elliptic}). Continuous blue lines 
($\blacktriangle$): uniformly triangulated grids (Fig.\ \ref{sfig: grid elliptic tri ordered}). 
Dash-dot red lines ($\vartriangle$): randomly triangulated grids (Fig.\ \ref{sfig: grid elliptic tri 
random}). Dashed black lines ($\blacksquare$): original structured (quadrilateral) grids (Fig.\ 
\ref{sfig: grid elliptic}).}
  \label{fig: errors elliptic tri}
\end{figure}

Mean and maximum errors on the triangulated elliptic grids are plotted in Fig.\ \ref{fig: errors 
elliptic tri}; some errors on the original (quadrilateral) elliptic grids are also included for 
comparison. In order not to clutter the plots, only the errors of a small subset of the gradients 
are plotted. The errors of all the gradients on the finest grids ($l = 8$) are plotted in Figs.\ 
\ref{fig: l8 errors elliptic mean} and \ref{fig: l8 errors elliptic max}, together with the errors 
on the original, quadrilateral, structured grids. Excluding for the moment the LSX and LSDX 
gradients, which will be defined and discussed in Sec.\ \ref{ssec: results extended stencils}, we 
see in Fig.\ \ref{fig: l8 errors elliptic mean} that the mean errors of the LS, LSA, LSD and TG 
gradients are very similar, and depend mostly on the exponent $q$ than on the gradient scheme. 
Hence in Fig.\ \ref{sfig: mean elliptic tri} only the TG(0) and LS(2) errors are plotted, which lie 
near the higher and lower ends of the spectrum, respectively. The LSA gradient appears to perform 
worst. The accuracy is usually better on the uniformly triangulated grids (Fig.\ \ref{sfig: grid 
elliptic tri ordered}) than on the randomly triangulated (Fig.\ \ref{sfig: grid elliptic tri 
random}), which is not surprising. The error difference between these two kinds of grids is larger 
when it comes to the maximum error -- Fig.\ \ref{fig: l8 errors elliptic max}.

Concerning the order of accuracy, the LS, LSA, LSD and TG gradients are first-order accurate on 
both kinds of triangulated structured grid, with respect to both the mean and maximum errors 
(Fig.\ \ref{fig: errors elliptic tri}). This contrasts with the original, quadrilateral structured 
grid, where all these gradients are second-order accurate with respect to the mean error, while the 
LS(3), LSA(3), LSD(3) and TG(2) are also second-order accurate with respect to the maximum error, 
as discussed previously. This is due to error cancellations between opposite faces, as discussed in 
Sec.\ \ref{ssec: accuracy special cases}, but in the case of triangulated grids there are no 
opposite faces and such cancellations do not occur, hence the accuracy remains of first order. 
Consequently, the gradients can produce 1-2 orders of magnitude larger errors on the finest 
triangulated grids than on the finest quadrilateral grid, despite the former having twice as many 
cells.

In fact, a more meticulous examination of the slopes of the maximum error lines in Fig.\ \ref{sfig: 
max elliptic tri} shows that the situation is slightly worse, with the order of convergence being 
below one, at about $0.6 - 0.9$ in most cases. This is another instance where the effect of 
refinement on the grid quality influences the observed order of convergence. The LS, LSA, LSD and 
TG gradients are formally first-order accurate, provided that grid refinement does not change the 
quality of the grid. If, however, the refinement procedure is such that the grid quality improves 
(e.g.\ opposite faces of quadrilaterals become more aligned, leading to larger error cancellations) 
or deteriorates then the observed order of convergence may be higher or lower than the formal one.

\begin{figure}[tb!]
    \centering
    \begin{subfigure}[b]{0.49\textwidth}
        \centering
        \includegraphics[width=0.99\linewidth]{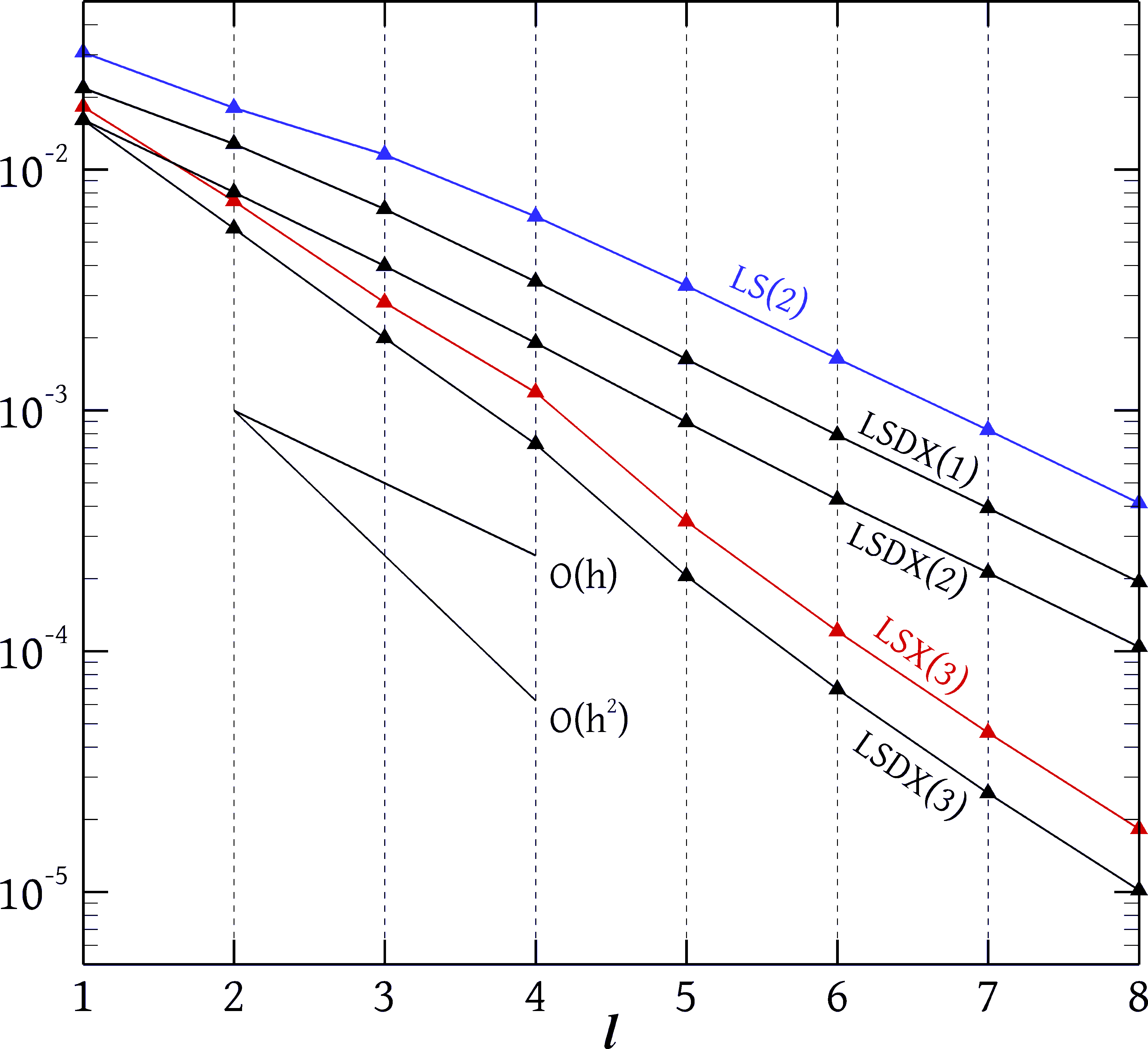}
        \caption{Mean errors}
        \label{sfig: mean gmsh}
    \end{subfigure}
    \begin{subfigure}[b]{0.49\textwidth}
        \centering
        \includegraphics[width=0.99\linewidth]{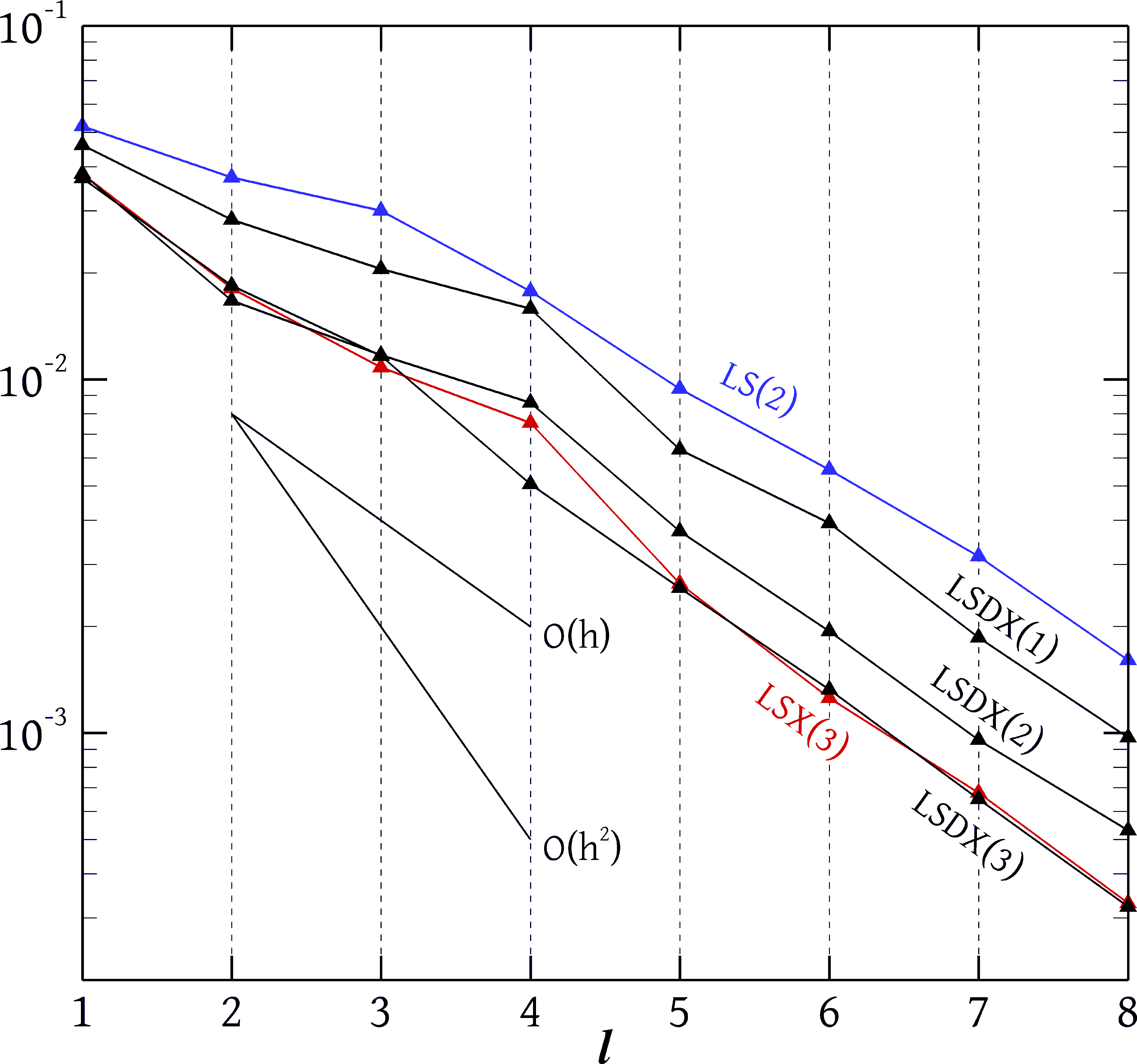}
        \caption{Maximum errors}
        \label{sfig: max gmsh} 
    \end{subfigure}
    \caption{Minimum and maximum errors of gradient schemes versus refinement level $l$, for the 
Gmsh grids (Fig.\ \ref{sfig: grid tri gmsh}).}
  \label{fig: errors gmsh}
\end{figure}

\begin{figure}[tb!]
  \centering
  \includegraphics[width=0.99\textwidth]{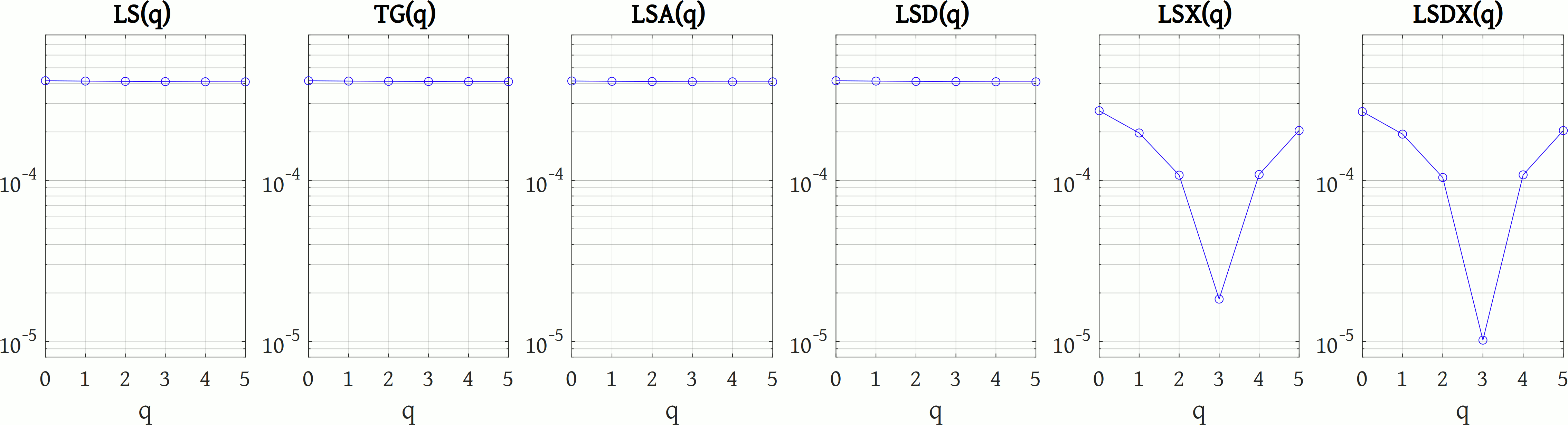}
  \caption{Average errors of various gradient schemes as a function of the exponent $q$, on the 
finest ($l=8$) Gmsh grid.}
  \label{fig: l8 errors gmsh mean}
\end{figure}

\begin{figure}[tb!]
  \centering
  \includegraphics[width=0.99\textwidth]{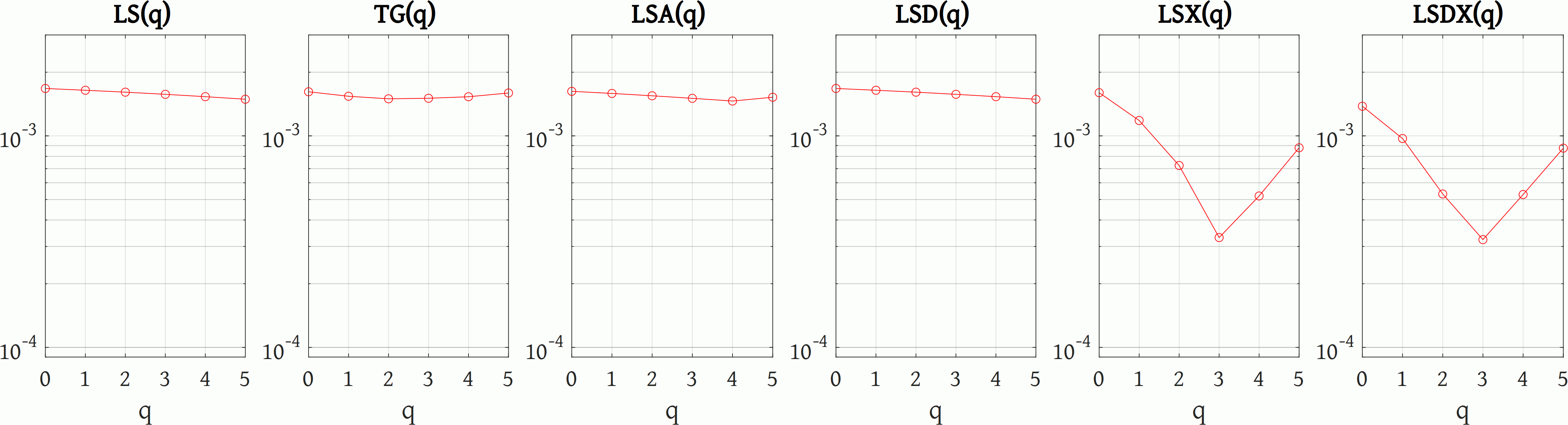}
  \caption{Maximum errors of various gradient schemes as a function of the exponent $q$, on the 
finest ($l=8$) Gmsh grid.}
  \label{fig: l8 errors gmsh max}
\end{figure}

Next, we consider a series of meshes constructed using the Gmsh mesh generator \cite{Geuzaine_2009}. 
The domain is the quarter-disc shown in Fig.\ \ref{sfig: grid tri gmsh}, centred at point (1,1) 
with additional corners at $(0,1)$ and $(1,0)$. The mesh spacing at each of these three points is 
set to $2^{-(l+1)}$, where $l$ is the grid level (again, we use 8 grids, corresponding to $l = 1, 
2, \ldots, 8$). Furthermore, we impose a mesh spacing of $2^{-(l-1)}$ at point $(0.5, 0.5)$, and a 
mesh spacing of $2^{-(l+3)}$ at point $(0.75, 0.75)$. This creates a 16-fold size (length) 
difference between cells close to $(0.5, 0.5)$ and cells close to $(0.75, 0.75)$. The $l = 3$ grid 
is shown in Fig.\ \ref{sfig: grid tri gmsh}. It was observed that on these Gmsh meshes the errors of 
the LS($q$), LSA($q$), LSD($q$) and TG($q$) gradients are nearly identical, for any value of $q$ 
(see Figs.\ \ref{fig: l8 errors gmsh mean} and \ref{fig: l8 errors gmsh max}), and hence we only 
plotted LS(2) in Fig.\ \eqref{fig: errors gmsh}. The methods are first-order accurate (again, the 
maximum errors fall at a rate that is slightly below first-order).


\subsection{Compact vs.\ extended stencils}
\label{ssec: results extended stencils}

\begin{figure}[tb!]
    \centering
    \begin{subfigure}[b]{0.49\textwidth}
        \centering
        \includegraphics[width=0.80\linewidth]{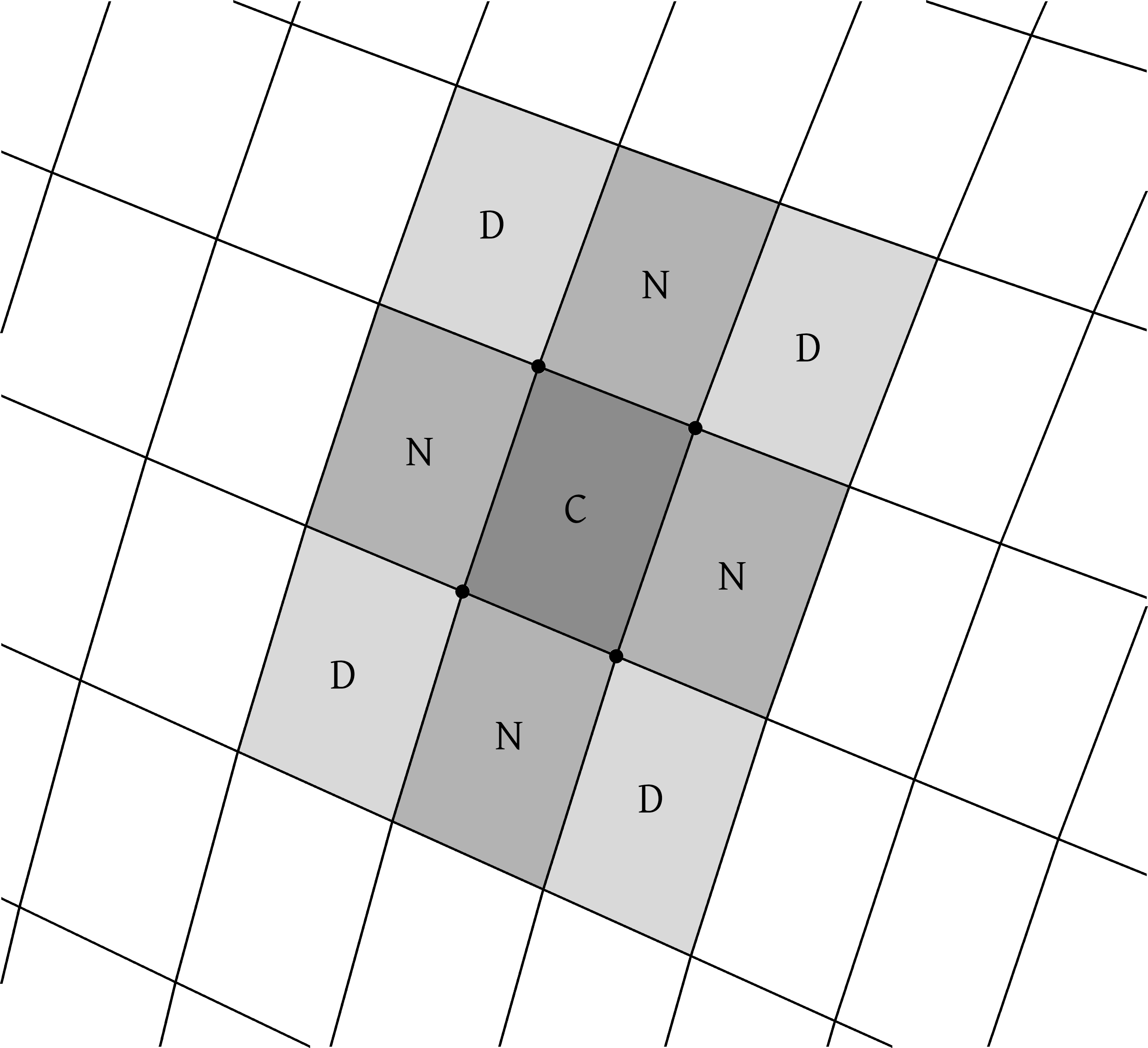}
        \caption{Quadrilateral mesh}
        \label{sfig: stencil quad}
    \end{subfigure}
    \begin{subfigure}[b]{0.49\textwidth}
        \centering
        \includegraphics[width=0.80\linewidth]{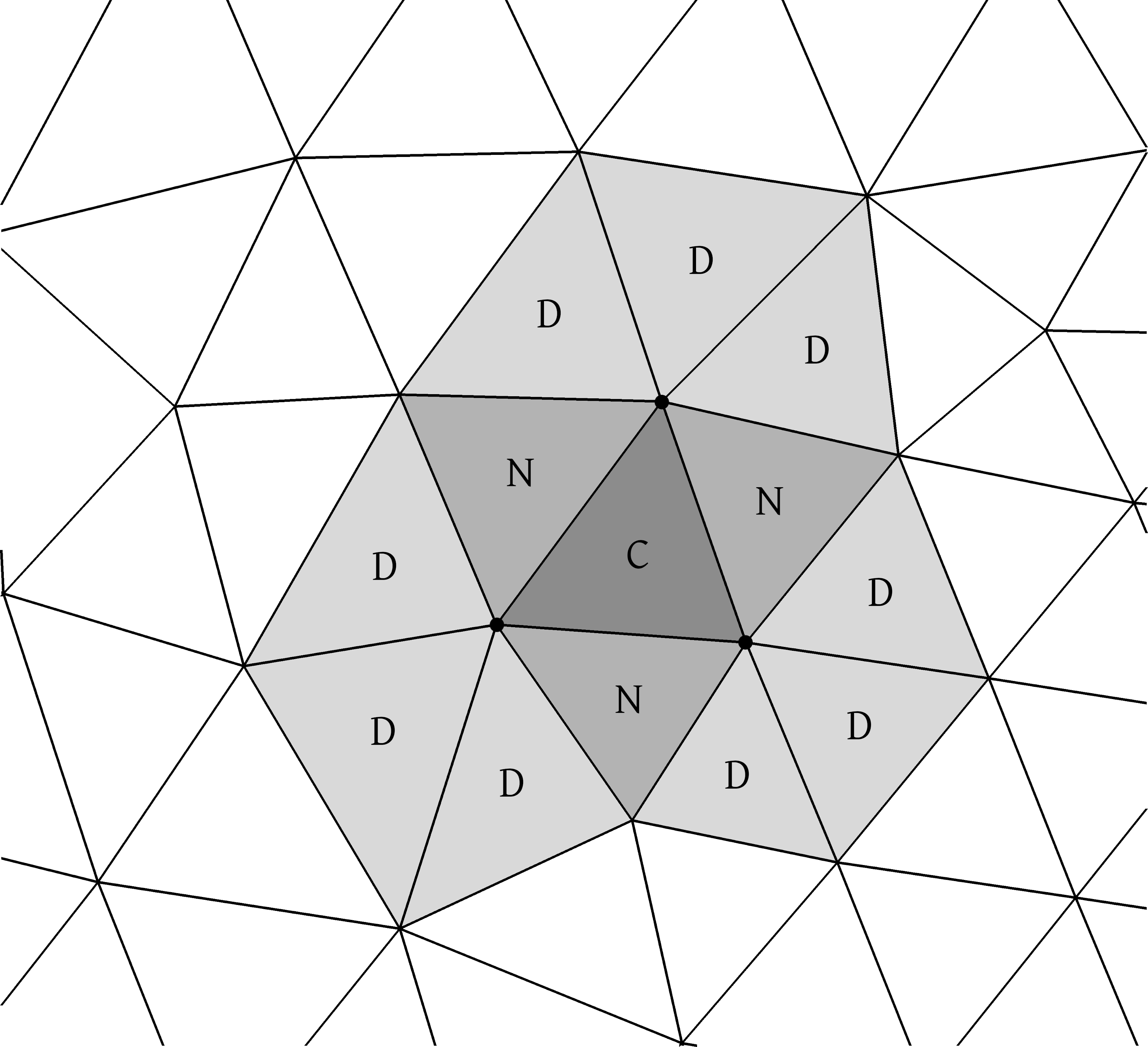}
        \caption{Triangular mesh}
        \label{sfig: stencil tri} 
    \end{subfigure}
    \caption{Compact and extended stencils for the calculation of the gradients, on quadrilateral 
\subref{sfig: stencil quad} and triangular \subref{sfig: stencil tri} grids. The gradient is 
calculated at the centroid of cell 'C', while 'N' marks its nearest neighbours (those that share a 
face with 'C') and 'D' its distant neighbours (those that share a single vertex with 'C').}
  \label{fig: stencils}
\end{figure}

In the previous results, the gradients (LS, LSA, LSD, TG, TGI) at the centroid of a cell were 
calculated using information only from the nearest neighbours of that cell (the cells that share a 
face with it). These cells are marked as ``N'' in Fig.\ \ref{fig: stencils}. Some studies have found 
it beneficial to include more than just the nearest neighbours in the calculation of the gradient, 
e.g.\ \cite{Diskin_2008, Sozer_2014, Wang_2019, Seo_2020}. Therefore, we also performed such
calculations. In particular, we employed extended stencils that include additionally all other cells 
(marked as ``D'' in Fig. \ref{fig: stencils}) that share a vertex with the cell where the gradient 
is calculated. Unfortunately, not all gradients examined here can employ extended stencils; those 
whose weight vectors involve geometrical features of the cell's faces, i.e.\ LSA, TG and TGI, can 
only use compact stencils. On the other hand, the LS and LSD gradients are face-free and do not have 
any restrictions on which neighbours to use. Hence, we applied them also in combination with the 
extended stencils of Fig.\ \ref{fig: stencils}, and for convenience we will refer to the resulting 
schemes as LSX($q$) and LSDX($q$), respectively (LS($q$) and LSD($q$) refer to the use of 
nearest-neighbour stencils).

Results for the LSX and LSDX gradients were also included in the Figures presented thus far. First, 
let us examine their performance on the elliptic quadrilateral grids (Fig.\ \ref{sfig: grid 
elliptic}). There, Figs.\ \ref{fig: errors elliptic}, \ref{fig: l8 errors elliptic mean} and 
\ref{fig: l8 errors elliptic max} show that these gradients perform well; nevertheless, they are 
outperformed by the TG(2) gradient. Also, Figs.\ \ref{fig: l8 errors elliptic mean} and \ref{fig: l8 
errors elliptic max} (quadrilateral grids) show that the optimal range of $q$ for LSX and LSDX is 
shifted to higher values compared to the compact-stencil gradients. This is not surprising, since 
the additional neighbours that LSX and LSDX involve will usually lie at greater distances than the 
nearest neighbours, and hence contribute larger distance-related truncation errors, which are 
minimised by increasing the distance-weighting exponent $q$. On the other hand, these additional 
neighbours contribute information from more directions, which may be beneficial for accuracy, 
provided that $q$ is large enough, so that, for example, for $q > 3$, LSX and LSDX are much more 
accurate than LS, LSA, LSD or TG. It is also noteworthy that the maximum errors of the LSX and LSDX 
gradients on the quadrilateral grids are, in general, larger than those of the nearest-neighbour 
gradients (Fig.\ \ref{fig: l8 errors elliptic max}), which can also be attributed to large distances 
of additional neighbours. Finally, we can note that, as seen in Figs.\ \ref{fig: errors elliptic} 
and \ref{fig: errors elliptic tri}, the extended-stencil gradients outperform the compact stencil 
gradients on coarse structured grids, where skewness and unevenness are still large and the 
benefits of structured refinement, which grant 2nd-order accuracy (Sec.\ \ref{ssec: accuracy 
special cases}), have not yet manifested.

So, the extra data structures and computational cost associated with the extended stencils is rather
unwarranted in the quadrilateral elliptic grids' case. However, the situation changes dramatically 
when these grids are triangulated. Figure \ref{fig: l8 errors elliptic mean} shows that extending 
the stencil can reduce the error by a factor of 50 on the finest orderly triangulated grid, and by 
a factor of 4.5 on the finest randomly triangulated grid. The LSDX gradient is slightly more 
accurate than the LSX. It is striking in Fig.\ \ref{fig: l8 errors elliptic mean} that the LSDX(3) 
error on the orderly triangulated grid is smaller than any other error, including that on the 
quadrilateral grids. Figure \ref{sfig: mean elliptic tri} shows that the LSDX(3) gradient is nearly 
2nd-order accurate on the orderly triangulated grids (so is the LSX(3) gradient, not shown). This is 
because, on the orderly triangulated grid, the cells of the extended stencil tend to align pairwise, 
giving rise to the situation described in Sec.\ \ref{ssec: accuracy special cases} where the $q = 3$ 
exponent causes error cancellations between opposite neighbours. On the randomly triangulated grids 
no such perfect alignment occurs, but nevertheless the multitude of directions associated with the 
many neighbours gives opportunities for partial cancellations of errors, so that the LSX(3) and 
LSDX(3) gradients are still substantially more accurate than the rest.

A similar large benefit is brought about by extending the stencil on the Gmsh grids, as Figs.\ 
\ref{fig: l8 errors gmsh mean} and \ref{fig: l8 errors gmsh max} show. The Gmsh algorithm appears 
to generate grids that are similar to the orderly triangulated elliptic grid over parts of the 
domain (Fig.\ \ref{sfig: grid tri gmsh}). The extended stencils give rise to error cancellations 
that increase the order of accuracy of the LSX(3) and LSDX(3) gradients to about 1.5, as the slopes 
of the corresponding curves of Fig.\ \ref{sfig: mean gmsh} reveal. Directional weighting again 
brings a modest increase in accuracy (LSDX(3) is more accurate than LSX(3)).

Finally, significant benefits of extending the stencil are also evident in Figs.\ \ref{sfig: mean 
random} and \ref{fig: l8 errors quad distorted mean}: on the randomly perturbed quadrilateral grids 
the LSDX(3) and LSX(3) gradients have mean errors that are approximately 3 and 2 times smaller, 
respectively, than those of the compact stencil gradients. The LSX(3) and LSDX(3) gradients are 
ultimately first order accurate, although, as Fig.\ \ref{sfig: mean random} shows, on coarser grids 
they exhibit second-order convergence. When it comes to the maximum error, Figs.\ \ref{sfig: max 
random} and \ref{fig: l8 errors quad distorted max} show that, on the contrary, the extended 
stencil gradients exhibit large maximum errors. This can again be attributed to distant cells being 
included in the stencil in some cases, and is remedied by increasing the exponent $q$ (Fig.\ 
\ref{fig: l8 errors quad distorted max}).

\section{Further numerical tests: curved geometry with grids of high aspect ratio}
\label{sec: tests hard}

\subsection{The tests and their results}
\label{ssec: tests hard}

\begin{figure}[thb]
 \centering
 \includegraphics[scale=0.90]{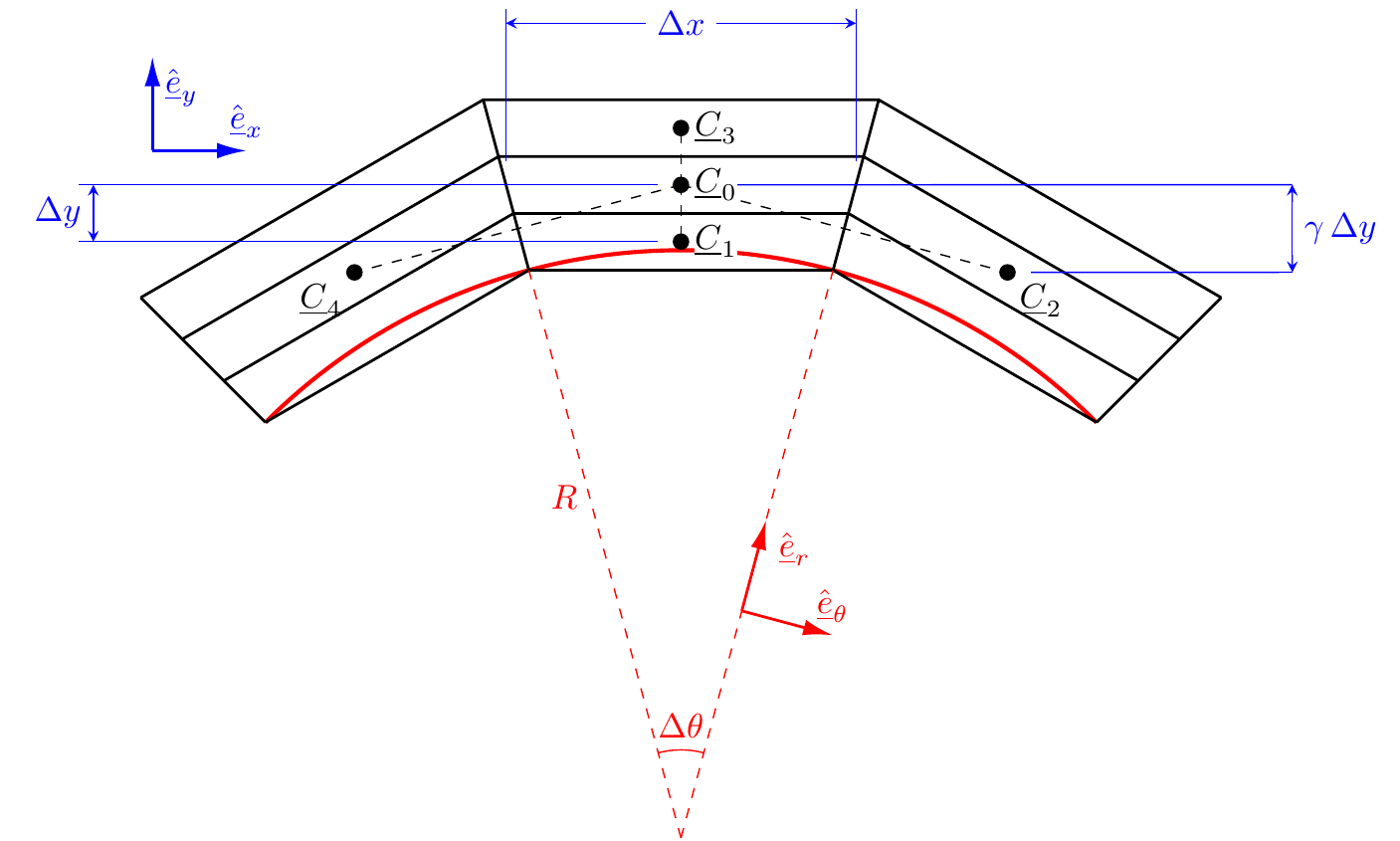}
 \caption{Structured grid of high aspect ratio cells over a curved boundary (the aspect ratio is 
shown greatly reduced for clarity).}
 \label{fig: HAR grid sketch}
\end{figure}

A situation that is quite challenging for gradient discretisation schemes and has gathered attention 
from researchers is that of domains with curved boundaries meshed with cells of very high aspect 
ratio, as typically used for the simulation of high-speed boundary layer flows in aerodynamics 
(e.g.\ \cite{Mavriplis_2003, Diskin_2008, Shima_2010, Sozer_2014, Wang_2019}). Hence it was included 
in the current set of experiments. Figure \ref{fig: HAR grid sketch} shows part of such a grid (the 
aspect ratio is reduced for clarity), which shall henceforth be referred to as HARC (High Aspect 
Ratio Curved) grid. The grid of Fig.\ \ref{fig: HAR grid sketch} is structured and a compact 
stencil is marked, but we will also examine triangulated grids and extended stencils. The 
structured grid case is important, because it is a popular strategy to use a structured grid near 
the boundary, in order to achieve high accuracy, even if farther from the boundary the grid is 
unstructured.

Oftentimes, the differentiated variable's contours more or less follow the shape of the boundary. 
In this case, the curvature introduces a nonlinearity that poses a challenge to gradient schemes 
like the ones considered here, which are founded on an assumption of linear variation of the 
variable in the neighbourhood of the cell. Furthermore, due to the large aspect ratio, the 
magnitudes of the contributions of different faces can differ by several orders of magnitude, 
depending on the weighting scheme. The unweighted LS gradient is particularly notorious. With 
reference to Fig.\ \ref{fig: HAR grid sketch}, LS(0) places equal emphasis on satisfying $\Delta 
\phi_f = \nabla \phi (\vf{C}_0) \cdot (\vf{C}_f - \vf{C}_0)$ for $f = 1$ (or $3$) as for $f = 2$ (or 
$4$). For $\gamma > 1$, where $\gamma$ is the ratio of the $y-$displacement of $\vf{C}_1$ to that of 
$\vf{C}_4$, both of them with respect to $\vf{C}_0$ (Fig.\ \ref{fig: HAR grid sketch}), this results 
in the LS(0) gradient underestimating the actual $\partial \phi/\partial y$ at $\vf{C}_0$ by a 
factor of approximately $\gamma$. The resulting inaccuracy can be very severe, as in practical 
applications $\gamma$ can be as high as $50$ or greater \cite{Mavriplis_2003}. Using proper 
weighting (inverse distance) greatly improves the accuracy.

So, we first construct a structured HARC grid over a circular arc of radius $R = 1$, like the one 
shown in Fig.\ \ref{fig: HAR grid sketch}. The grid spans from $\theta=-0.256$ to $\theta=+0.256$ 
radians in the azimuthal direction, with a corresponding spacing of $\Delta \theta_l = 0.256 / 2^l$ 
radians, for levels of refinement $l = 0, 1, \ldots, 9$. The radial spacing is $\Delta r_l = R \, 
\Delta \theta_l / A$ where $A = 1000$ is the cell aspect ratio. Grid level $l = 0$ has $2 \times 2$ 
cells, and grid level $l = 9$ has $1024 \times 1024$ cells in the $(r,\theta)$ directions.

The first function to be differentiated is selected to vary only in the radial direction:
%^c
\begin{equation} \label{eq: test function radial}
 \phi(r) \;=\; \tanh \left( f(r) \right)
 \qquad \text{where} \qquad
 f(r) \;=\; f_{\min} \;+\; (f_{\max} - f_{\min}) \frac{r - r_{\min}}{r_{\max} - r_{\min}}
\end{equation}
%^c
The function $f$ varies linearly in the radial direction, from $f = f_{\min} = 1$ at $r_{\min} = R 
= 1$, to $f = f_{\max} = 3$ at $r_{\max} = 1.0005$ ($r_{\max}$ is close to the outer radius of the 
grid, which is $1.000512$). Thus the differentiated function $\phi$ varies from $\tanh(1)$ to 
$\tanh(3)$ across the radial width of the grid. An average value for the magnitude of $\nabla \phi$ 
is therefore $G = (\tanh(3) - \tanh(1)) / 0.0005 = 466.9$.

\begin{figure}[thb]
 \centering
 \includegraphics[scale=1.25]{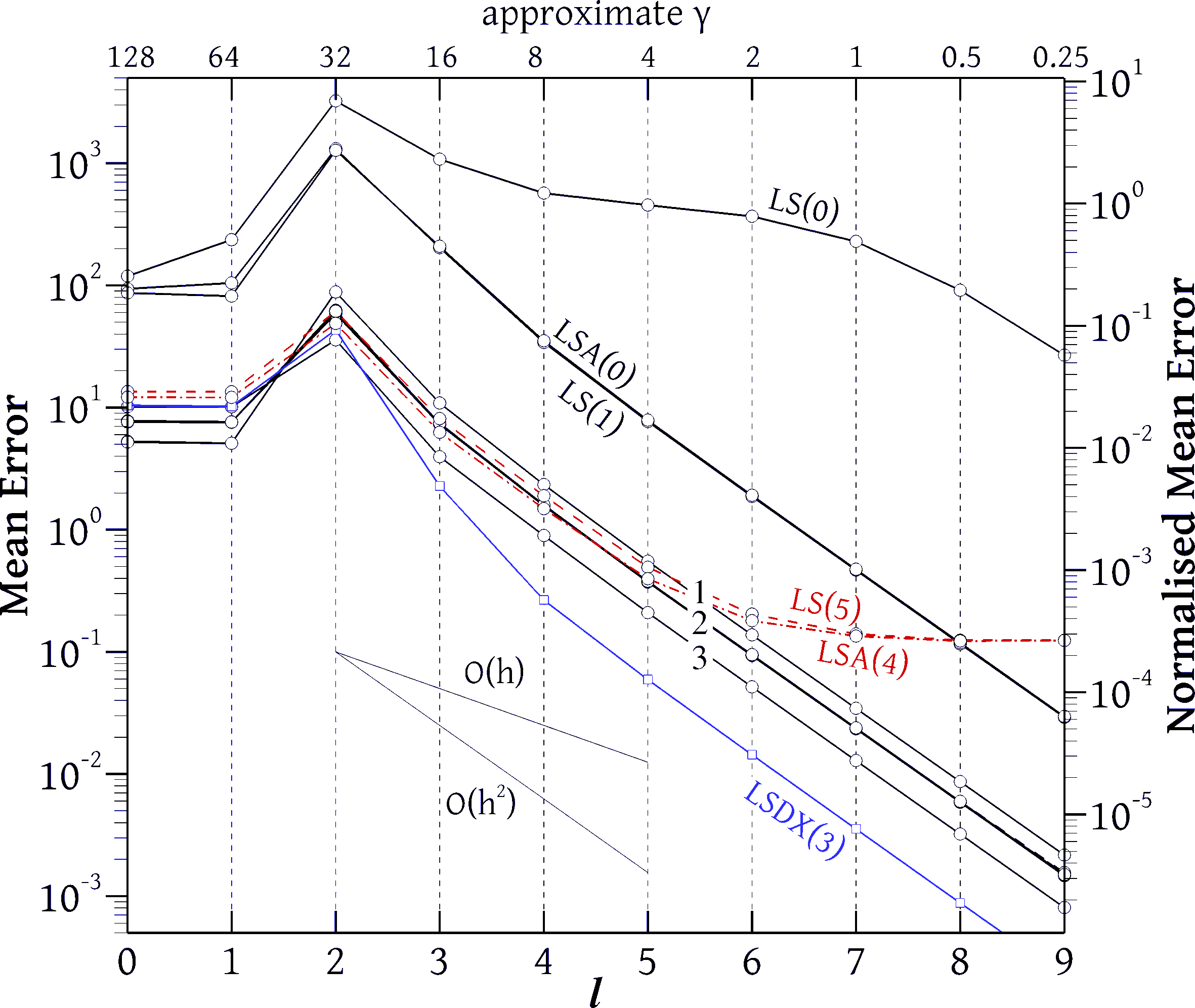}
 \caption{Mean errors of several gradient schemes when differentiating the radial function 
\eqref{eq: test function radial} on the structured HARC grid. Line group 1: TG(0), LSA(1). Line 
group 2: TGI(0), TGI(1), TG(1), TGI(2), GG, LS(2), LSA(2), LSX(2). Line group 3: TG(2), LS(3), 
LSA(3), LSDX(4).}
 \label{fig: errors HARC}
\end{figure}

Figure \ref{fig: errors HARC} shows the mean errors of several gradient schemes, at each refinement 
level $l$. The left vertical axis shows the mean error, while the right vertical axis shows this 
error normalised by the average gradient $G$. Many error curves collapse onto the lines labelled 
``1'' (TG(0) and LSA(1)), ``2'' (TGI(0), TGI(1), TG(1), TGI(2), GG, LS(2), LSA(2), LSX(2)) and ``3'' 
(TG(2), LS(3), LSA(3), LSDX(4)). All lines exhibit 2nd-order convergence, which is expected since 
the grid is structured. The most accurate among these groups is group ``3'' which includes gradients 
such as LS(3), LSA(3) and TG(2) that retain 2nd-order accuracy even at boundary cells. Looking at 
the normalised error, most schemes achieve adequate accuracy except the LS(0) gradient which only 
gives meaningful results on grids $l \geq 7$, for which $\gamma \leq 1$, in accordance with the 
theory outlined above. Approximate values of $\gamma$ are printed on the top axis, calculated as 
$\gamma \approx A \Delta \theta / 2$, a valid approximation for small $\Delta \theta$ 
\cite{Mavriplis_2003}. On the other extreme, the LSDX(3) gradient stands out as the most accurate by 
a significant margin. The LSA(0) and LS(1) gradients form a small group of their own; their 
accuracy is relatively poor, albeit much better than that of the LS(0) gradient. In general, it was 
observed that the LSA($q$) gradient is almost identical to the LS($q+1$) gradient on this grid, 
which is not surprising since, for this geometry, weighting by $\|\vf{S}_f\|$ is almost equivalent 
to weighting by $\|\vf{R}\|^{-1}$.

The plot \ref{fig: errors HARC} includes the error curves for LS(5) and LSA(4), shown in red dashed 
and dash-dot lines, respectively. As expected, these curves nearly coincide; however, what stands 
out is that these errors appear to diverge beyond level $l = 5$. This seems to stem from a numeric 
instability, as repeating the calculations in higher precision arithmetic, 10-byte (extended 
precision) or 16-byte (quadruple precision), eliminates the divergence and the gradients exhibit 
second-order convergence down to the finest grid. Depending on the algorithm used to calculate the 
cell centroids, the instability can be much worse than that shown on Fig.\ \ref{fig: errors HARC}. 
Hence, this issue deserves a deeper investigation, and we devote a separate section to it, Sec.\ 
\ref{ssec: stability}.

\begin{figure}[thb]
 \centering
 \includegraphics[scale=1.50]{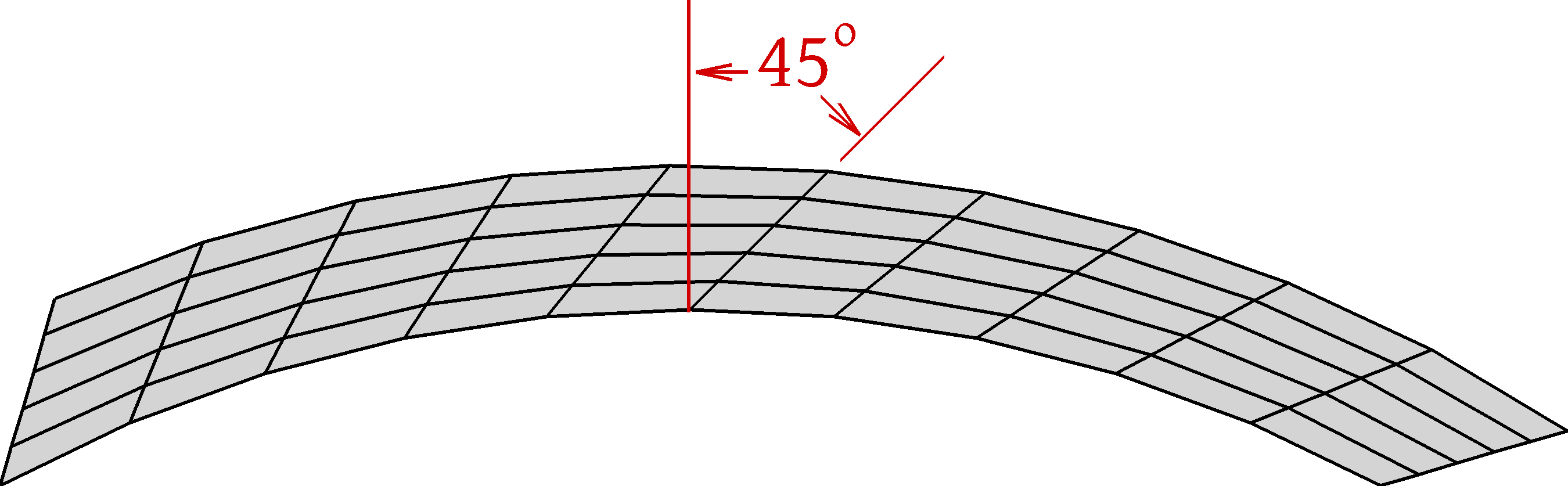}
 \caption{A High Aspect Ratio Curved Oblique (HARCO) grid.}
 \label{fig: HARCO grid sketch}
\end{figure}

To get a more complete picture, in addition to the structured HARC grid of Fig.\ \ref{fig: HAR grid 
sketch}, other kinds of high-aspect-ratio grids were used as well. This includes a version where 
the radial grid lines of HARC grids are replaced by oblique ones, at an angle of 45\degree\ to the 
radial direction, as in the sketch of Fig.\ \ref{fig: HARCO grid sketch}; this family of grids will 
be referred to as HARCO grids (HARC grids with Oblique grid lines). The other grid families employed 
are orderly and randomly triangulated versions of the HARC and HARCO grids, in exactly the same 
fashion that the elliptic grid of Fig.\ \ref{sfig: grid elliptic} gave rise to the triangulated 
grids of Figs.\ \ref{sfig: grid elliptic tri ordered} and \ref{sfig: grid elliptic tri random}.

\begin{figure}[tb!]
    \centering
    \begin{subfigure}[b]{0.99\textwidth}
        \centering
        \includegraphics[width=0.99\linewidth]{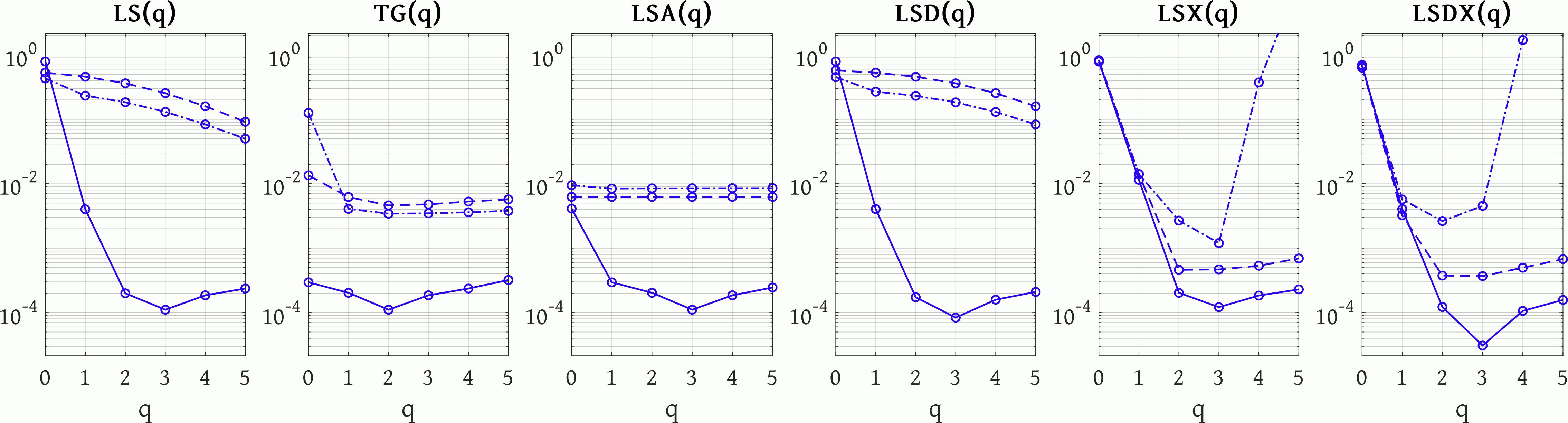}
        \caption{HARC meshes}
        \label{sfig: errors HARC mean l6}
    \end{subfigure}
    \\[0.5cm]
    \begin{subfigure}[b]{0.99\textwidth}
        \centering
        \includegraphics[width=0.99\linewidth]{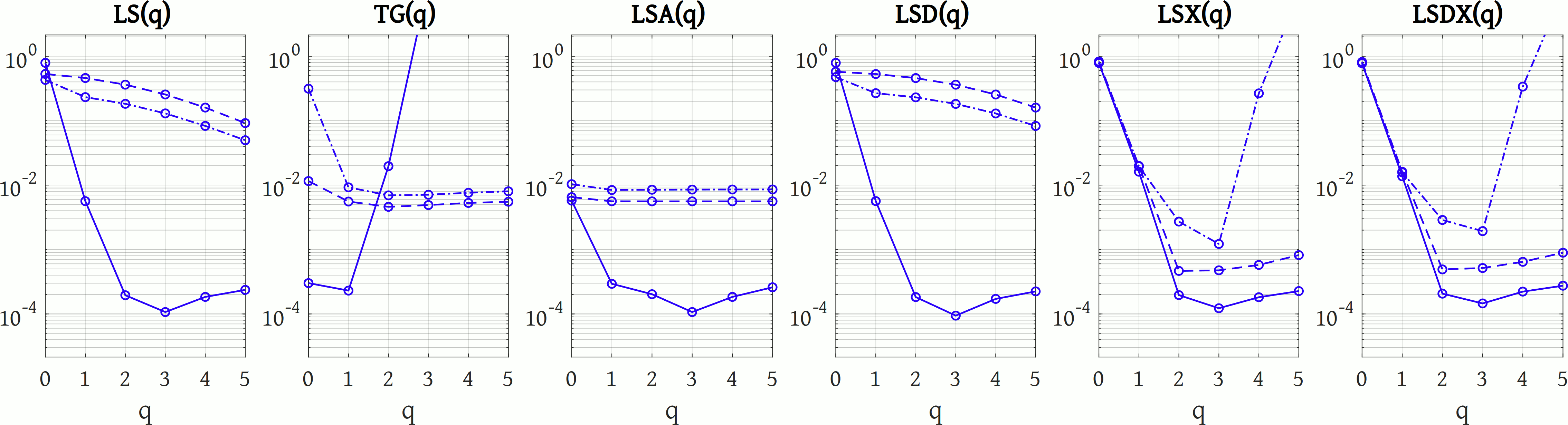}
        \caption{HARCO meshes}
        \label{sfig: errors HARCO mean l6} 
    \end{subfigure}
    \caption{Average normalised errors of several gradient schemes, as a function of the exponent 
$q$, for computing the gradient of the radially-varying function \eqref{eq: test function radial} 
on the $l=6$ HARC \subref{sfig: errors HARC mean l6} and HARCO \subref{sfig: errors HARCO mean l6} 
grids. The errors on structured grids are drawn with continuous lines, those on orderly triangulated 
grids with dashed lines, and those on randomly triangulated grids with dash-dot lines.}
  \label{fig: errors HARC HARCO mean l6}
\end{figure}

Figure \ref{fig: errors HARC HARCO mean l6} shows the mean normalised (by $G$) errors on the level 
6 grids of all these mesh families, as a function of the weighting exponent $q$. The 10-byte 
calculation results are shown, in order to avoid any pollution by roundoff errors. One can observe 
the following. The errors on the structured grids are, in general, 1-3 orders of magnitude smaller 
than on the triangulated grids. The difference is especially huge for the compact-stencil LS and LSD 
gradients, which perform poorly on the triangulated grids. The difference is smallest for the 
extended stencil gradients, LSX and LSDX; extending the stencil can increase the accuracy of the LS 
and LSD gradients on triangulated grids by several orders of magnitude, especially on the orderly 
triangulated grids. On the quadrilateral grids, the gains from extending the stencil are small, 
although the best accuracy overall is observed for the LSDX(3) gradient on the HARC quadrilateral 
grid.

The performance of the least-squares gradients is very similar on the HARC and HARCO grids. This is 
not surprising, given that the least-squares gradients' geometric input is only or mostly the cell 
centres, which do not differ much between these two grids. On the other hand, the TG gradient relies 
heavily also on the face normal vectors, which differ significantly between the HARC and HARCO 
grids; hence, in the quadrilateral case, Fig.\ \ref{fig: errors HARC HARCO mean l6} shows that, 
whereas the TG gradient is quite efficient on the HARC grid, its performance suffers tremendously, 
to the point of TG becoming unusable, on the HARCO grid if $q > 1$. Such impact on the TG gradient 
performance is not exhibited in the triangulated grids case; one only notices a slight deterioration 
in the randomly triangulated case.

On triangulated grids, the accuracy is usually better when the triangulation is orderly, but, unlike 
in the elliptic grid case of Sec.\ \ref{sec: tests} (Figs.\ \ref{sfig: grid elliptic tri ordered} 
and \ref{sfig: grid elliptic tri random}), the LS and LSD gradients appear to be more accurate on 
the randomly triangulated grids (as is the TG gradient on the HARC grid). Nevertheless, the 
accuracies on both kinds of grids are usually comparable. The extended-stencil gradients have a 
larger advantage on the orderly triangulated grids than on the randomly triangulated ones, as for 
the elliptic grids of Sec.\ \ref{sec: tests}. Surprisingly, on the randomly triangulated grids, the 
errors of the extended-stencil gradients skyrocket when $q$ is increased beyond 3.

\begin{figure}[tb!]
    \centering
    \begin{subfigure}[b]{0.49\textwidth}
        \centering
        \includegraphics[width=0.99\linewidth]{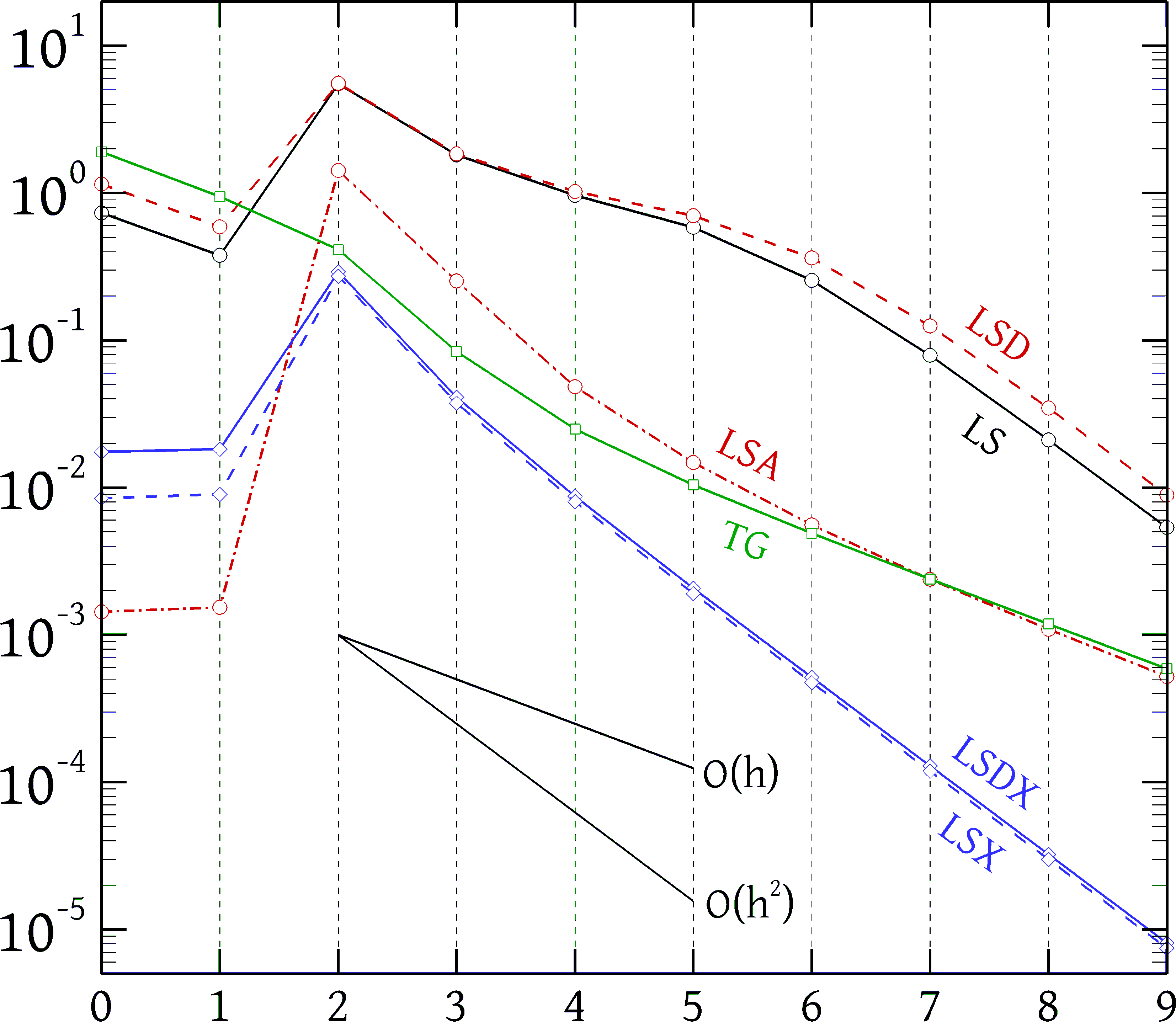}
        \caption{Orderly triangulated HARCO grids}
        \label{sfig: HARCO triO q3 R}
    \end{subfigure}
    \begin{subfigure}[b]{0.49\textwidth}
        \centering
        \includegraphics[width=0.99\linewidth]{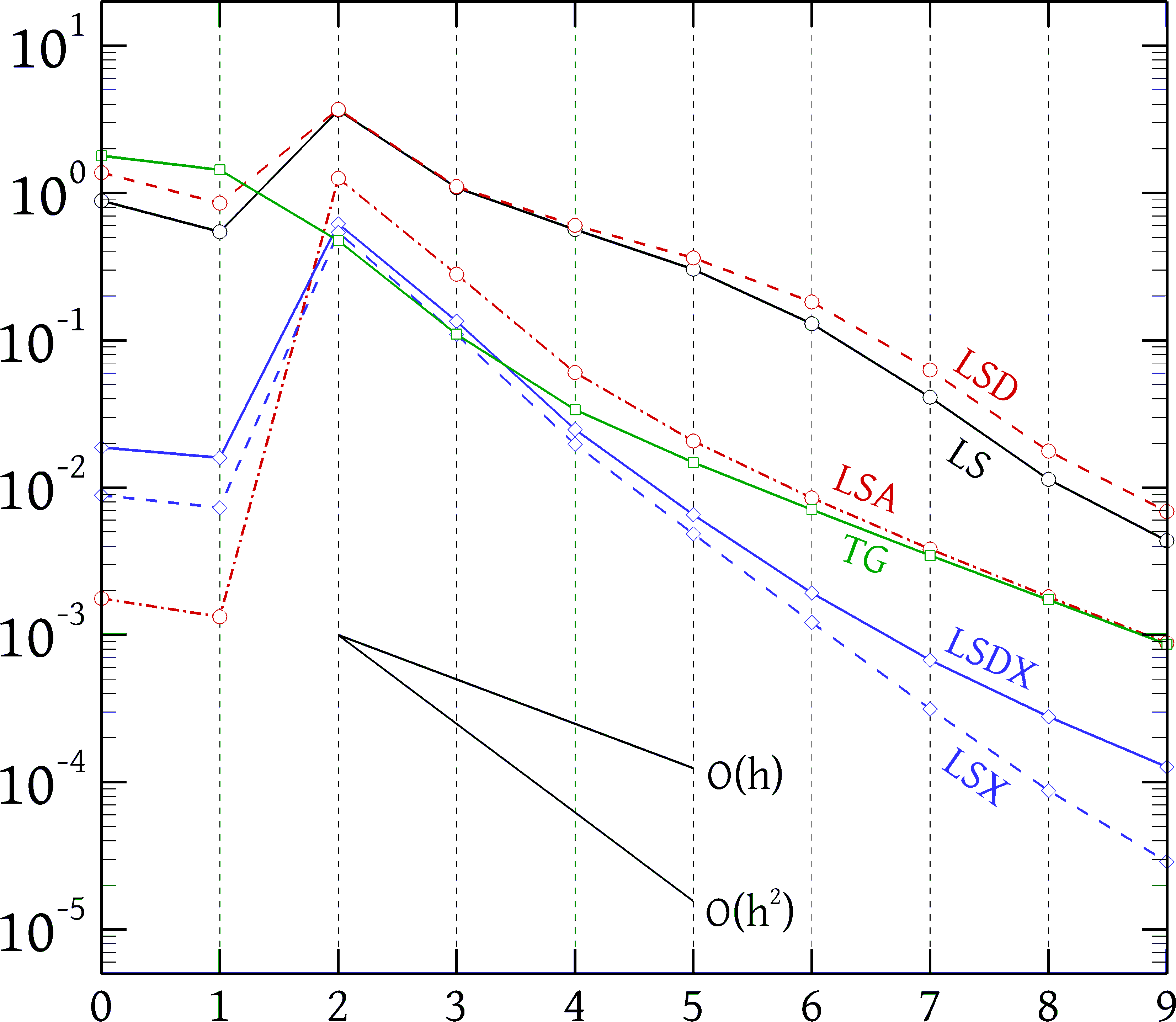}
        \caption{Randomly triangulated HARCO grids}
        \label{sfig: HARCO triR q3 R} 
    \end{subfigure}
    \caption{Mean normalised errors of various gradients with $q = 3$ weight exponent, on 
\subref{sfig: HARCO triO q3 R} orderly triangulated HARCO grids and \subref{sfig: HARCO triR q3 R}, 
as a function of the grid level. The function differentiated is the radial function \eqref{eq: test 
function radial}.}
  \label{fig: errors HARCO q3 tri}
\end{figure}

Figure \ref{fig: errors HARCO q3 tri} shows the convergence of the mean errors (normalised by $G$) 
of several gradients, all of which have a weighting exponent of $q = 3$, on the triangulated HARCO 
grids. As noted, the gradients are slightly more accurate on the orderly triangulated grids. On the 
orderly grids, we can observe the following: compared to the other gradients, the LS(3) and LSD(3) 
gradients have poor accuracy, but appear to converge at a second-order rate, in contrast with the 
LSA(3) and TG(3) gradients, which are more accurate but exhibit first-order convergence. Since, on 
this type of grid, first-order convergence is expected, presumably the convergence of the LS(3) and 
LSD(3) gradients would downgrade to first-order on even finer levels, as indeed we can see happening 
on the randomly triangulated grids (Fig.\ \ref{sfig: HARCO triR q3 R}). The LSDX(3) and LSX(3) 
gradients exhibit second-order convergence on the orderly grids, like in the elliptic grid case.
On the randomly triangulated grids, the LSX(3) and LSDX(3) convergence rates are of second order on 
intermediate levels, but taper off towards first-order on the finest levels. Figure \ref{fig: errors 
HARCO q3 tri} (and also Fig.\ \ref{sfig: HARCO quad q3 R}, to be discussed later) shows that on the 
HARC and HARCO grids the asymptotic convergence rates may not yet be attained on intermediate 
levels, and hence a comparison of the gradients on the same grid level, as in Fig.\ \ref{fig: errors 
HARC HARCO mean l6}, may not be completely fair. Finally, we note that there are no visible signs of 
arithmetic instabilities in Fig.\ \ref{fig: errors HARCO q3 tri}, even at the finest levels (all 
calculations were performed in 8-byte arithmetic). As will be shown in Sec.\ \ref{ssec: stability}, 
the instability on the quadrilateral grids is due to the closeness and alignment of neighbour 
centroids in the radial direction, conditions that do not pertain in the case of triangulated grids.

Further tests were conducted where the differentiated function varies in the azimuthal direction; 
for example, in a boundary layer flow simulation the velocity would vary predominantly in the 
radial direction whereas the pressure would vary mostly in the azimuthal direction. The function 
differentiated is
%^c
\begin{equation} \label{eq: test function azimuthal}
 \phi(\theta) \;=\; \tanh \left( f(\theta) \right)
 \qquad \text{where} \qquad
 f(\theta) \;=\; f_{\min} \;+\; (f_{\max} - f_{\min}) 
                                \frac{\theta - \theta_{\min}}{\theta_{\max} - \theta_{\min}}
\end{equation}
%^c
where $f_{\min} = 1$ and $f_{\max} = 3$ as before, while $\theta_{\min} = - 0.256$ \si{rad} and 
$\theta_{\max} = + 0.256$ \si{rad} are the extents of the domain in the azimuthal direction. Thus, 
since our grids have an equal number of cells in the radial and azimuthal directions, again the 
value of $\phi$ given by \eqref{eq: test function azimuthal} varies from $\tanh(1)$ to $\tanh(3)$ 
across the same number of cells (1024 for $l = 9$) as when $\phi$ was given by \eqref{eq: test 
function radial}. Due to the high aspect ratio though, the distance over which $\phi$ given by 
\eqref{eq: test function azimuthal} varies is $A = 1000$ times larger than that over which function 
\eqref{eq: test function radial} varies, which means that $\nabla \phi$ of \eqref{eq: test function 
azimuthal} is about $A = 1000$ times smaller than that of \eqref{eq: test function radial}. In 
particular, the average gradient $G$ is now defined as $G = (\tanh(3) - \tanh(1)) / ((\theta_{\max} 
- \theta_{min})R) = 0.456$, which is about 1000 times smaller than for $\phi$ given by \eqref{eq: 
test function radial}.

\begin{figure}[tb!]
    \centering
    \begin{subfigure}[b]{0.99\textwidth}
        \centering
        \includegraphics[width=0.99\linewidth]{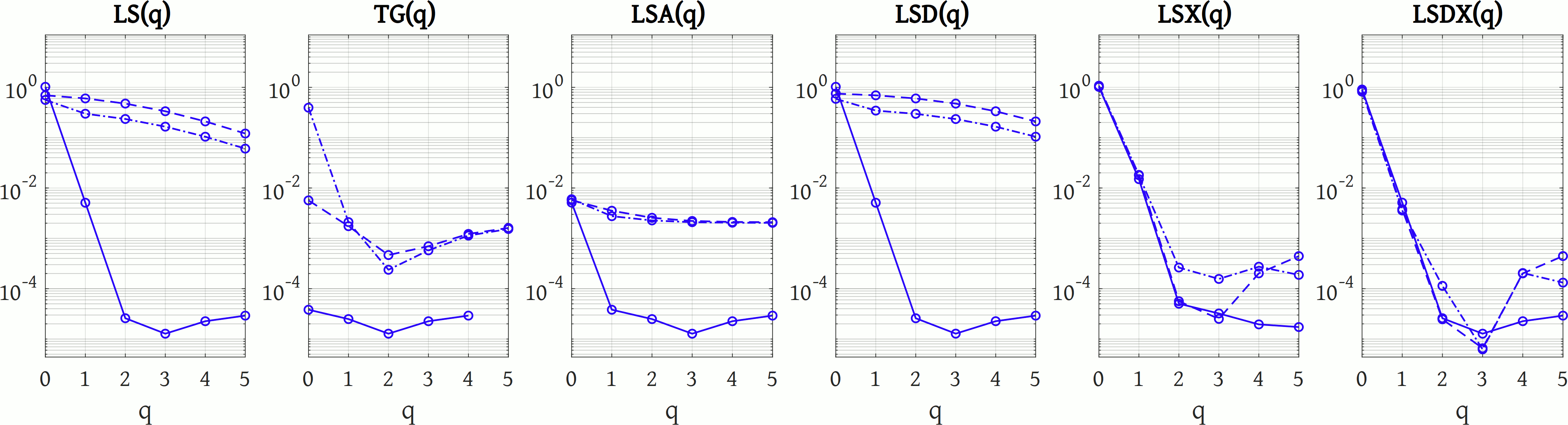}
        \caption{HARC meshes}
        \label{sfig: errors HARC mean l6 C}
    \end{subfigure}
    \\[0.5cm]
    \begin{subfigure}[b]{0.99\textwidth}
        \centering
        \includegraphics[width=0.99\linewidth]{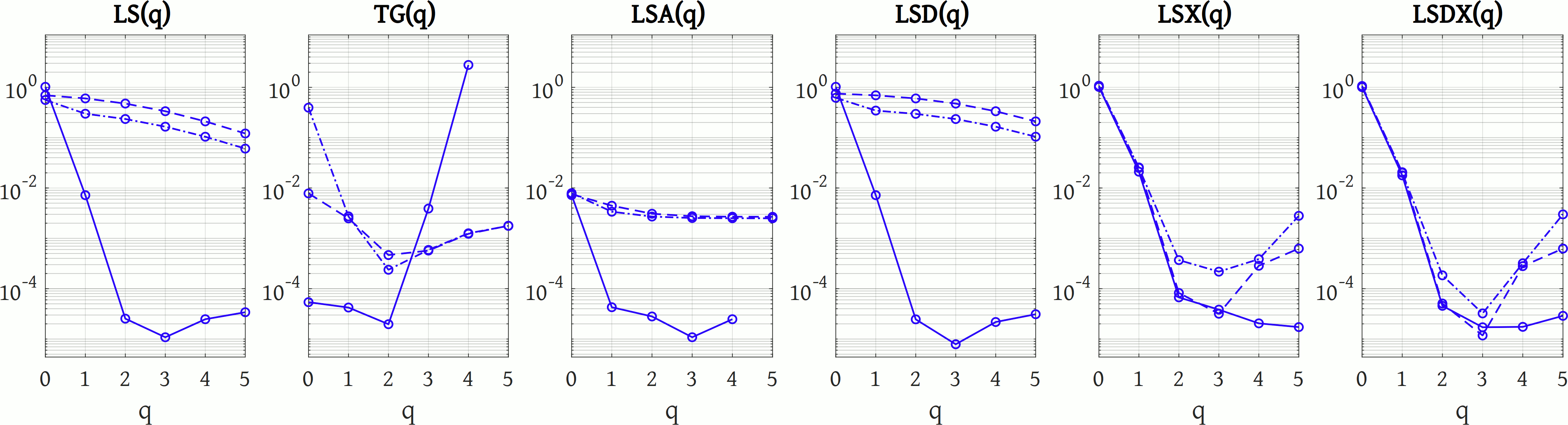}
        \caption{HARCO meshes}
        \label{sfig: errors HARCO mean l6 C} 
    \end{subfigure}
    \caption{Average normalised errors of several gradient schemes, as a function of the exponent 
$q$, for computing the gradient of the azimuthally-varying function \eqref{eq: test function 
azimuthal} on the $l=6$ HARC \subref{sfig: errors HARC mean l6} and HARCO \subref{sfig: errors 
HARCO mean l6} grids. The errors on structured grids are drawn with continuous lines, those on 
orderly triangulated grids with dashed lines, and those on randomly triangulated grids with 
dash-dot 
lines.}
  \label{fig: errors HARC HARCO mean l6 C}
\end{figure}

Figure \ref{fig: errors HARC HARCO mean l6 C} shows the normalised (by the new $G$) errors on level 
$l = 6$ of the HARC and HARCO meshes. Results are given only for the cases where the 8-byte 
calculations are arithmetically stable. This appears to be an easier test case for the gradients, 
as the normalised errors are about an order of magnitude smaller than those for the radial function 
\ref{eq: test function radial}.  The accuracy improvement compared to the differentiation of the 
radial function is particularly large for the extended-stencil gradients LSX and LSDX with high $q$ 
exponents on the randomly triangulated grids. On the other hand, the accuracy of the 
nearest-neighbour stencil gradients LS and LSD on the triangle grids is still terrible.

\subsection{Numerical stability}
\label{ssec: stability}

\begin{figure}[tb!]
    \centering
    \begin{subfigure}[b]{0.49\textwidth}
        \centering
        \includegraphics[width=0.99\linewidth]{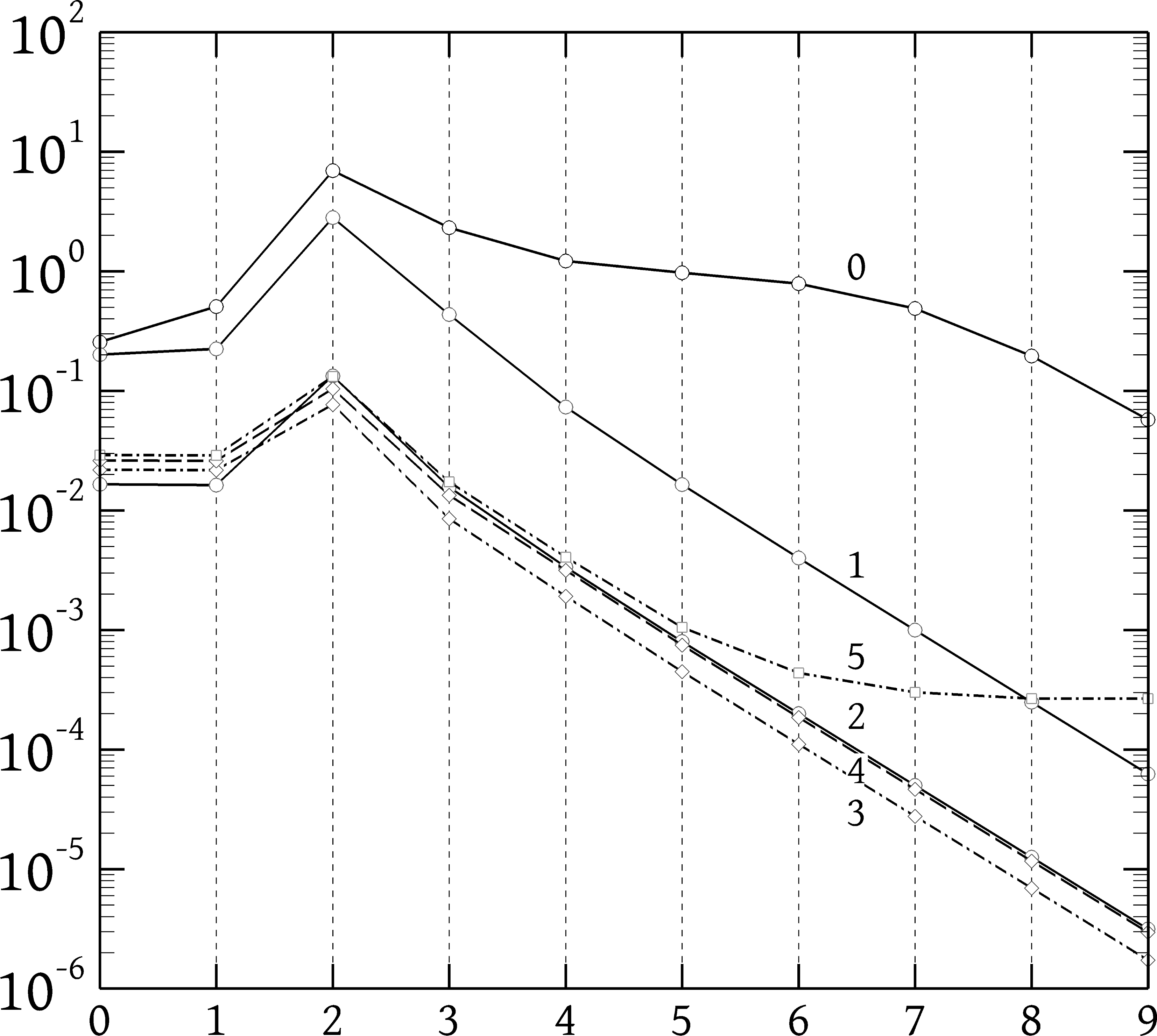}
        \caption{}
        \label{sfig: HARC LS errors NA}
    \end{subfigure}
    \begin{subfigure}[b]{0.49\textwidth}
        \centering
        \includegraphics[width=0.99\linewidth]{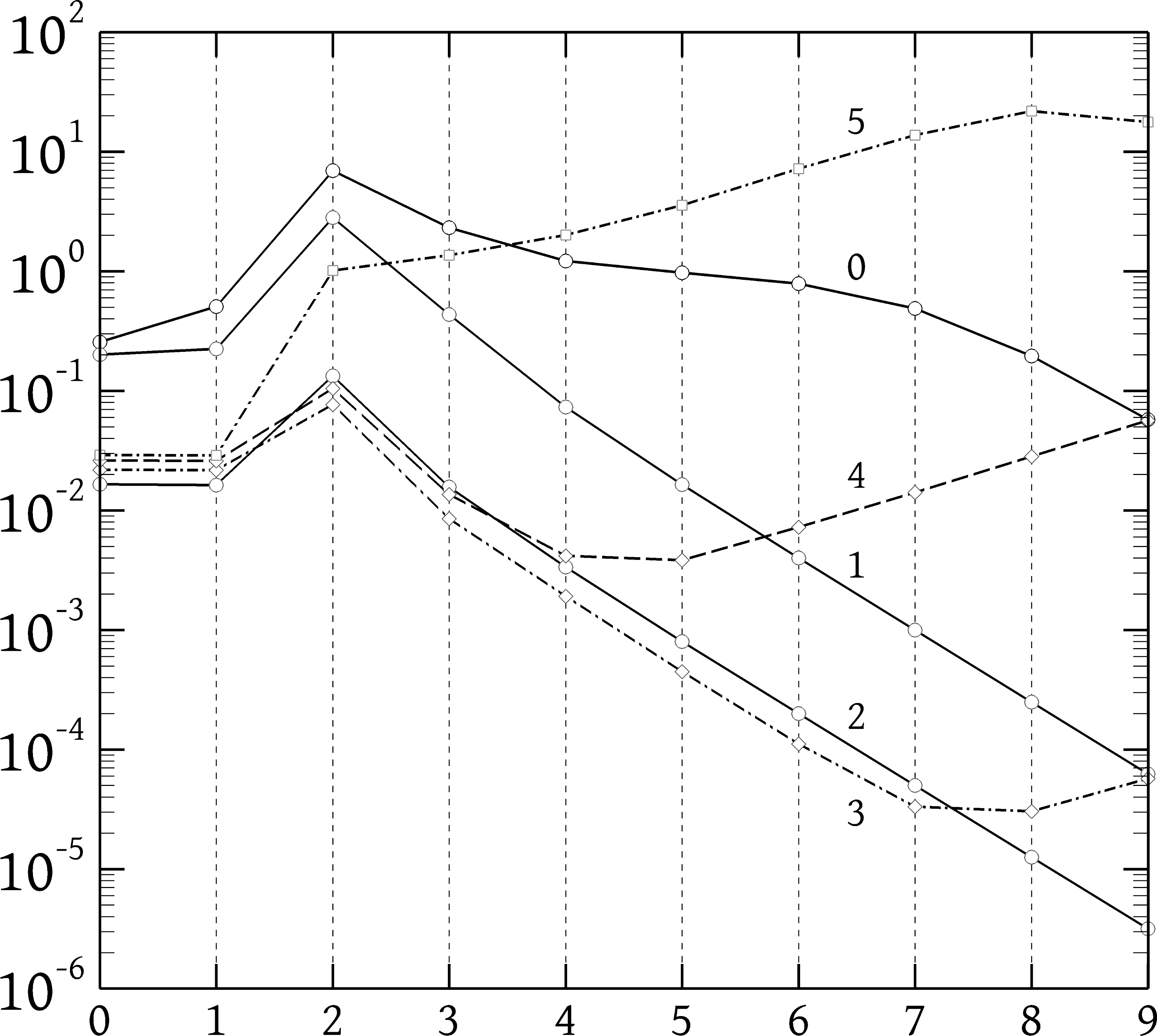}
        \caption{}
        \label{sfig: HARC LS errors G} 
    \end{subfigure}
    \caption{Mean normalised errors of the LS($q$) gradients, with $q$ indicated on each curve, as 
a function of the refinement level $l$, on the HARC quadrilateral grids. \subref{sfig: HARC LS 
errors NA} Cell centroids calculated as simple averages of the cells' nodes. \subref{sfig: HARC LS 
errors G} Cell centroids calculated via formulae \eqref{eq: centroid}. The function \eqref{eq: test 
function radial} is differentiated. All calculations were performed in 8-byte precision.}
  \label{fig: HARC quad errors NA G}
\end{figure}

As mentioned, the results plotted in Fig.\ \ref{fig: HARC quad errors NA G} revealed that some 
gradients exhibited a numerical instability at finer meshes. The instability disappeared when the 
calculations were performed in higher precision arithmetic, which indicates that roundoff errors 
play a role. Factors that were observed to either trigger or aggravate the instability are the grid 
fineness and the weighting exponent $q$. Of course, it is reasonable to assume that the high aspect 
ratio is also a necessary ingredient, since no such instabilities were observed in the tests of 
Sec.\ \ref{sec: tests}.

When it comes to least-squares methods, it is well known \cite{Trefethen_1997} that solution of the 
over-determined system by means of the normal equations, which is what is practiced in the present 
work, can lead to numerical instabilities. For orthogonal projections (Eq.\ \eqref{eq: projection 
orthogonal})) our solution method, Eq.\ \eqref{eq: projected system solution M}, gives $\mf{x} = 
(\Tr{\mf{A}}\mf{A})^{-1} \Tr{\mf{A}} \mf{y}$. These operations return $\mf{x}$ but can amplify 
round-off errors by a factor of $\kappa(\mf{A})^2$, where $\kappa(\mf{A})$ is the condition number 
of the matrix $\mf{A}$, so depending on this number the error of the calculation can be very large 
\cite{Trefethen_1997}. The remedy, in the orthogonal projection case, is to do a QR or SVD 
decomposition of the matrix $\mf{A}$ which leads to alternative algorithms \cite{Trefethen_1997} 
that do not solve the normal equations but involve cancellations between orthogonal matrices done 
beforehand in exact arithmetic (not applicable in oblique projection cases). These algorithms 
amplify roundoff errors in $\mf{A}$ and $\mf{y}$ only by a factor $\kappa(\mf{A})$ and are therefore 
much more stable.

The effect of the condition number on the accuracy of least squares gradients has not been 
thoroughly investigated in the literature, but some studies do exist. In \cite{Jalali_2013} a 
least-squares procedure was devised to construct gradients of first- and higher-order accuracy, 
using SVD; the higher-order variants had much larger condition numbers (up to $10^{12}$ reported for 
aspect ratios of 1000), but this could be brought down significantly by scaling the columns of the 
matrix $\mf{A}$. Seo et al.\ \cite{Seo_2020} compared what is effectively our LS(2) and LSX(2) 
gradients, computed via solution of the normal equations, against a couple of Green-Gauss gradients. 
Tests were performed on grids of aspect ratios up to 500, and it was found that there is a positive 
correlation between the condition number $\kappa(\Tr{\mf{A}} \mf{A}) \sim \kappa(\mf{A})^2$ and the 
gradient error. Condition numbers of nearly up to 20,000 were reported for the product $\Tr{\mf{A}} 
\mf{A}$ of the LS(2) gradient, but those of the LSX(2) gradient were much smaller; this gave rise to 
the idea of using $\kappa(\Tr{\mf{A}} \mf{A})$ as a criterion for switching between the cheaper 
LS(2) gradient and the more expensive LSX(2) gradient.

To find out whether our instability is related to the normal equations, we repeated the LSA gradient 
calculations (the most unstable ones) using the SVD algorithm instead. The DGELSD subroutine of the 
LAPACK \cite{LAPACK} library was used for this purpose; it returns also the singular values of 
$\mf{A}$, from which $\kappa(\mf{A})$ can be easily computed as the ratio of largest to smallest 
singular value. The LSA gradient can be brought into the form \eqref{eq: projection orthogonal} with 
$\mf{A} = \mf{D}^{\frac{1}{2}} \mf{R}$ and $\mf{y} = \mf{D}^{\frac{1}{2}} \mf{b}$, in the notation 
of Sec.\ \ref{ssec: linear algebra} (Eqs.\ \eqref{eq: overdetermined system ME} and \eqref{eq: 
overdetermined system M}), where $\mf{D}^{\frac{1}{2}}$ is a diagonal matrix with elements $D_{ff} = 
(\|\vf{S}_f\| / \|\vf{R}_f\|^q)^{\frac{1}{2}}$. Unfortunately, using the SVD did not improve the 
stability: the results were exactly the same as for the normal equations. Furthermore, while the 
condition numbers $\kappa(\mf{A})$ were indeed found to increase with increasing $q$ (e.g.\ 
$\kappa(\mf{A})$ has mean values of roughly 1, 32 and 1000 for $q$ = 1, 2 and 3, respectively, and 
the maximum $\kappa(\mf{A})$ are not much larger than the mean), they were not large enough to 
explain the instability (even the LSA(3) became unstable under certain conditions, described 
below). Finally, the condition numbers were found to be nearly independent of the grid fineness, 
whereas the instability shows a strong dependence on it.

So, the origin of the instability must be sought elsewhere. In fact, experiments showed a very 
strong dependence on the algorithm used to calculate the cell centroids. For all the experiments of 
Sec.\ \ref{sec: tests}, and for the triangular grid experiments of Sec.\ \ref{sec: tests hard}, the 
coordinates of the cell centroids $\vf{C}_0 = (x_0, y_0)$ were calculated with formulae derived 
using the Gauss theorem:
\begin{equation} \label{eq: centroid}
 x_0 \;=\; \frac{1}{\Omega_0} \iiint_{\Omega_0} x \, \mathrm{d}\Omega
     \;=\; \frac{1}{\Omega_0} \iint_{S_0} \frac{x^2}{2} \, \hat{\vf{e}}_x \cdot \mathrm{d} \vf{S}
     \;=\; \frac{1}{6\Omega_0} \sum_{f=1}^4 (x_{f,1}^2 + x_{f,1} x_{f,2} + x_{f,2}^2) 
           \hat{\vf{e}}_x \cdot \vf{S}_f
\end{equation}
and similarly for $y_0$, where $\Omega_0$ and $S_0$ are the volume and bounding surface of the 
cell, $\hat{\vf{e}}_x$ is the unit vector in the $x$ direction, and $x_{f,1}$ and $x_{f,2}$ are the 
$x$-coordinates of the two vertices of face $f$. Another method that was tried was simply averaging 
the coordinates of the cell's vertices. Strictly speaking, this is not equal to the geometric 
centroid, but the two converge with grid refinement on structured quadrilateral grids such as the 
present ones. The great difference in stability resulting from these two methods can be seen in 
Fig.\ \ref{fig: HARC quad errors NA G}. The HARC and HARCO quadrilateral grid results of Sec.\ 
\ref{ssec: tests hard} were obtained with the simple averaging method. Yet another method that was 
tried was splitting each quadrilateral into two triangles, and computing the centroid as an 
area-weighted average of the centroids of the two triangles. In exact precision arithmetic, this 
method would give exactly the same result as formula \eqref{eq: centroid}. This method resulted in 
better stability than Eq.\ \ref{eq: centroid}, but worse than simple averaging.

To provide a better feel for the potential severity of this instability, Fig.\ \ref{fig: errors 
HARCO q3 quad} shows the errors of all gradients, with $q=3$, on the HARCO quadrilateral grids, for 
both the radial and azimuthal functions, with centroids calculated via \eqref{eq: centroid}. To 
better test the stability of the gradients, two additional refinement levels are included ($l = 10$ 
and $11$, with $2048$ and $4096$ cells in each direction, respectively). Both 8- and 10-byte 
calculations are included, so that the effects of the instability can be evident. The instability is 
more prominent in the radial function case, but the azimuthal function case is not immune to it 
either. In the radial function case, the 8-byte LSA gradient becomes unstable already at level $l = 
4$, and even the 10-byte version does not manage to remain stable all the way down to the finest 
grid. All the other gradients become unstable at $l = 8$, except the TG gradient which, although 
stable, has a huge error.

\begin{figure}[tb!]
    \centering
    \begin{subfigure}[b]{0.49\textwidth}
        \centering
        \includegraphics[width=0.99\linewidth]{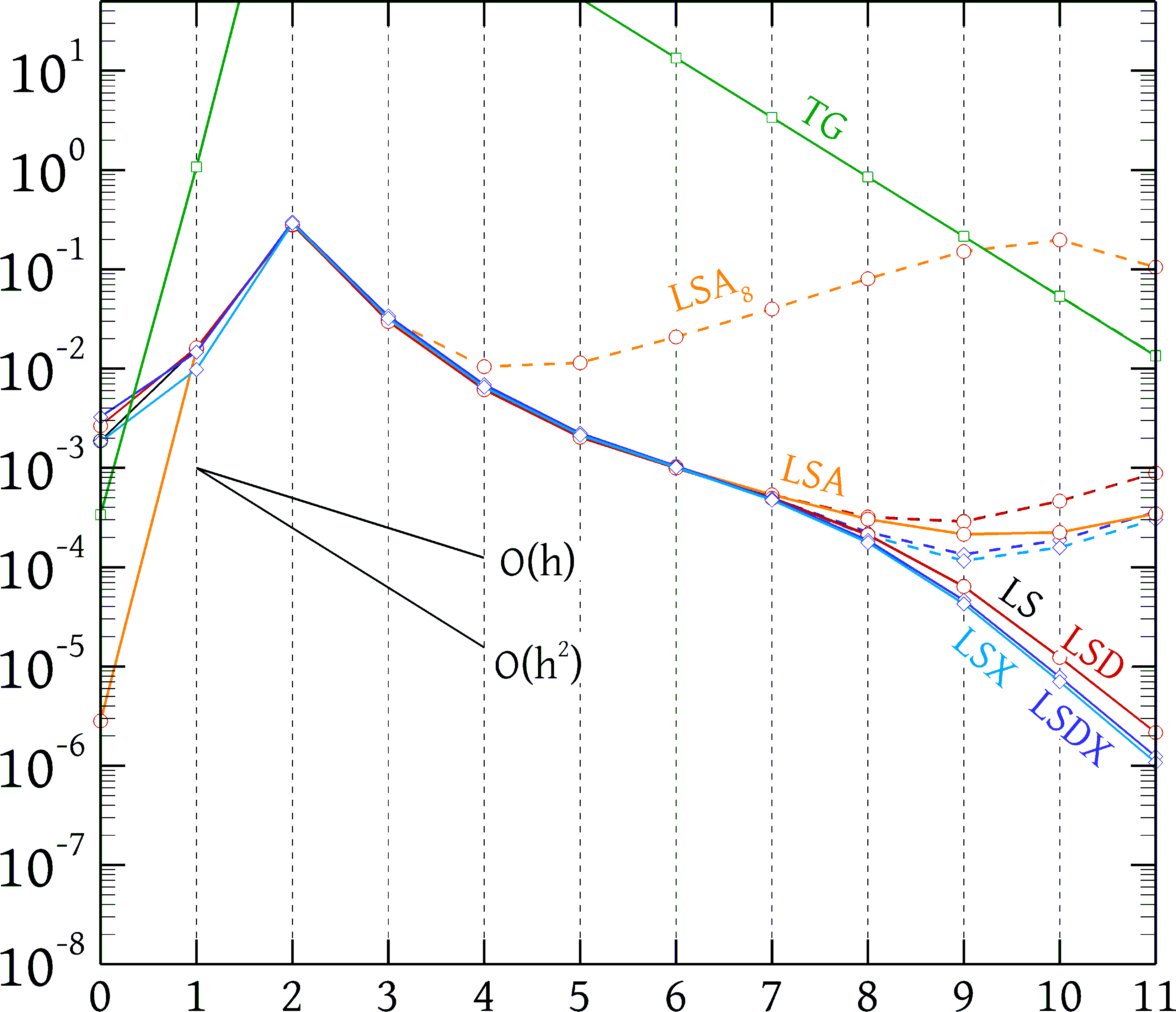}
        \caption{Radial function \eqref{eq: test function radial}}
        \label{sfig: HARCO quad q3 R}
    \end{subfigure}
    \begin{subfigure}[b]{0.49\textwidth}
        \centering
        \includegraphics[width=0.99\linewidth]{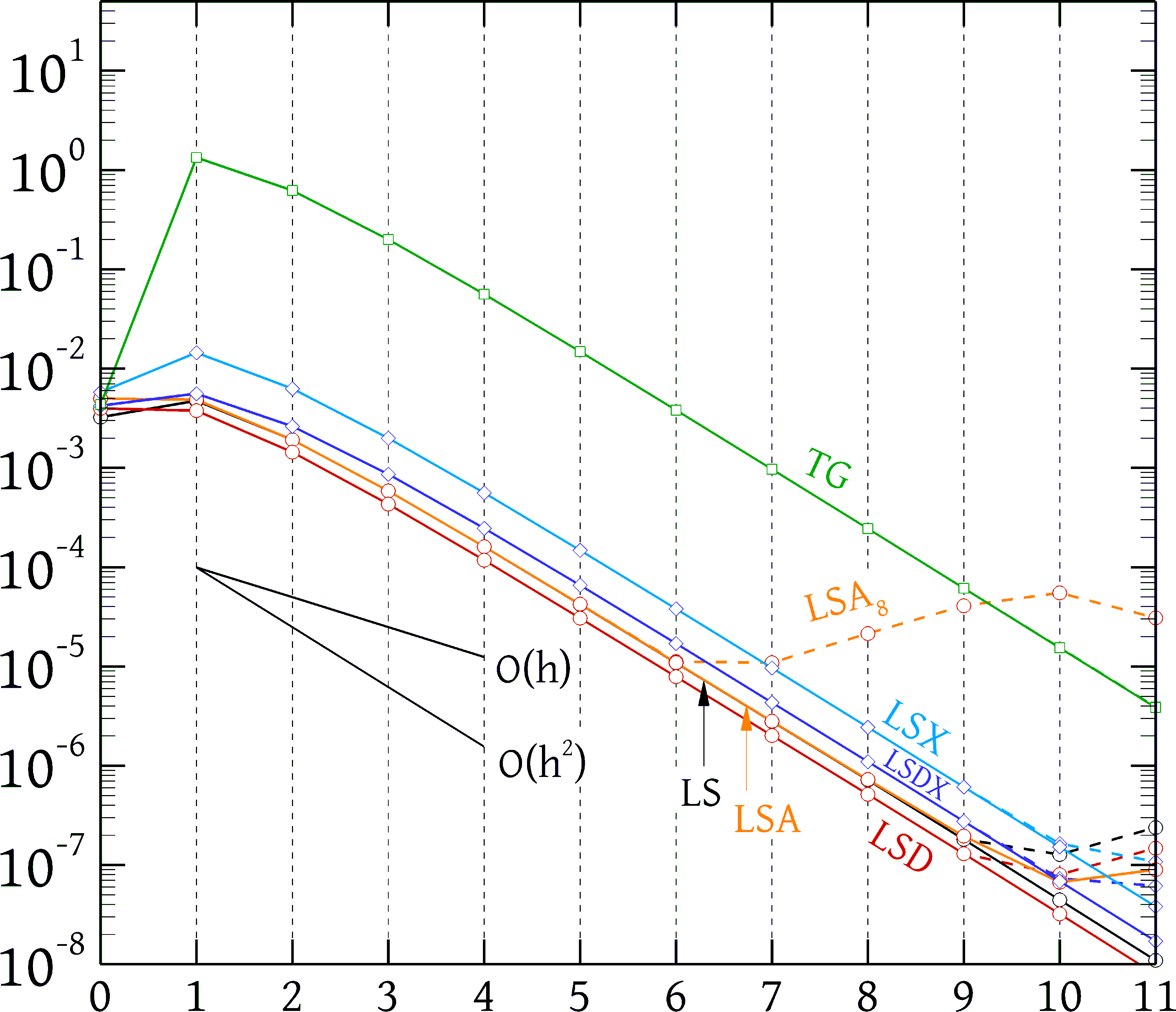}
        \caption{Azimuthal function \eqref{eq: test function azimuthal}}
        \label{sfig: HARCO quad q3 C} 
    \end{subfigure}
    \caption{Mean normalised errors of the LS(3), LSD(3), LSA(3), TG(3), LSX(3) and LSDX(3) 
gradients on the quadrilateral HARCO grids. Dashed lines: 8-byte calculations; continuous lines: 
10-byte calculations.}
  \label{fig: errors HARCO q3 quad}
\end{figure}

To locate the source of the instability, we performed downscaling tests \cite{Thomas_2008}, which 
use only a tiny fraction of the grid (the stencil of Fig.\ \ref{fig: HAR grid sketch}) and allow 
refinement down to very fine levels. The cell configuration was representative of the HARC 
quadrilateral grid, as in Fig.\ \ref{fig: HAR grid sketch}, and the radial function was considered. 
The findings that helped pinpoint the source of the instability are the following.
\begin{enumerate}
 \item \label{enum: obs 1}
 If all calculations are performed in quadruple (16 byte) precision, then the instabilities 
vanish.
\item \label{enum: obs 2}
 The formula \eqref{eq: centroid}, in 8-byte arithmetic, which results in more intense instability, 
introduces larger misalignment between the ideally colinear nodes $\vf{C}_1$, $\vf{C}_0$ and 
$\vf{C}_3$ (Fig.\ \ref{fig: HAR grid sketch}), than the more stable node averaging method.
\item \label{enum: obs 3}
 If only the centroid calculations are performed in 8-byte precision and all other operations 
(calculation of the $\vf{R}_f$ and $\vf{V}_f$ vectors, assembly and solution of the normal 
equations etc.) are performed in 16-byte precision, then there is no improvement in stability 
compared to the case that all calculations are performed in 8-byte precision.
\item \label{enum: obs 4}
 Misalignment between the colinear centroids correlates with the production of a large spurious 
azimuthal component of the gradient, whereas the error in the radial component remains small.
\end{enumerate}

Observation \ref{enum: obs 3} is important because it means that the errors that we labeled as 
stability errors thus far, seen in Figs.\ \ref{fig: errors HARC}, \ref{fig: HARC quad errors NA G} 
and \ref{fig: errors HARCO q3 quad}, are actually not round-off errors but truncation errors, since 
they occur even if all gradient-related operations are essentially exact (quadruple precision). 
Calculation of the cell centroids cannot be considered to be part of the gradient calculation. If 
the gradient schemes have a small truncation error then they should accurately compute the gradient 
whether the centroids are aligned or not. So, all these observations taken together point to the 
following explanation for the instabilities.

\begin{figure}[thb]
 \centering
 \includegraphics[scale=1]{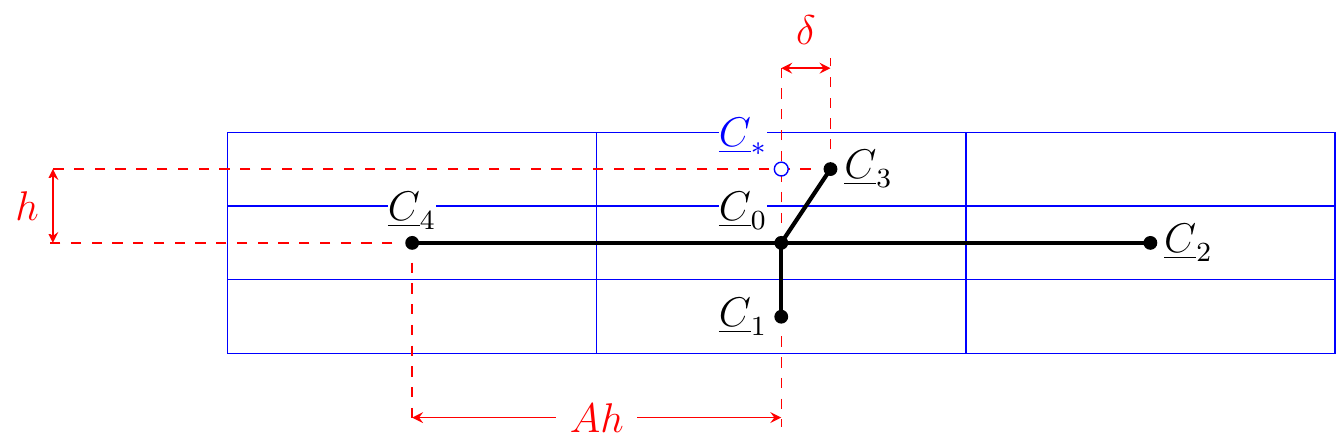}
 \caption{A model stencil for a theoretical analysis of the instability.}
 \label{fig: stability sketch}
\end{figure}

To keep things simple, consider the model stencil of Fig.\ \ref{fig: stability sketch}. Grid 
curvature is not necessary for the instability to occur, so it has been omitted. We have perturbed 
only one centroid, $\vf{C}_3$, a distance $\delta$ horizontally. The $\vf{R}_f$ vectors are 
therefore $\vf{R}_1 = (0, -h)$, $\vf{R}_2 = (Ah,0)$, $\vf{R}_3 = (\delta, h)$ and $\vf{R}_4 = 
(-Ah,0)$. Hence the $\mf{R}$ and $\mf{V}$ matrices for the LS($q$) gradient are:
\begin{equation} \label{eq: analysis R}
 \mf{R} \;=\; 
 \Tr{
 \begin{bmatrix}
    0  &  Ah  &  \delta  &  -Ah  \\
   -h  &   0  &  h       &    0
 \end{bmatrix}
 }
\end{equation}
\begin{equation} \label{eq: analysis V}
 \mf{V} \;=\;
 \Tr{
 \begin{bmatrix}
    0        &  A^{1-q} h^{1-q}  &  \delta h^{-q}  &  -A^{1-q} h^{1-q} \\
   -h^{1-q}  &  0                &  h^{1-q}        &  0
 \end{bmatrix}
 }
\end{equation}
where, for simplicity, we assumed in $\vf{V}_3$ that $\|\vf{R}_3\| \approx h$.

Now, assume that we are differentiating a radially-varying function such as \eqref{eq: test function 
radial}, with the origin of the polar coordinate system located on the line joining $\vf{C}_0$ and 
$\vf{C}_1$, as in Fig.\ \ref{fig: HAR grid sketch}. Then $\phi_2 = \phi_4$ and the vector $\mf{b}$ 
in Eq.\ \eqref{eq: projected system solution M} is ($\phi_f$ being short for $\phi(\vf{C}_f)$):
\begin{equation} \label{eq: analysis b}
  \mf{b} \;=\; 
  \Tr{
  \left[ \; \phi_1 - \phi_0 \quad 0 \quad \phi_3 - \phi_0 \quad 0 \; \right]
  }
\end{equation}
Then, the solution \eqref{eq: projected system solution M} to the system \eqref{eq: projected system 
M} is:
\begin{align}
 \label{eq: analysis x1}
 \mathsf{x}_1 \;&=\; \frac{\delta (\phi_3 -2\phi_0 + \phi_1)}{4A^{2-q}h^2 + \delta^2}
 \\[0.25cm]
 \label{eq: analysis x2}
 \mathsf{x}_2 \;&=\; \frac{4A^{2-q}h^2}{4A^{2-q}h^2 + \delta^2} \; \frac{\phi_3-\phi_1}{2h}
            \;+\;
            \frac{\delta^2} {4A^{2-q}h^2 + \delta^2} \; \frac{\phi_0-\phi_1}{h}
\end{align}
($\mathsf{x}_1$ and $\mathsf{x}_2$ must not be confused with the $x$ coordinate; they are the two 
components of the vector $\mf{x}$ of Eq.\ \eqref{eq: projected system solution M} which are intended 
to hold approximations to $\partial \phi / \partial x$ and $\partial \phi / \partial y$, 
respectively). It is noted that if $\phi$ were a linear function, $\phi = \phi_0 + Gy$ (it must have 
this form for $\phi(\vf{C}_2)$ to equal $\phi(\vf{C}_4)$), then $\mathsf{x}_1 = 0$ and $\mathsf{x}_2 
= G$, which are the exact values. If $\phi$ is a non-linear, radial function ($\phi = \phi(r)$) but 
$\delta = 0$, then $\mathsf{x}_1 = 0$ (the exact value) and $\mathsf{x}_2 = (\phi_3 - \phi_1) / 2h$ 
(a very good approximation).

Now suppose that $\phi = \phi(r)$ and $\delta \neq 0$. The value of $\mathsf{x}_2$ can be seen from 
\eqref{eq: analysis x2} to lie between $(\phi_3 - \phi_1) / 2h$ and $(\phi_0 - \phi_1) / h$; when 
$\delta$ is small such that $\delta^2 \ll 4A^{2-q} h^2$, then $\mathsf{x}_2 \rightarrow (\phi_3 - 
\phi_1) / 2h$, which, in this case, is a good approximation to $\partial \phi / \partial y$. At the 
other extreme, when $\delta^2 \gg 4A^{2-q} h^2$ then $\mathsf{x}_2 \rightarrow (\phi_0 - \phi_1) / 
h$, which is still a good approximation to $\partial \phi / \partial y$. In general, no matter what 
the value of $\delta$, $\mathsf{x}_2$ will return acceptable approximations to $\partial \phi / 
\partial y$.

However, this is not the case for $\mathsf{x}_1$, the approximation to the azimuthal component of 
the gradient, which should ideally be zero. Let $\bar{\phi}_*$ be the linearly extrapolated value 
of $\phi$ at point $\vf{C}_* = (0,h)$ (Fig.\ \ref{fig: stability sketch}), assuming $\partial \phi 
/ \partial y \approx (\phi_0 - \phi_1) / h$:
\begin{equation} \label{eq: analysis phi star}
  \bar{\phi}_* \;=\; \phi_0 \;+\; \frac{\phi_0 - \phi_1}{h} \, h
               \;=\; 2\phi_0 \;-\; \phi_1
\end{equation}
Noting that $\phi_3 -2\phi_0 + \phi_1 = \phi_3 - \bar{\phi}_*$, Eq.\ \eqref{eq: analysis x1} can be 
written in a form similar to \eqref{eq: analysis x2} as
\begin{equation} \label{eq: analysis x1b}
  \mathsf{x}_1 \;=\; \frac{4A^{2-q}h^2}{4A^{2-q}h^2 + \delta^2} \; 0
               \;+\;
               \frac{\delta^2} {4A^{2-q}h^2 + \delta^2} \; \frac{\phi_3-\bar{\phi}_*}{\delta}
\end{equation}
The first term on the right-hand side is zero, but it has been included because it represents the 
finite difference $(\phi_2 - \phi_4) / (2Ah) = 0$. Thus, the value of $\mathsf{x}_1$ lies somewhere 
between the finite difference $(\phi_2 - \phi_4) / (2Ah)$ and the finite difference $(\phi_e - 
\bar{\phi}_*) / \delta$. To which one it is closer is determined by the relative magnitude of 
$\delta^2$ compared to $4A^{2-q}h^2$. Inverse-distance weighting can lead to the contributions of 
points $\vf{C}_2$ and $\vf{C}_4$ to the LS gradient being suppressed, if the aspect ratio $A$ and 
the weight exponent $q$ are large enough so that $4A^{2-q}h^2 < \delta^2$.  In that case 
$\mathsf{x}_1$ will be closer to $(\phi_e - \bar{\phi}_*) / \delta$. In the extreme case of very 
large $A$ and $q$ the contributions of $\vf{C}_2$ and $\vf{C}_4$ are completely eliminated and we 
are left only with $\vf{C}_0$, $\vf{C}_1$ and $\vf{C}_3$. These are three non-colinear points, so 
the system \eqref{eq: overdetermined system ME} is not overdetermined but has a unique solution. 
This solution is the same as that of Eqs.\ \eqref{eq: analysis x1b} and \eqref{eq: analysis x2} if 
we set $4A^{2-q}h^2 \rightarrow 0$:
\begin{equation} \label{eq: analysis x1 x2 extreme}
 \mathsf{x}_1 \;=\; \frac{\phi_3-\bar{\phi}_*}{\delta}
 \qquad\qquad
 \mathsf{x}_2 \;=\; \frac{\phi_0-\phi_1}{h}
\end{equation}
As mentioned, the error is small for $\mathsf{x}_2$. However, it can be huge for $\mathsf{x}_1$; the 
problem is that, if $\delta \rightarrow 0$, $\phi_3$ tends to the value $\phi(\vf{C}_*)$, which is 
different from $\bar{\phi}_*$ because the latter is just an extrapolated value. Hence, as $\delta 
\rightarrow 0$ the difference $\phi_3-\bar{\phi}_*$ remains finite and the ratio $\mathsf{x}_1 = 
(\phi_3-\bar{\phi}_*) / \delta$ goes to infinity.

For a quantitative estimate of this error, we can expand $\phi(r)$ as Taylor series in $(x,y)$ 
coordinates around point $\vf{C}_0$. If $\partial \phi / \partial x = 0$ and $\partial \phi / 
\partial y = G$ is the exact gradient, it is not hard to show\footnote{One needs that, at 
$\vf{C}_0$, $\partial r/ \partial x = 0$, $\partial r / \partial y = 1$, $\partial^2 r / \partial 
x^2 = 1/r$, $\partial^2 r / \partial y^2 = 0$, and $\partial^2 r / \partial x \partial y = 0$; $r 
= \sqrt{x^2 + y^2}$.} that also $\partial^2 \phi / \partial x^2 = G/r$, $\partial^2 \phi / \partial 
y^2 = \mathrm{d}^2 \phi / \mathrm{d}r^2$ and $\partial^2 \phi / \partial x \partial y = 0$ (all 
derivatives evaluated at $\vf{C}_0$). Expanding $\phi_1$ and $\phi_3$ thus we obtain
\begin{align}
  \label{eq: analysis Taylor phi star}
  \bar{\phi}_* \;&=\;
    \phi_0 \;+\; Gh \;-\; \frac{1}{2} \phi'' h^2
\\[0.25cm]
  \label{eq: analysis Taylor phi 3}
  \phi_3 \;&=\; 
    \phi_0 \;+\; Gh \;+\; \frac{1}{2} \phi'' h^2 \;+\; \frac{1}{2r} G \delta^2
\end{align}
where $\phi'' = \partial^2 \phi / \partial r^2$. It is obvious that these two values do not converge 
if $\delta \rightarrow 0$. Substituting in \eqref{eq: analysis x1 x2 extreme} we get for 
$\mathsf{x}_1$:
\begin{equation} \label{eq: analysis x1 extreme Taylor}
  \mathsf{x}_1 \;=\; \frac{\phi''h^2}{\delta} \;+\; \frac{G \delta}{2r}
\end{equation}
If $\delta \rightarrow 0$ then the second term on the right-hand side tends to zero, but the first 
term tends to infinity, hence we get very large errors in the azimuthal component of the gradient. 
It is noteworthy that, perhaps counterintuitively at first sight, the smaller $\delta$ is, i.e.\ 
the better the alignment of the cell centroids, the larger the error; it is only when $\delta$ is 
exactly zero that the error becomes zero too. In practice though, we observed that better alignment 
of the centroids results in smaller error (e.g.\ node averaging vs.\ the formulae \eqref{eq: 
centroid}). This is because if $\delta$ becomes too small then the assumption on which Eq.\ 
\eqref{eq: analysis x1 extreme Taylor} is based, that $\vf{C}_2$ and $\vf{C}_4$ do not contribute to 
the gradient, is invalidated. As seen from Eq.\ \eqref{eq: analysis x1b}, formula \eqref{eq: 
analysis x1 extreme Taylor} is an accurate approximation to the solution $\mathsf{x}_1$ if $\delta^2 
\gg 4A^{2-q}h^2$. So, the peak error would be expected to occur roughly at a value $\delta_*$ for 
which $\delta_*^2 = 4A^{2-q}h^2$, or
\begin{equation} \label{eq: analysis delta star}
 \delta_* \;=\; 2Ah A^{-q/2}
 \qquad
 \text{or}
 \qquad
 \tilde{\delta}_* \;\equiv\; \frac{\delta_*}{2Ah} \;=\; A^{-q/2}
\end{equation}
where $\tilde{\delta}_*$ is a non-dimensional version of $\delta_*$ that compares it to the 
distance $2Ah$ between $\vf{C}_2$ and $\vf{C}_4$. If $\delta$ is increased beyond $\delta_*$ then 
equation \eqref{eq: analysis x1 extreme Taylor} describes the error well, but since $\delta$ is 
increasing, the error is falling; on the other hand, if $\delta$ is decreased below $\delta_*$, 
then Eq.\ \ref{eq: analysis x1 extreme Taylor} does not describe the error accurately, and 
$\mathsf{x}_1$ becomes increasingly dominated by the finite difference $(\phi_2 - \phi_4) / (2Ah)$, 
which has a small error (zero, in this case).

This theoretical analysis is confirmed by the downscaling tests. Figure \ref{fig: stability 
analysis} shows results of such tests, which include grid curvature as in Fig.\ \ref{fig: HAR grid 
sketch}, with $A = 1000$ and $r = 1$. The maximum errors do indeed occur at about $\delta_*$. The 
larger the values of $q$, the larger the possible errors. The errors can be huge for $q = 4$ and 
$5$, and large for $q=3$. Use of $q \leq 2$ is recommended for safety. The errors decrease with 
grid refinement; in Figs.\ \ref{fig: HARC quad errors NA G} and \ref{fig: errors HARCO q3 quad} the 
errors are seen to actually increase with refinement up to a certain grid level, and decrease 
subsequently. This can be attributed to the fact that grid refinement also affects $\delta$, and 
can bring it closer to $\delta_*$.

\begin{figure}[tb!]
    \centering
    \begin{subfigure}[b]{0.49\textwidth}
        \centering
        \includegraphics[width=0.99\linewidth]{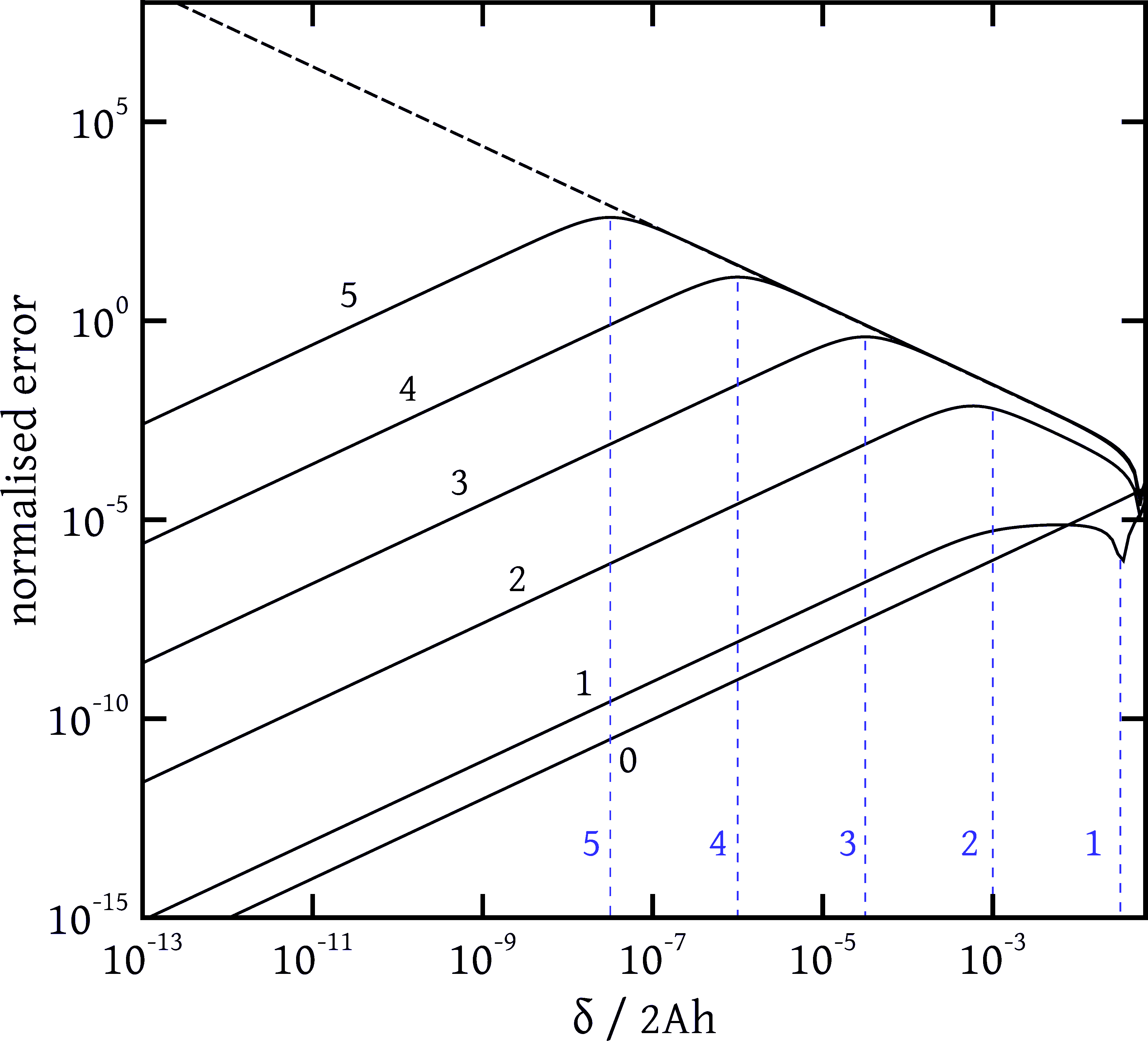}
        \caption{$l = 5$}
        \label{sfig: stability l05}
    \end{subfigure}
    \begin{subfigure}[b]{0.49\textwidth}
        \centering
        \includegraphics[width=0.99\linewidth]{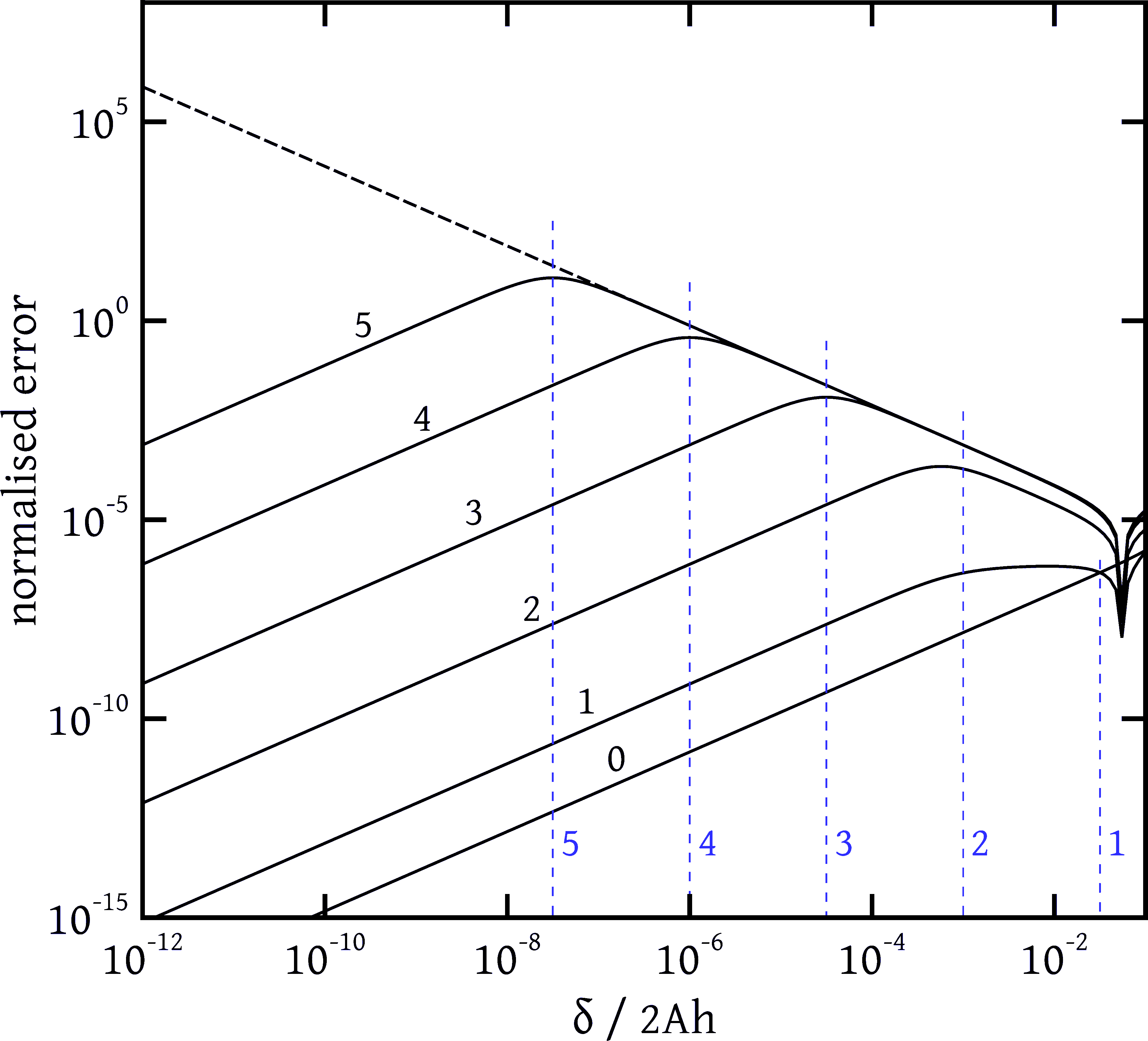}
        \caption{$l = 10$}
        \label{sfig: stability l10} 
    \end{subfigure}
    \caption{Solid lines: normalised errors ($|\mathsf{x}_1| / G$) of the LS($q$) gradients ($q$ 
shown next 
to each curve) as a function of the horizontal perturbation $\delta$ of point $\vf{C}_3$, in 
downscaling tests (Fig.\ \ref{fig: HAR grid sketch}) with $A = 1000$, at $\vf{C}_0 = (r=1, 
\theta=0)$. The function \eqref{eq: test function radial} is differentiated ($G = 1679.9$, $\phi'' 
= -1.0235\cdot 10^7$). Dashed back line: the error predicted by Eq.\ \eqref{eq: analysis x1 extreme 
Taylor}. Vertical blue dashed lines: the values of $\delta_*$ (Eq.\ \eqref{eq: analysis delta 
star}) for each $q$. Left plot \subref{sfig: stability l05}: grid level $l = 5$ ($h = 8 \cdot 
10^{-6}$). Right plot: grid level $l = 10$ ($h = 2.5 \cdot 10^{-7}$).}
  \label{fig: stability analysis}
\end{figure}

\section{Conclusions}
\label{sec: conclusions}

The current study presented a general framework for formulating gradient reconstruction schemes 
based on the solution, via projection onto a selected subspace, of an over-determined system, 
derived from Taylor series expansion of the differentiated variable in a neighbourhood of cells. 
Two particular sub-families were the focus of investigation: least-squares gradients, based on 
orthogonal projection, and Taylor-Gauss gradients, based on oblique projection on a subspace formed 
by the cells' face normal vectors, which turns out to be equivalent to a self-corrected Green-Gauss 
gradient. Of course, these sub-groups do not exhaust the general framework, and other projection 
subspaces, possibly even based on vectors that depend on the solution and vary dynamically, may be 
devised and tested in the future.

The hope going into this study was to find a gradient scheme that performs well under all 
circumstances, with the ultimate goal of finding a gradient that can be safely applied in general 
purpose finite volume solvers, without worrying about the particularities of each given problem. 
Eventually, each of the schemes tested turned out to have some advantages and disadvantages.

The LS($q$) gradients usually perform well on quadrilateral grids with $q = 2$ or $3$. On triangular 
grids, their performance suffers but can be greatly improved by extending the stencil (LSX($q$) 
gradients). On high aspect ratio quadrilateral grids, the weighting exponent $q$ should not be 
increased beyond 2, or 3 with some caution, if the risk of instability is to be avoided -- however, 
the value $q = 0$ should also not be used.

The LSA and LSD gradients were designed to offer improved performance on grids with directional 
clustering, compared to the LS gradients. The LSD gradient is slightly more expensive than the LSA, 
but it has the advantage that it can be used also with extended stencils; indeed, the LSDX gradient 
was the most accurate in most cases (followed closely by LSX). Furthermore, the LSD gradient is more 
stable than the LSA on high aspect ratio grids.

The TG and TGI gradients usually performed on par with the compact least squares gradients, LS, LSA 
and LSD, or even outperformed them (on the elliptic and locally refined grids). However, their 
inapplicability to extended stencils significantly limits their accuracy on triangular grids. 
Furthermore, they performed very poorly on the HARCO grids, unless $q = 0$ or $q = 1$ was used.

In general, on the grids of Sec.\ \ref{sec: tests} most gradients would perform satisfactorily, 
even though some are better than others. An exception may be triangular grids, on which extended 
stencils should be used. For more complex unstructured grids with varying types of cell geometry, 
the technique proposed in \cite{Seo_2020} for switching between compact and extended stencils may 
be useful. On the other hand, on high aspect ratio grids such as those of Sec.\ \ref{sec: tests 
hard}, which present extreme cases, more caution is warranted, as any one of the gradients we 
examined could fail catastrophically under some set of conditions. High aspect ratio combined with 
relatively high distance weighting exponent $q$ can give rise to the instability discussed in Sec.\ 
\ref{ssec: stability}, particularly for the LSA gradient, but also for the others. Even without 
this instability, the TG gradient fails catastrophically on quadrilateral HARCO meshes if $q > 1$. 
The same is true also of the LSX and LSDX gradients on randomly triangulated grids for $q > 3$. The 
LS and LSD gradients have quite poor accuracy on triangulated high aspect ratio grids, for all 
values of $q$, although there is a slight improvement as $q$ is increased.

Summarising, for normal aspect ratio grids such as those of Sec.\ \ref{sec: tests}, if the grid is 
composed of quadrilaterals, the TG(2) gradient may be a good choice that combines accuracy with low 
computational cost; it inherently accounts for directional clustering, and offers 2nd order accuracy 
at boundaries. If the grid is composed of triangles, then the LSDX(3) (or LSX(3)) gradient would be 
a better choice. The LSDX(3) gradient would also be a good choice for quadrilateral grids, if the 
additional computational cost can be afforded, as it proved to be the most accurate on distorted 
grids such as that of Fig.\ \ref{sfig: grid random}. For better efficiency, one could use a 
criterion to switch between compact and extended stencils such as that used in \cite{Seo_2020}. On 
high-aspect ratio grids, if they are composed of quadrilaterals, the LS(2) gradient is a good 
choice; LS(3) would be more accurate, but it is best to avoid large $q$ exponents due to the fear 
of instability. If the grid is triangular, then LSX(2) should be preferred instead. Although we did 
not test this idea, theoretically on HARC quadrilateral grids one could use the LS(3) gradient to 
benefit by its higher accuracy at boundaries, and mitigate the danger of instability by multiplying 
the weight vectors $\vf{V}_2$ and $\vf{V}_4$ (Fig.\ \ref{fig: HAR grid sketch}) by appropriate 
factors so that their magnitudes do not become too small in comparison to those of $\vf{V}_1$ and 
$\vf{V}_3$ (or, equivalently, by applying a smaller $q' < q$ exponent in the directions 2 and 4 
than the $q = 3$ exponent in the directions 1 and 3).

Finally, it must be noted that the present study is a necessary first step for assessing these 
gradient schemes for the purpose of employing them in Finite Volume Methods, but this must be 
followed by a study where they are tested in the actual FVM solution of partial differential 
equations. The present study evaluated the truncation error of the gradients schemes themselves, 
but how this will affect the truncation error of the overall FVM and the discretisation error of 
the solution remains to be seen. For example, the property of gradients such as the LS(3), TG(2) 
etc.\ that, on quadrilateral grids, they retain second-order accuracy even at boundary cells, may 
reduce the discretisation error throughout the domain, as this error is produced locally at sites of 
high truncation error and convected and diffused along with the flow \cite{Syrakos_2012, 
Syrakos_2014}. Furthermore, the choice of gradient scheme may have an impact on the iterative 
convergence of the algebraic solver \cite{Diskin_2011}. Such a study is planned for the future.

\section*{Acknowledgements}
AS, YD and JT gratefully acknowledge funding from the LIMMAT Foundation during the initial stages 
of this work, under the Project ``MuSiComPS''. We also wish to thank Ganesh Natarajan of the Indian 
Institute of Technology Palakkad for bringing to our attention the ``quasi-Green'' gradient and its 
connection to the Taylor-Gauss gradient.

\section*{Appendix}

In this appendix, it will be shown that the second term on the right-hand side of Eq.\ \eqref{eq: QG 
manipulation 1} is equal to the identity tensor. Consider a Cartesian coordinate system with origin 
at $\vf{C}_0$, with $x_i$ being the $i$-th coordinate direction and $\hat{\vf{e}}_i$ the 
corresponding unit vector. Then
%^c
\begin{equation} \label{eq: appendix derivation 1}
 \nabla \cdot (x_i \hat{\vf{e}}_j)
 \;=\;
 \delta_{ij}
 \;\Rightarrow\;
 \int_{\Omega_0} \nabla \cdot (x_i \hat{\vf{e}}_j) \mathrm{d} \Omega
 \;=\;
 \int_{\Omega_0} \delta_{ij} \mathrm{d} \Omega
 \;=\; \delta_{ij} \,\Omega_0
\end{equation}
%^c
The integral of Eq.\ \eqref{eq: appendix derivation 1} can also be evaluated using the divergence 
(Gauss) theorem:
%^c
\begin{equation} \label{eq: appendix derivation 2}
 \int_{\Omega_0} \nabla \cdot (x_i \hat{\vf{e}}_j) \mathrm{d} \Omega
 \;=\;
 \int_{S_0} x_i \hat{\vf{e}}_j \cdot \hat{\vf{n}} \, \mathrm{d}S
 \;=\;
 \sum_f \hat{\vf{e}}_j \cdot \hat{\vf{n}}_f \int_{S_f} x_i  \, \mathrm{d}S
 \;=\;
 \sum_f \hat{\vf{e}}_j \cdot \vf{S}_f R_{f,i}
\end{equation}
%^c
where $S_0$ is the surface of cell $\Omega_0$, $\mathrm{d}S$ is an infinitesimal element of that 
surface, and $\hat{\vf{n}}$ is the outward normal unit vector, which is constant and equal to 
$\hat{\vf{n}}_f$ over each face $f$. In the last equality of Eq.\ \eqref{eq: appendix derivation 2} 
we have used that $\int_{S_0} x_i \, \mathrm{d}S = S_f c_{f,i}$, with $c_{f,i}$ being the 
$i$-th component of the centroid $\vf{c}_f$, by definition of the centroid. The latter also equals 
$R_{f,i}$, the $i$-th component of $\vf{R}_f = \vf{c}_f - \vf{C}_0$, because $\vf{C}_0 = \vf{0}$ 
is the coordinates' origin . Thus Eq.\ \eqref{eq: appendix derivation 2} becomes $\sum_f S_{f,j} 
R_{f,i}$, i.e.\ the $(j,i)$ component of the tensor $\sum_f \vf{S}_f \vf{R}_f$. This is equal to 
$\delta_{ji} \Omega_P$, the $(j,i)$ component of the tensor $\Omega_P \tf{I}$, by Eq.\ \eqref{eq: 
appendix derivation 1}. Thus the tensors $\Omega_P \tf{I}$ and $\sum_f \vf{S}_f \vf{R}_f$ are 
identical, having all their components equal, which is the sought result. Placing the origin 
elsewhere ($\vf{C}_0 \neq \vf{0}$) does not change the result, because $\sum_f \vf{S}_f \vf{C}_0 = 
(\sum_f \vf{S}_f) \vf{C}_0 = \vf{0} \vf{C}_0 = \tf{0}$.



 \bibliographystyle{elsarticle-num} 
 \bibliography{newGradients}


% \end{thebibliography}

\end{document}